\documentclass{article}

\usepackage{neurips_2020}
\title{Causal mediation analysis with one or multiple mediators: a comparative study}
\usepackage{times}
\usepackage[dvipsnames]{xcolor}
 \usepackage{graphicx}
 \usepackage{placeins}
\usepackage[utf8]{inputenc} % allow utf-8 input
\usepackage[T1]{fontenc}    % use 8-bit T1 fonts
\usepackage{hyperref}       % hyperlinks
\usepackage{url}            % simple URL typesetting
\usepackage{booktabs}       % professional-quality tables
\usepackage{amsfonts}       % blackboard math symbols
\usepackage{nicefrac}       % compact symbols for 1/2, etc.
\usepackage{microtype}      % microtypography
\usepackage{multicol}
\usepackage[htt]{hyphenat}
\usepackage{array}
\usepackage{makecell}
\usepackage{longtable} % for 'longtable' environment
\usepackage{pdflscape} % for 'landscape' environment
\usepackage{multirow}
\usepackage{listings}
\usepackage{adjustbox}

\usepackage{mathrsfs}
\usepackage{tikz}

\usetikzlibrary{shapes,decorations,arrows,calc,arrows.meta,fit,positioning}
\tikzset{
    -Latex,auto,node distance =1 cm and 1 cm,semithick,
    state/.style ={ellipse, draw, minimum width = 0.7 cm},
    point/.style = {circle, draw, inner sep=0.04cm,fill,node contents={}},
    bidirected/.style={Latex-Latex,dashed},
    el/.style = {inner sep=2pt, align=left, sloped}
}
\usepackage{witharrows}

\WithArrowsOptions{tikz={font={\normalfont\footnotesize}}}

\usepackage{color, colortbl}
\usepackage{caption}
\usepackage{subcaption}
\usepackage{annotate-equations}

\usepackage{amsthm}
\theoremstyle{definition}

\newtheorem{definition}{Definition}

\newtheorem{assumption}{Assumption}

\usepackage{amsmath}

\newcommand\independent{\protect\mathpalette{\protect\independenT}{\perp}}
\def\independenT#1#2{\mathrel{\rlap{$#1#2$}\mkern2mu{#1#2}}}

\author{%
 Judith~Abécassis\\
 Soda project-team,\\Mind project-team,\\INRIA Saclay,\\Institut Polytechnique de Paris,\\France.\\
 \scriptsize{\texttt{judith.abecassis@inria.fr}}\\
 \And
  Houssam~Zenati\\
  Mind project-team,\\INRIA Saclay, CEA,\\Université Paris-Saclay,\\France.\\
 \scriptsize{\texttt{houssam.zenati@inria.fr}}\\
 \And
 Sami~Boumaïza\\
 Soda project-team,\\INRIA Saclay,\\Institut Polytechnique de Paris,\\France.\\
\And
 Julie~Josse\\
 PreMeDICaL project team,\\INRIA Sophia-Antipolis-Inserm,\\Université de Montpellier,\\France.\\
 \scriptsize{\texttt{julie.josse@inria.fr}}\\
 \And
 Bertrand~Thirion\\
 Mind project-team,\\INRIA Saclay, CEA,\\Université Paris-Saclay,\\France.\\
  \scriptsize{\texttt{bertrand.thirion@inria.fr}}\\
}

\usepackage[colorinlistoftodos,textwidth=2.1cm]{todonotes}

\begin{document}

\maketitle

{\centering{\textit{\large{Working version - \today - This paper has not been peer reviewed.}}}}\vspace{1cm}
\begin{abstract}
 Mediation analysis breaks down the causal effect of a treatment on an outcome into an indirect effect, acting through a third group of variables called mediators, and a direct effect, operating through other mechanisms.
  Mediation analysis is hard because confounders between treatment, mediators, and outcome blur effect estimates in observational studies.
  Many estimators have been proposed to adjust on those confounders and provide accurate causal estimates.
  We consider parametric and non-parametric implementations of classical estimators and provide a thorough evaluation for the estimation of the direct and indirect effects in the context of causal mediation analysis for binary, continuous, and multi-dimensional mediators. 
  We assess several approaches in a comprehensive benchmark on simulated data.
  Our results show that advanced statistical approaches such as the multiply robust and the double machine learning estimators achieve good performances in most of the simulated settings and on real data.
  As an example of application, we propose a thorough analysis of factors known to influence cognitive functions to assess if the mechanism involves modifications in brain morphology using the UK Biobank brain imaging cohort.
This analysis shows that for several physiological factors, such as hypertension and obesity, a substantial part of the effect is mediated by changes in the brain structure.
This work provides guidance to the practitioner from the formulation of a valid causal mediation problem, including the verification of the identification assumptions, to the choice of an adequate estimator.
\end{abstract}

% \begin{keywords}%
%   Causal inference, mediation analysis, non-parametric estimation, cross-fitting, multi-dimensional mediators
% \end{keywords}

\section{Introduction}

Causal inference in observational studies is primarily used to measure the causal effect of a treatment on an outcome.
Nevertheless, in a lot of fields, disentangling the mechanism of action is just as important, as it allows us to identify potential intermediate intervention targets, and more generally, deepen our understanding of the processes that lead to the observed outcome. 

Causal mediation analysis aims at separating the (total) causal effect into two components: an indirect effect through a third (group of) variable(s) called mediator(s), and a direct effect through alternative path(s)~\citep{pearl2001direct}.
A central issue in solving this question is confounding, as even in the "ideal" case of a randomized controlled trial, only the exposure is randomized, and not the mediator.
Further control on the mediator is thus necessary to identify the direct and indirect effects, as outlined in~\cite{robins1992identifiability} in the potential outcomes framework~\citep{imbens2015causal}.
Estimation in causal mediation analysis aims to provide a robust statistical framework for adjusting for confounding variables by accounting for their relationships with the treatment, the mediator, and the outcome, respectively.
Previous work on mediation was mostly based on parametric structural equations~\citep{cochran1957analysis,judd1981process,baron1986moderator}, that may not fully account for the confounders' impact, and neglected identifiability assumptions.
Further work has clarified identification assumptions and developed new estimation approaches in parametric and non-parametric settings~\citep{robins1992identifiability,pearl2001direct,robinssemantics,petersen2006estimation,imai2010identification,hong2010ratio,albert2011generalized,tchetgen2012semiparametric,huber2014identifying,zheng2012targeted,farbmacher2020causal,hines2021robust}.
This has led to the development of more complex estimators, that proceed in two steps:
first they fit intermediate variables through classification or regression models based on the data, and then they use the predictions of these models to compute the effects of interest.
The models of the first step are called nuisance models because they do not directly estimate the quantities of interest.
Most existing applicative studies generally resort to parametric linear models for the nuisance functions, that may fail to reflect the role of confounding variables.
Yet, the consistency of the different estimators relies on the consistency of at least a subset of those nuisance models.
Hence, model misspecification, i.e. the inadequacy between a parametric model and the actual underlying phenomenon, can lead to erroneous estimates.

Another limitation of classical estimators is that most methods are dedicated to the case of a one-dimensional binary mediator, while the problem of handling multiple binary or continuous mediators is increasingly considered in the literature. 
\citet{vanderweele2014mediation,jerolon2020causal,avin2005identifiability} have specified identifiability assumptions for direct and indirect effects, either jointly through all mediators, or for a path-specific effect through one particular mediator.
An additional difficulty lies in potential causal relations between mediators~\citep{vanderweele2014mediation,jerolon2020causal,huber2019review,nguyen2021clarifying}.
Recent work \citep{chen_high-dimensional_2018,zhao2020sparse,huang2016hypothesis} has focused on high-dimensional settings, where there are numerous mediators, potentially highly correlated, such as gene expression or medical imaging.
In~\citet{huang2016hypothesis}, dimensionality reduction is performed first, using Principal Component analysis (PCA) while in~\citet{zhao2020sparse}, Independent Component Analysis (ICA) or a sparse version of PCA is used instead.
The objective was to compute a smaller number of orthogonal "directions of mediation".
 Alternatively,~\citet{zhang2021mediation,djordjilovic2019global} propose statistical tests to select only a relevant subset of mediators.
After dimension reduction, the authors of those works conduct an analysis of mediation using structural equations models with the classic approach of the product of coefficients~\citep{baron1986moderator,mackinnon2012introduction}, which relies on a linear parametric relation between the directions of mediation, the covariates and the treatment, and between the directions of mediation, the covariates and the outcome, even if confounding covariates are often omitted.

In this work, we focus on the problem of jointly estimating the direct and indirect effects of one or more mediators (no path-specific effect) using either classical approaches or more recent approaches that are robust to model misspecification.
We propose new variants of existing estimators, with more flexible machine learning models to account for complex relations between the variables of interest, and assess the relevance of using such models over various contexts.
We provide a comprehensive evaluation of classical and more recent methods on simulated data, extending the work of~\citep{huber2016finite,rudolph2020peril} for a binary mediator to the continuous and multi-dimensional mediators, with additional estimators.
We rely on a variety of simulation settings to explore the practical implications of violations of parametric model specifications, violations of the overlap assumption, and variations in the number of observations.
This benchmark provides an overview of available estimators, their validity conditions, and limitations which constitutes a valuable guide to the practitioner.
We also provide a Python implementation of all the estimators for an easy off-the-shelf use for Python users, as existing implementations are available only in other programming languages such as R or SAS~\cite{valente2020causal}.

To go beyond performance analysis on simulated data, we conduct several mediation analyses on data from the UK Biobank to explore cognitive functions in a cohort of middle-aged adults.
UK Biobank~\citep{sudlow2015uk} is a prospective cohort of about 500,000 healthy participants in the UK with very thorough socio-demographic, medical, lifestyle, physical, and cognitive assessments.
A subset of nearly 40,000 participants also underwent a more enhanced functional exploration including brain structural and functional Magnetic Resonance Imaging (MRI).
This unprecedentedly large imaging database allows us to assess the potential role of brain structure in the shaping of cognitive functions while observing potential confounders.
Specifically, we build on the view that brain characteristics do not represent an independent source of information about cognitive scores but rather integrate individual characteristics that are accessible from questionnaires.
For this reason, we handle brain morphological characteristics as a mediator between socio-demographic variables and behavioral outcomes.
The results obtained for several potential exposures further illustrate the properties of the different estimators considered, in particular the accuracy and the stability of the classical coefficient product, approach and the robust double machine learning estimator in a variety of complex real-world data settings.
Moreover, this real-data study offers a comprehensive example encompassing all facets of causal mediation analysis. 
In contrast, most existing works focus either on the correct framing of a causal mediation question~\citep{lee2021guideline,stuart2021assumptions,rohrer2022sa}, or the estimation~\citep{rijnhart2017comparison,huber2016finite,rudolph2020peril,nguyen2023causal}.
However, both aspects are essential for achieving rigorous results.

The rest of the article is organized as follows.
In section~\ref{problem_setting} we formalize the causal mediation analysis problem, defining the natural direct and indirect effects, and the associated identifiability assumptions.
In Section~\ref{estimators}, we present the assessed estimators, as well as the underlying models and implementations.
We propose a companion Github package\footnote{\label{github_url}\url{https://github.com/judithabk6/med_bench}} to this paper, with Python implementation or wrapping of the R implementation for all those estimators, as well as tutorials for the practitioner.
Section~\ref{simu} presents in detail the simulation process and the main trends in results.
An application of mediation analysis to decipher some aspects of the effect of education on middle-age cognitive functions is proposed in Section~\ref{application}.
We highlight a group of physiological exposures for which an significant part of their effect on cognitive function is mediated through the brain structure.
Finally, we discuss the overall results and the limits of our work in Section~\ref{discussion_conclusion}.

\section{Problem setting: causal mediation analysis} \label{problem_setting}
In this section, we introduce the potential outcome framework to define the causal quantities of interest, also called estimands.
We then specify the required assumptions to identify those quantities, i.e. estimate them with the data at our disposal.
\subsection{Natural direct and indirect effects}

\begin{figure}
\centering
\begin{tikzpicture}
    \node[thick, state, align=center,inner sep=1pt] (1) {\Large{$T$}\\\small{(treatment)}};
    \node[thick, state, align=center,inner sep=1pt] (2) [right =3 of 1] {\Large{\textbf{$M$}}\\\small{(mediator(s))}};
    \node[thick, state, align=center,inner sep=1pt] (3) [right =3 of 2] {\Large{$Y$}\\\small{(outcome)}};
    \node[thick, state, align=center,inner sep=1pt] (4) [above right =4.3 of 1,xshift=-1.5cm,yshift=-1.5cm] {\Large{$X$}\\\small{(confounder(s))}};

    \path (1) edge[ultra thick] node[above, align=center] {\textbf{indirect}} (2);
    \path (1) edge[ultra thick, below, bend right=30, align=center] node[above] {\textbf{direct effect}} (3);
    \path (2) edge[ultra thick] node[above, align=center] {\textbf{effect}} (3);

    \path (4) edge[ultra thick] node[el,above] {} (1);
    \path (4) edge[ultra thick] node[el,above] {} (2);
    \path (4) edge[ultra thick] node[el,above] {} (3);

\end{tikzpicture}
\caption{\textbf{A general directed acyclic graph for causal mediation analysis.} Each node represents a group of variables, and arrows denote causal relations between them.
}
\label{mediation_graph}
\end{figure}
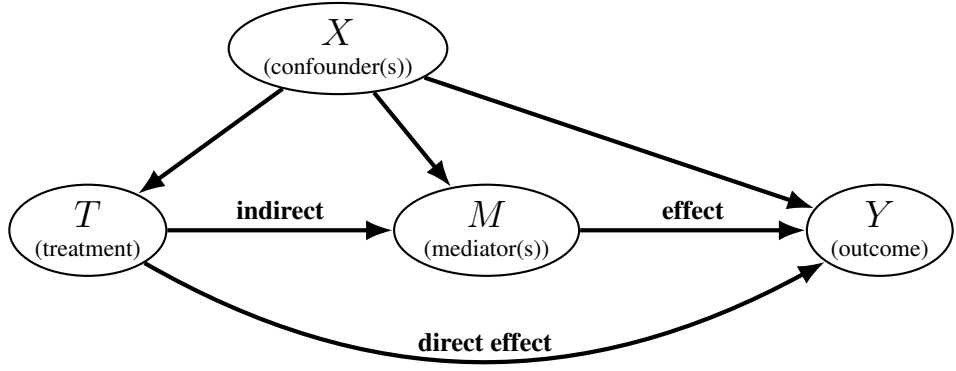

The objective of mediation analysis is to quantify the part of the total effect achieved through the mediator, i.e. the indirect effect, and the effect of the treatment without further intermediate, i.e. the direct effect.
For each individual, we denote the (binary) treatment $T$, the observed outcome $Y$, the observed mediator(s) $M$, and the covariate(s) $X$; covariates associated with at least two variables among the treatment, mediator(s) and outcome are confounders and should be adjusted for.
Let us note $\mathcal{T}, \mathcal{Y}, \mathcal{M}, \mathcal{X}$  the respective supports of $T, Y, M$ and $X$.
The relations between those variables are illustrated in Figure~\ref{mediation_graph}.
Extending the potential outcomes framework~\citep{imbens2015causal}, in the case of a binary treatment, we can define $M(t)$ and $Y(t, M(t))$ the potential mediator and the potential outcome under the treatment value $t \in \{0, 1\}$. 
For each unit, only one potential outcome and mediator state are observed.
We assume that the observed outcome and mediator are the potential outcome and mediator under the actual assigned treatment.

\begin{assumption}[SUTVA (Stable Unit Treatment Values) and consistency]\label{a:sutva_con}
\begin{align}
M = T M(1) + (1-T)M(0) \;\mathrm{ and }\; Y=T Y(1, M(1)) + (1-T) Y(0, M(0))
\end{align}
\end{assumption}

We then define the total average treatment effect (ATE) as:
\begin{definition}[Total average treatment effect]
\begin{equation}
\tau = \mathbb{E}[Y(1, M(1)) - Y(0, M(0))] \label{ate_equation}
\end{equation}
\end{definition}

To further decompose the total effect into natural direct and indirect effects~\citep{pearl2001direct,huber2019review,nguyen2021clarifying}, we define cross-world potential outcomes that correspond to varying the treatment, while maintaining the value of the mediator to the value it would have without changing the treatment, and the converse.
Contrary to the previously mentioned potential outcomes, where one of them is observed, cross-world outcomes can never be observed.
Those additional terms allow us to define the natural direct effect as
\begin{definition}[Natural direct effect]
\begin{equation*}
\theta(t) = \mathbb{E}[Y(1, M(t)) - Y(0, M(t))], \;\;\;\;t \in \{0, 1\},
\end{equation*}
\end{definition}

and the natural indirect effect

\begin{definition}[Natural indirect effect]\label{def_nie}
\begin{equation*}
\delta(t) = \mathbb{E}[Y(t, M(1)) - Y(t, M(0))], \;\;\;\;t \in \{0, 1\}.
\end{equation*}
\end{definition}

We hence define two natural direct effects, the natural direct effect on the exposed, $\theta(1)$ or the total direct effect, while $\theta(0)$ is called the natural direct effect on the unexposed, or pure direct effect~\citep{vansteelandt2012natural}.
Similarly, we can distinguish the natural indirect effect on the exposed, $\delta(1)$, also called the total indirect effect, and the natural indirect effect on the unexposed, $\delta(0)$, or pure indirect effect.
In the case where there is no interaction between the treatment and the mediator, i.e. effect heterogeneity, the effects are equal among the exposed or the unexposed (i.e. $\theta(1) = \theta(0)$ and $\delta(1) = \delta(0)$)~\citep{huber2016finite,rohrer2022sa}.
The ATE in equation ~\eqref{ate_equation} is the sum of the direct and indirect effects of opposite treatment states: $\tau = \theta(0) + \delta(1) = \theta(1) + \delta(0)$. 
In the rest of this work, we will use the first decomposition.

\subsection{Identification assumptions}
As mentioned above, the parameters of interest include both unobserved and unobservable terms.
In addition to the SUTVA and consistency assumptions, already introduced in the previous paragraph, identification requires additional assumptions, stronger than in standard causal inference settings:

\begin{assumption}[Sequential conditional independence of the treatment~\citep{imai2010identification}]\label{a:seq-treatment}
\begin{align}
\{Y(t',m), M(t)\} &\independent T | X \;\;\;\;\mbox{ for all } t', t \in \{0, 1\} \mbox{ and } m \in \mathcal{M},
\end{align}
where $\independent$ denotes independence.
\end{assumption}

Assumption~\ref{a:seq-treatment} imposes the absence of unobserved confounders of the treatment and the outcome on the one hand, and of the treatment and the mediator on the other hand.

\begin{assumption}[Sequential conditional independence of the mediator~\citep{imai2010identification}]\label{a:seq-mediator}
\begin{align*}
Y(t', m) &\independent M|T=t, X=x\;\;\;\; \;\mbox{ for all } t', t \in \{0, 1\}, m \in \mathcal{M}  \mbox{ and }  x \in \mathcal{X}.
\end{align*}
where $X$ contains no variable affected by the treatment (post-treatment variable).
\end{assumption}

Assumption~\ref{a:seq-mediator} forbids the existence of unobserved confounders affecting both the mediator and the outcome.

Finally, a common support assumption (also called positivity or overlap) is needed.
Informally, this assumption ensures that for each individual, we can find a comparable individual with the opposite treatment or mediator value in the dataset, ensuring that the effect of interest can be computed.

\begin{assumption}[Positivity assumption]\label{a:seq-positivity}
\begin{equation*} \begin{split}
    &\mbox{P}\left(T=t | X=x\right)>0 \mbox{ and }0<\mbox{P}\left(M(t)=m | T=t, X=x\right) \mbox{ for all } t \in \{0, 1\},\; m \in \mathcal{M}  \mbox{ and }  x \in \mathcal{X}.\end{split}
    \end{equation*} 
\end{assumption}

Relying on Assumptions~\ref{a:seq-treatment}, \ref{a:seq-mediator}, and \ref{a:seq-positivity} the total effect, along with the natural direct and indirect effects are identifiable. 
We give the demonstration in the Appendix section~\ref{identifiability_proof} that the mean potential outcomes and cross-world potential outcomes needed to compute the effects of interest can be identified non-parametrically, that is without any form restriction such as linearity~\citep{imai2010identification,huber2019review}.

\subsection{The case of several mediators}
\label{sec:several}

If we now consider several mediators of interest, $\textbf{M}=(M_1, \dots, M_K)$, and aim at computing the indirect effect through all the mediators jointly, all the definitions and assumptions above can be written similarly, by just replacing the mediator $M$ by a mediator vector $\textbf{M}$.
The identification of path-specific effects through one particular mediator among several mediators of interest requires additional assumptions~\citep{vanderweele2014mediation,huber2019review} and is not the objective of this work.
However, Assumption~\ref{a:seq-mediator} forbids the existence of any post-treatment confounder of the mediator of interest and the outcome.
Such a variable would actually be an additional mediator.

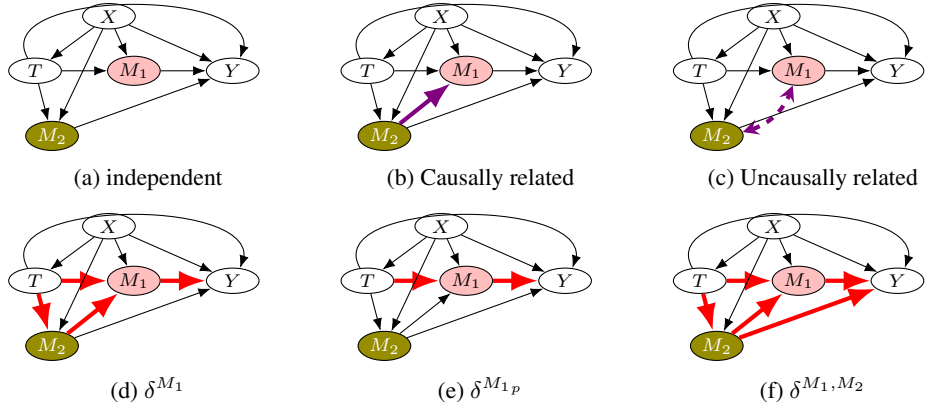
\begin{figure}
\centering
\begin{subfigure}{0.31\textwidth}
\begin{tikzpicture}
    \node[state, align=center,inner sep=1pt] (1) {\scriptsize{$T$}};
    \node[state, align=center,inner sep=1pt, fill=pink] (2) [right =0.6 of 1] {\scriptsize{$M_1$}};
    \node[state, align=center,inner sep=1pt] (3) [right =0.6 of 2] {\scriptsize{$Y$}};
    \node[state, align=center,inner sep=1pt] (4) [above right =1.1 of 1,xshift=-0.3cm,yshift=-0.3cm] {\scriptsize{$X$}};
    \node[state, align=center,inner sep=1pt, fill=olive, text=white] (5) [below left =0.4cm of 2,xshift=-0.3cm,yshift=-0.3cm] {\scriptsize{$M_2$}};

    \path (1) edge node[above, align=center, anchor=center] {} (2);
    \path (1) edge[bend left=120, align=center, anchor=center] node[below] {} (3);
    \path (2) edge node[above, align=center, anchor=center] {} (3);

    \path (4) edge node[el,above] {} (1);
    \path (4) edge node[el,above] {} (2);
    \path (4) edge node[el,above] {} (3);
    \path (4) edge node[el,above] {} (5);
    \path (1) edge node[el,above] {} (5);
    %\path (5) edge node[el,above] {} (2);
    \path (5) edge node[el,above] {} (3);
\end{tikzpicture}
\caption{independent}
\end{subfigure}
\begin{subfigure}{0.31\textwidth}
\begin{tikzpicture}
    \node[state, align=center,inner sep=1pt] (1) {\scriptsize{$T$}};
    \node[state, align=center,inner sep=1pt, fill=pink] (2) [right =0.6 of 1] {\scriptsize{$M_1$}};
    \node[state, align=center,inner sep=1pt] (3) [right =0.6 of 2] {\scriptsize{$Y$}};
    \node[state, align=center,inner sep=1pt] (4) [above right =1.1 of 1,xshift=-0.3cm,yshift=-0.3cm] {\scriptsize{$X$}};
    \node[state, align=center,inner sep=1pt, fill=olive, text=white] (5) [below left =0.4cm of 2,xshift=-0.3cm,yshift=-0.3cm] {\scriptsize{$M_2$}};

    \path (1) edge node[above, align=center, anchor=center] {} (2);
    \path (1) edge[bend left=120, align=center, anchor=center] node[below] {} (3);
    \path (2) edge node[above, align=center, anchor=center] {} (3);

    \path (4) edge node[el,above] {} (1);
    \path (4) edge node[el,above] {} (2);
    \path (4) edge node[el,above] {} (3);
    \path (4) edge node[el,above] {} (5);
    \path (1) edge node[el,above] {} (5);
    \path (5) edge[ultra thick, violet] node[el,above] {} (2);
    \path (5) edge node[el,above] {} (3);
\end{tikzpicture}
\caption{Causally related}
\end{subfigure}
\begin{subfigure}{0.31\textwidth}
\begin{tikzpicture}
    \node[state, align=center,inner sep=1pt] (1) {\scriptsize{$T$}};
    \node[state, align=center,inner sep=1pt, fill=pink] (2) [right =0.6 of 1] {\scriptsize{$M_1$}};
    \node[state, align=center,inner sep=1pt] (3) [right =0.6 of 2] {\scriptsize{$Y$}};
    \node[state, align=center,inner sep=1pt] (4) [above right =1.1 of 1,xshift=-0.3cm,yshift=-0.3cm] {\scriptsize{$X$}};
    \node[state, align=center,inner sep=1pt, fill=olive, text=white] (5) [below left =0.4cm of 2,xshift=-0.3cm,yshift=-0.3cm] {\scriptsize{$M_2$}};

    \path (1) edge node[above, align=center, anchor=center] {} (2);
    \path (1) edge[bend left=120, align=center, anchor=center] node[below] {} (3);
    \path (2) edge node[above, align=center, anchor=center] {} (3);

    \path (4) edge node[el,above] {} (1);
    \path (4) edge node[el,above] {} (2);
    \path (4) edge node[el,above] {} (3);
    \path (4) edge node[el,above] {} (5);
    \path (1) edge node[el,above] {} (5);
    \path (5) edge[stealth-stealth, bend right=30,dashed, ultra thick, violet] node[el,above] {} (2);
    \path (5) edge node[el,above] {} (3);
\end{tikzpicture}
\caption{Uncausally related}
\end{subfigure}
% \caption{Possible relations between two mediators}
% \end{figure}

% \begin{figure}
\centering
\begin{subfigure}{0.31\textwidth}
\begin{tikzpicture}
    \node[state, align=center,inner sep=1pt] (1) {\scriptsize{$T$}};
    \node[state, align=center,inner sep=1pt, fill=pink] (2) [right =0.6 of 1] {\scriptsize{$M_1$}};
    \node[state, align=center,inner sep=1pt] (3) [right =0.6 of 2] {\scriptsize{$Y$}};
    \node[state, align=center,inner sep=1pt] (4) [above right =1.1 of 1,xshift=-0.3cm,yshift=-0.3cm] {\scriptsize{$X$}};
    \node[state, align=center,inner sep=1pt, fill=olive, text=white] (5) [below left =0.4cm of 2,xshift=-0.3cm,yshift=-0.3cm] {\scriptsize{$M_2$}};

    \path (1) edge[ultra thick, red] node[above, align=center, anchor=center] {} (2);
    \path (1) edge[bend left=120, align=center, anchor=center] node[below] {} (3);
    \path (2) edge[ultra thick, red] node[above, align=center, anchor=center] {} (3);

    \path (4) edge node[el,above] {} (1);
    \path (4) edge node[el,above] {} (2);
    \path (4) edge node[el,above] {} (3);
    \path (4) edge node[el,above] {} (5);
    \path (1) edge[ultra thick, red] node[el,above] {} (5);
    \path (5) edge[ultra thick, red] node[el,above] {} (2);
    \path (5) edge node[el,above] {} (3);
\end{tikzpicture}
\caption{$\delta^{M_1}$}
\end{subfigure}
\begin{subfigure}{0.31\textwidth}
\begin{tikzpicture}
    \node[state, align=center,inner sep=1pt] (1) {\scriptsize{$T$}};
    \node[state, align=center,inner sep=1pt, fill=pink] (2) [right =0.6 of 1] {\scriptsize{$M_1$}};
    \node[state, align=center,inner sep=1pt] (3) [right =0.6 of 2] {\scriptsize{$Y$}};
    \node[state, align=center,inner sep=1pt] (4) [above right =1.1 of 1,xshift=-0.3cm,yshift=-0.3cm] {\scriptsize{$X$}};
    \node[state, align=center,inner sep=1pt, fill=olive, text=white] (5) [below left =0.4cm of 2,xshift=-0.3cm,yshift=-0.3cm] {\scriptsize{$M_2$}};

    \path (1) edge[ultra thick, red] node[above, align=center, anchor=center] {} (2);
    \path (1) edge[bend left=120, align=center, anchor=center] node[below] {} (3);
    \path (2) edge[ultra thick, red] node[above, align=center, anchor=center] {} (3);

    \path (4) edge node[el,above] {} (1);
    \path (4) edge node[el,above] {} (2);
    \path (4) edge node[el,above] {} (3);
    \path (4) edge node[el,above] {} (5);
    \path (1) edge node[el,above] {} (5);
    \path (5) edge node[el,above] {} (2);
    \path (5) edge node[el,above] {} (3);
\end{tikzpicture}
\caption{$\delta^{{M_1}_p}$}
\end{subfigure}
\begin{subfigure}{0.31\textwidth}
\begin{tikzpicture}
    \node[state, align=center,inner sep=1pt] (1) {\scriptsize{$T$}};
    \node[state, align=center,inner sep=1pt, fill=pink] (2) [right =0.6 of 1] {\scriptsize{$M_1$}};
    \node[state, align=center,inner sep=1pt] (3) [right =0.6 of 2] {\scriptsize{$Y$}};
    \node[state, align=center,inner sep=1pt] (4) [above right =1.1 of 1,xshift=-0.3cm,yshift=-0.3cm] {\scriptsize{$X$}};
    \node[state, align=center,inner sep=1pt, fill=olive, text=white] (5) [below left =0.4cm of 2,xshift=-0.3cm,yshift=-0.3cm] {\scriptsize{$M_2$}};

    \path (1) edge[ultra thick, red] node[above, align=center, anchor=center] {} (2);
    \path (1) edge[bend left=120, align=center, anchor=center] node[below] {} (3);
    \path (2) edge[ultra thick, red] node[above, align=center, anchor=center] {} (3);

    \path (4) edge node[el,above] {} (1);
    \path (4) edge node[el,above] {} (2);
    \path (4) edge node[el,above] {} (3);
    \path (4) edge node[el,above] {} (5);
    \path (1) edge[ultra thick, red] node[el,above] {} (5);
    \path (5) edge[ultra thick, red] node[el,above] {} (2);
    \path (5) edge[ultra thick, red] node[el,above] {} (3);
\end{tikzpicture}
\caption{$\delta^{M_1,M_2}$}
\end{subfigure}
\caption{\textbf{Configurations to consider with two mediators $M_1$ and $M_2$.} In panels a, b and c, we show the possible relations between two distinct mediators (adapted from~\citet{jerolon2020causal}). In panels d, e and f, we show in bold red arrows several indirect effects in the case of causally related mediators that can be of interest (adapted from~\citet{huber2019review}). The indirect effect shown in panel d, the indirect effect mediated by $M_1$ is not identified, as it involves counterfactual outcomes with two different potential states for the mediator $M_2$, which is impossible to estimate. The indirect effect shown in panel e (the indirect effect going through $M_1$ but not through $M_2$) is identified, as well as the total mediated effect going through both $M_2$ and $M_1$ (panel f).}
\label{multi_med}
\end{figure}

To illustrate what happens in this case, let us consider two mediators $M_1$ and $M_2$, with the causal graph defined in Figure~\ref{multi_med}.
Huber~\citep{huber2019review} defines several path-specific (indirect) effects:
\begin{definition}[Natural indirect effect (with respect to $M_1$, operating through $M_2$) Figure~\ref{multi_med}, panel d]
\begin{equation*}
\delta^{M_1}(t) = \mathbb{E}[Y(t, M_1(1, M_2(1)), M_2(t)) - Y(t, M_1(0, M_2(0)), M_2(t))], \;\;\;\;t \in \{0, 1\}.
\end{equation*}
\end{definition}

\begin{definition}[Path-specific indirect effect (with respect to $M_1$, not operating through $M_2$) Figure~\ref{multi_med}, panel e]
\begin{equation*}
\delta^{{M_1}_p}(t) = \mathbb{E}[Y(t, M_1(1, M_2(t)), M_2(t)) - Y(t, M_1(0, M_2(t)), M_2(t))], \;\;\;\;t \in \{0, 1\}.
\end{equation*}
\end{definition}

\begin{definition}[Joint indirect effect through $M_1$ and $M_2$, Figure~\ref{multi_med}, panel f]
\begin{equation*}
\delta^{M_1, M_2}(t) = \mathbb{E}[Y(t, M_1(1, M_2(1)), M_2(1)) - Y(t, M_1(0, M_2(0)), M_2(0))], \;\;\;\;t \in \{0, 1\}.
\end{equation*}
\end{definition}

Of these three indirect effects of interest, the most interesting one, $\delta^{M_1}(t)$ is not identifiable without restrictive additional assumptions, as it gives rise to potential outcomes with two different states of the mediator $M_2$, which is impossible to estimate~\citep{robins1992identifiability}.
The effect $\delta^{{M_1}_p}(t)$ is identifiable, but provides only a partial view of the role of $M_1$.
The effect $\delta^{M_1, M_2}(t)$ is also identifiable, but does not represent the contribution of a particular mediator to the total effect.

Therefore, if we are interested in the indirect effect through a mediator $M_1$, we should still consider all other mediators placed on the path between the treatment node and the mediator $M_1$ node in the causal graph (i.e. post-treatment variables affecting both the mediator $M_1$ and the outcome $Y$).
In the simple example of the causal diagram of Figure~\ref{multi_med}, if we ignore entirely the mediator $M_2$, and consider the indirect effect from Definition~\ref{def_nie}, the considered estimand is $\delta^{{M_1}_p}(t)$, which does not capture entirely the indirect effect mediated by $M_1$.
As a consequence, identification of the direct and indirect effects is challenging, as one should unpack all the causal chain between the treatment and the mediator(s) of interest, which is practically infeasible~\citep{rohrer2022sa}.
In this paper, we only consider the joint effect of all mediators, hence $\delta^{M_1, M_2}(t)$.

\section{Overview of estimators for mediation analysis}\label{estimators}
In this study, we consider several estimators for the natural direct and indirect effects that can be used for binary, continuous and/or multi-dimensional mediators.
We compare their properties, conditions of application and performances on simulated data with binary and continuous multi-dimensional mediators in Section~\ref{simu}.
This section introduces the different estimators, and describes the implementations of the estimators used in the experiments.
Some estimators include a procedure for estimating uncertainty, but not all, so we have limited our comparison to estimating direct and indirect effects and assessing the variance of estimators with several independent data draws.
We distinguish \emph{parametric estimators} that rely solely on the output of nuisance models to compute the effects of interest, and \emph{semi-parametric} estimators where the original observed values for the treatment and the outcome variables are also used in the computation of the model.
We note that we can however use non-parametric models for predictions even in the so-called "parametric estimators".

\subsection{Parametric estimators}
\subsubsection{Coefficient product}
\label{coeff_prod}
The method of coefficient product developed by~\cite{baron1986moderator} is the first  estimator developed for mediation analysis. It consists in assuming the following structural equation model:
\begin{align}
    M(x, t) &= \beta_0 + \beta_T t + x^t \beta_X + \epsilon_M \\
    Y(x, t, m) &= \gamma_0 + \gamma_T t + x^t \gamma_X + \gamma_M m + \epsilon_Y
\end{align}

where $\epsilon_M$ and $\epsilon_Y$ are independent centered normal random variables, and the superscript $t$ represents the transpose operation.

Then we have, after estimation of the regression parameters
\begin{align*}
    \hat{\theta}(0) &= \hat{\theta}(1) = \hat{\gamma}_T \\
    \hat{\delta}(0) &= \hat{\delta}(1) = \hat{\gamma}_M \hat{\beta}_T \\
    \hat{\tau} &= \hat{\theta}(1) + \hat{\delta}(0) = \hat{\theta}(0) + \hat{\delta}(1).
\end{align*}

This estimator is consistent if the models for the outcome and for the mediator are both correctly specified. The models can be easily extended to situations with interactions by adding interaction terms to the structural equations~\citep{vanderweele2015explanation}.
The coefficient product exists in many variations to account for binary mediators or outcomes, or different parametric models, as detailed in~\citet{vanderweele2015explanation}, which requires to vary the formulas to obtain the effects and the implementation.
For the sake of simplicity and to avoid overwhelming the reader with those slight variations, we will consider this version even for the case of a binary mediator, where we observe exactly the expected behavior of this estimator on our simulations.

\subsubsection{G-computation}
\label{g_comput}
The following formula is directly derived from the identifiability demonstration step~\ref{mediation_formula_origin} in section~\ref{identifiability_proof}.
It is very intuitive: it relies on regressing the outcome $Y$ on the mediator(s) $M$, the treatment $T$, and the covariates $X$, and consider differences between the predictions obtained for different values of the treatment and/or the mediator(s) to derives estimates for the natural direct and indirect effects.
Hence, it is slightly more sophisticated than the coefficient product in the choice of the model allowed for this regression, and in the estimation, based on the predictions rather than the coefficient values.
We denote $f_{A=a}$ and $f_{A=a|B=b}$ the probability density function of a random variable $A$, unconditionally or conditionally given other random variable(s) $B$.
The factual and counterfactual treatments are denoted $t$ and $t^{\prime}$ respectively.
$$ \mathbb{E}\left[Y\left(t, M\left(t^{\prime}\right)\right)\right] = \iint \mathbb{E}[Y | T=t, M=m, X=x] f_{M=m | T=t^{\prime}, X=x} f_{X=x} d m d x $$
It directly yields the mediation formula~\citep{pearl2001direct,imai2010identification}, 
$$
\begin{aligned}
&\theta(t)=\iint\{\mathbb{E}[Y \mid T=1, M=m, X=x]-\mathbb{E}[Y \mid T=0, M=m, X=x]\} f_{M=m \mid T=t, X=x} d m f_{X=x} d x\\
&\delta(t)=\iint \mathbb{E}[Y \mid T=t, M=m, X=x]\left\{f_{M=m \mid T=1, X=x}-f_{M=m \mid T=0, X=x}\right\} d m f_{X=x} d x.%
\end{aligned}
$$
In practice, we perform parametric or non-parametric estimation of $\hat{\mu}_{Y}(t, m, x)$ for the conditional mean outcome $\mathbb{E}[Y \mid T=t, M=m, X=x]$ by regressing  and $\hat{f}(m|t, x)$ for the conditional mediator density $f_{M=m \mid T=t, X=x}$  (or conditional probability if the mediator is discrete), and the final effects are estimated as follows:
\begin{align} \label{explicit_g_computation}
\begin{split}
    \hat{\theta}(t) &= \frac{1}{n} \sum_{i=1}^n \sum_{m=0}^1 \left\{(\hat{\mu}_{Y}(1, m, X_i) - \hat{\mu}_{Y}(0, m, X_i)) \hat{f}(m|t, X_i)\right\}, \\
    \hat{\delta}(t) &= \frac{1}{n} \sum_{i=1}^n \sum_{m=0}^1 \left\{\hat{\mu}_{Y}(t, m, X_i)(\hat{f}(m|1, X_i) - \hat{f}(m|0, X_i)) \right\}.
    \end{split}
\end{align}
  
The G-computation estimator is consistent if both plug-in models are correctly specified.
The estimation of the conditional mediator density is challenging for multi-dimensional mediators.
Indeed, one can perform parametric density estimation, which relies on strong assumptions about the distribution (e.g., assuming a normal distribution), or non-parametric density estimation, which requires a number of observations that increases exponentially with dimensionality~\citep{Scott04multi-dimensionaldensity}.
Considering the case of independent mediators facilitates density estimation.

Alternatively, following the implementation strategy in the R~package \texttt{causalweight}~\citep{bodory2018causalweight}, this mediator density step can also be ignored altogether with the following estimation approach.
The derivation of this strategy is detailed in Supplementary~\ref{sup:g_comput}, and will be referred to as implicit integration G-computation.

In this work, this estimator is implemented with the explicit integration over the mediator for a binary mediator as outlined in equation~\ref{explicit_g_computation}, and with the implicit integration for other cases (continuous or multi-dimensional mediator) with the formula of equation~\ref{implicit_g_computation}, following the best-performing strategy for each situation.

\subsection{Semi-parametric estimators}
\subsubsection{Inverse probability weighting estimator (IPW)}
\label{ipw}
The identifiability computation in step~\eqref{propensity_weighting} in Appendix~\ref{identifiability_proof} provides the ground for an inverse probability weighting-type approach (IPW), developed in~\cite{huber2014identifying}.
It consists in the extension to the mediation analysis of a very classical strategy in causal inference~\citep{horvitz1952generalization,huber2014identifying}.
Computation requires the estimation of two nuisance parameters, the conditional probability of treatment given $X$ covariates, $p(X)=\mathbb{P}(T=1 \mid X)$, and the conditional probability of treatment given covariates and mediator(s), $\rho(M, X)=\mathbb{P}(T=1 \mid X, M)$. 
In practical implementation, the weights are normalized for more stability; we can write the normalized estimator as follows
    \begin{align*}
        \hat{\theta}(0) &= \frac{\sum_{i=1}^n Y_iT_i(1-\hat{\rho}(M_i, X_i)) / [\hat{\rho}(M_i, X_i)(1-\hat{p}(X_i))]}{\sum_{i=1}^n T_i(1-\hat{\rho}(M_i, X_i)) / [\hat{\rho}(M_i, X_i)(1-\hat{p}(X_i))]} - \frac{\sum_{i=1}^n Y_i(1-T_i)/(1-\hat{p}(X_i))}{\sum_{i=1}^n (1-T_i)/(1-\hat{p}(X_i))} \\
        \hat{\delta}(1) &= \frac{\sum_{i=1}^n Y_iT_i/\hat{p}(X_i)}{\sum_{i=1}^n T_i/\hat{p}(X_i)} - \frac{\sum_{i=1}^n Y_iT_i(1-\hat{\rho}(M_i, X_i)) / [\hat{\rho}(M_i, X_i)(1-\hat{p}(X_i))]}{\sum_{i=1}^n T_i(1-\hat{\rho}(M_i, X_i)) / [\hat{\rho}(M_i, X_i)(1-\hat{p}(X_i))]}
    \end{align*}

The IPW estimator is consistent if models for $p$ and $\rho$ are consistently estimated.
The absence of estimation of the density of the estimator(s) allows us to use this estimator when the mediator is continuous and/or multi-dimensional.
The total effect, which is the sum of the direct effect $\hat{\theta}(0)$, and the indirect effect $\hat{\delta}(1)$, does not involve $\hat{\rho}$.
Hence it is robust to misspecification of the mediator model.

\subsubsection{Multiply-robust estimator}
\label{multiply_robust}
The multiply-robust estimator has been derived in the framework of semi-parametric theory, that aims at minimizing the variation of the causal estimate for a variation (or an error) on the nuisance functions~\citep{kennedy2024semiparametric}.
Hence it gives a property of robustness to misspecification of the nuisance models to the obtained estimator, which is a highly desirable property.
The inverse mediator density weighting implemented in the multiply-robust estimator~\citep{tchetgen2012semiparametric} is motivated by the last step~\ref{density_weighting} of the identifiability computation in Appendix~\ref{identifiability_proof}.
We can then write~\citep{tchetgen2012semiparametric,huber2016finite}

\begin{align}
    \hat{\theta}(0) &= \frac{1}{n}\sum_{i=1}^n \left\{ \left[\frac{T_i \hat{f}(M_i|0, X_i)}{\hat{p}(X_i)\hat{f}(M_i|1, X_i)} - \frac{1-T_i}{1-\hat{p}(X_i)}\right]  (Y_i-\hat{\mu}_{Y}(T_i, M_i, X_i)) \nonumber \right.\\ & \left.+  \frac{1-T_i}{1-\hat{p}(X_i)} (\hat{\mu}_{Y}(1, M_i, X_i) - \hat{\mu}_{Y}(0, M_i, X_i) - \hat{\omega}_Y(1, 0, X_i) + \hat{\omega}_Y(0, 0, X_i)) \nonumber \right.\\ & \left.- \hat{\omega}_Y(1, 0, X_i) + \hat{\omega}_Y(0, 0, X_i)\right\} \label{multiply_robust_direct}\\
    \hat{\delta}(1) &= \frac{1}{n} \sum_{i=1}^n \left\{ \frac{T_i}{\hat{p}(X_i)}\left[Y_i - \hat{\psi}(1, X_i) - \frac{\hat{f}(M_i|0, X_i)}{\hat{f}(M_i|1, X_i)}(Y-\hat{\mu}_{Y}(1, M_i, X_i))\right] \nonumber \right.\\ & \left. - \frac{1-T_i}{1-\hat{p}(X_i)} (\hat{\mu}_{Y}(1, M_i, X_i) - \hat{\omega}_Y(1, 0, X_i)) + \hat{\omega}_Y(1, 1, X_i) - \hat{\omega}_Y(1, 0, X_i)\right\} \label{multiply_robust_indirect}
\end{align}

cross-world conditional mean outcomes, $\omega_Y(t, t', x)=\mathbb{E}\left[\mu_{Y}(t, m, x)|T=t', X=x\right]$

with $\hat{\mu}_{Y}(t, m, x)$ estimating the conditional mean outcome $\mathbb{E}[Y|T=t, M=m, X=x]$, $\hat{f}(m|t, x)$ estimating the conditional mediator density $f_{M|T=t, X=x}(m)$ (or conditional probability if the mediator is discrete), $\hat{p}(x)$ estimating the treatment propensity score $P(T=1|X=x)$, and $\hat{\omega}_Y(t, t', x)$ estimating the cross-world conditional mean outcome $\mathbb{E}\left[\mu_{Y}(t, m, x)|T=t', X=x\right]$, which can be estimated using two nested regressions, as introduced for the implicit integration of the G-computation, and detailed in Appendix~\ref{sup:g_comput}.

A very interesting property of this estimator is that it is triply-robust, as it remains consistent if at least two of the three following models are well specified: (i) the conditional mean outcome, $\mathbb{E}[Y|T, M, X]$, (ii) the conditional density of $M$ given $T$, $X$, and (iii) the treatment propensity score.
This estimator is also asymptotically semi-parametrically efficient if (i), (ii), and (iii) are all correctly specified, meaning that it optimally uses the available data to obtain (theoretically) the smallest variance possible.
The estimation of the mediator conditional density makes the use of this estimator difficult for a continuous and/or multi-dimensional mediator without further assumptions, however the same technique for implicit integration described in Appendix~\ref{sup:g_comput} can also be applied here, but was not used in this work as it is redundant with the Double-Machine Learning estimator presented in the next section.
As for the IPW estimator, the weights have been normalized to improve numerical stability.

\subsubsection{Double-Machine Learning for mediation estimator (medDML)}
\label{medDML}
The multiply-robust estimator (Section~\ref{multiply_robust}) was extended in~\cite{farbmacher2020causal}, by removing the requirement to estimate the conditional mediator density using Bayes' law.
Indeed, it can be shown that
\begin{align*}
\frac{f(M|T=1-t, X)}{p(T=t|X) f(M|T=t, X)} &= \frac{(1-p(T=t|X, M))f(M|X)}{1-p(T=t|X)} \frac{p(T=t|X)}{p(T=t|X, M)f(M|X)p(T=t|X)}\\ &= \frac{1-p(T=t|X, M)}{p(T=t|X, M)(1-p(T=t|X))}
\end{align*}

This new expression can be directly used in equations~\eqref{multiply_robust_direct} and~\eqref{multiply_robust_indirect}, lifting the previously mentioned restrictions for continuous and/or multi-dimensional mediators.
Additionally, the authors show that this new estimator is Neyman orthogonal, which is crucial to the application of the double machine learning framework, in addition to the use of cross-fitting~\citep{chernozhukov2018double}.
Compared to the semi-parametric framework used in the multiply-robust estimator presented in the previous section (Section~\ref{multiply_robust}), the property that the model can reach a fast convergence rate, even with complex machine learning models for the plug-in estimator, by using cross-fitting.
As for the IPW estimator, the weights have been normalized to improve numerical stability.

\subsection{Implementation considerations}
\label{implementation}
We re-implemented some of the considered estimators, using the Python package \texttt{scikit-learn}~\citep{scikit-learn}.
This re-implementation has allowed us to centralize the main mediation estimators within a unified framework, with more flexibility in the choice of algorithms for nuisance parameters estimation, in particular the use of random forests~\citep{breiman2001random}.
We provide a flexible interface for those estimators allowing to use any model or combination of models from \texttt{scikit-learn}, including regularization, probability calibration~\citep{zadrozny2002transforming,niculescu2005predicting}, ensemble models.
The main characteristics of available implementations for each estimator are presented in Table~\ref{estimators_implementations}.
We refer the reader to this review of existing implementations in languages other than Python for a complete view~\citep{valente2020causal}.
In the experiments presented here, we used the Python implementation when it was available.
Our implementations are available as a Python package at \url{https://github.com/judithabk6/med_bench}, with examples and guidance for the practitioner.

\begin{table}[h!!!!!!!!!!]
\begin{adjustbox}{width=1.2\textwidth,center=\textwidth}
\begin{tabular}{p{2.5cm}cccp{5cm}p{5cm}}

  \thead{estimator} & \thead{binary} & \thead{continuous} & \thead{multi\\dimensional}& \thead{reference\\R implementation} & \thead{Python\\implementation\\(package \texttt{med\_bench})}  \\ \hline
  \makecell[t]{coefficient\\product \ref{coeff_prod}} & x & x & x & no & \textbf{function \texttt{mediation\_coefficient\_product} with linear regression models with L2 regularization (cross-validation or fixed at $\alpha=10^{-12}$)}. Also estimation method \texttt{mediation.two\_stage\_regression} in \texttt{DoWhy} package. \\ \hline
  g-Computation or g-formula \ref{g_comput} & x & x & x & no & \textbf{function \texttt{mediation\_g\_formula} with linear and logistic models (with or without L2 regularization) or random forests to estimate $\hat{\mu}_Y$ and  $\hat{f}$} \\ \hline
  IPW \ref{ipw} & x & x & x & function \texttt{medweight} in package \texttt{causalweight}~\citep{bodory2018causalweight} with logit or probit regression to estimate conditional probabilities, and package \texttt{twangMediation} with gradient boosting & \textbf{function \texttt{mediation\_IPW} with linear models (with or without L2 regularization) or non-parametric estimators (random forests)} \\ \hline
  \makecell[t]{multiply-\\robust\\estimator} \ref{multiply_robust} & x & - & - & no & \textbf{function \texttt{mediation\_multiply\_robust} with linear models (with or without L2 regularization) or non-parametric estimators (random forests)} \\ \hline
  medDML \ref{medDML} & x & x & x & function \texttt{medDML} in package \texttt{causalweight}~\citep{bodory2018causalweight}, with parametric linear models & \textbf{function \texttt{mediation\_DML} with linear models (with or without L2 regularization) or non-parametric estimators (random forests)}\\ \hline
\end{tabular}
\end{adjustbox}
\caption{\textbf{Characteristics of the estimator implementations.} The columns "binary", "continuous" and "multi-dimensional" refer to the ability of methods to handle mediators of this kind.
Symbols "x" and "-" mean available or not respectively.
Most estimators have an implementation in R.
We have reimplemented them in Python to make it easier for Python users to perform mediation analysis.}
\label{estimators_implementations}
\end{table}

The R package, \texttt{CMAverse}~\citep{shi2021cmaverse} is currently developed with a similar objective of unification.
However, the implemented estimators are not exactly the same as the ones we consider: the definitions of some estimators slightly differ (\texttt{g-formula}, \texttt{weighting-based approach}), and the most recent  robust approaches with the best performances in our work, such as the multiply robust and the double machine learning estimators are not available.

Another robust estimator for mediation is derived from the Targeted Maximum Likelihood Estimation (TMLE) framework~\citep{zheng2012targeted}.
From a theoretical point of view, TMLE also relies on influence functions from the semi-parametric theory~\citep{kennedy2024semiparametric}, like the double-machine learning and the multiply robust estimator, but with a different bias-correction approach.
Implementations are available as R packages \texttt{tmle3mediate}~\citep{hejazi2021tmle3mediate-rpkg} and \texttt{medoutcon}~\citep{hejazi2022medoutcon}.
The empirical performances and properties are very close to the other robust estimators~\citep{huber2016finite}, so we have not reimplemented the TMLE estimator in Python.

\section{Performances on simulations}
\label{simu}
\subsection{Generation of simulated data}
\label{simulation_generation}
We have generated simulated datasets with a number of varying characteristics: 
\begin{itemize}
\item the number of observations $n \in \{500, 1000, 10000\}$ (3 possible values)
\item the dimension (and variable type - binary or continuous) of the mediator $M$, either $1$ (binary and continuous) or $5$ (continuous) (3 possible values)
\item the linearity (or not) of the mediator and outcome models ; we introduced non-linearities with interactions between the input variables. (4 possible options)
\item the mediated proportion, through varying the mediator parameter in the outcome model, and the common support assumption, through varying the strength of the association between treatment and mediator (3 sets of association strengths).
\end{itemize}

Overall, this yields a total of 108 ($3\times3\times4\times3$) simulated settings. Supplementary Table~\ref{sim_setting} summarizes the characteristics of the 36 main simulation settings (excluding the variation of the number of observations).

We note $K$ the number of dimensions of confounder covariates $X$.
In this set of simulations, $K$ is fixed at 5.
We define a confounder as a variable causally associated with at least two of the three variable types:  treatment, outcome and mediator.
We follow guidelines from~\citet{nguyen2021clarifying} and associate the (potentially disjoint) sets $T-Y$, $T-M$ and $M-Y$ confounders in a single set $X$.
Similarly for mediators, $L$ is the true number of dimensions of mediators (a true mediator is characterized by a non-zero causal effect from the treatment on the mediator, and a non-zero effect of the mediator on the outcome).

We have used the following simulation framework

\begin{align}
    X &\sim \mathcal{N}(0, I_p) \\
    T|X &\sim \mbox{Bernoulli}(\mbox{expit}(\alpha_0 + X^t\alpha_X)) \\
    M|X, T &\sim
    \begin{cases}
     \mbox{Bernoulli}(\mbox{expit}(\beta_0 + X^t\beta_X + \eqnmarkbox[Plum]{mm}{(TX)^t\beta_{TX}} + \eqnmarkbox[OliveGreen]{k1}{\omega_T}\beta_T T)) &  \text{if } M \text{ is binary}\\
    \beta_0 + X^t\beta_X + \eqnmarkbox[Plum]{mm2}{(TX)^t\beta_{TX}} + \eqnmarkbox[OliveGreen]{k2}{\omega_T}\beta_T T + \mathcal{N}(0, \sigma_M^2)  &  \text{if } M \text{ is continuous}\\
    \end{cases} \\
    Y|X, T, M &\sim \gamma_0 + \gamma_T T+ X^t \gamma_X + \eqnmarkbox[NavyBlue]{m1}{\omega_M} \gamma_M M + \eqnmarkbox[WildStrawberry]{MT}{\gamma_{MT} MT} + \mathcal{N}(0, \sigma_Y^2)
\end{align}
\annotate[yshift=-1em]{below}{MT}{non-linearity in the Y model}
\annotate[yshift=-3em]{below}{m1}{adjusting the strength of the indirect effect through M-Y edge}
\annotatetwo[yshift=3em]{above}{mm}{mm2}{non-linearity in the M model}
\annotatetwo[yshift=5em]{above}{k1}{k2}{adjusting the strength of the indirect effect through T-M edge}

\vspace{1cm}
For all simulations, we have set $\sigma_M = \sigma_Y = 0.5$.
The coefficients $\alpha_0, \beta_0$ and $\gamma_0$ are set to zero; $\alpha_X, \beta_X, \beta_T$, and $\beta_{TX}$ are drawn from a standard normal distribution, scaled by the total size of the coefficient, with $\beta_{TX}$ set to zero for simulations without non-linearity in the mediator model. %\ja{TODO: rather scale by the sq root of the size, remark from Bertrand}
For each simulation setting, a random seed is set, so that the coefficients do not vary among the repetitions, only the drawn data points.
The coefficients for the outcome model are respectively set to
\begin{itemize}
    \item $\gamma_0=0$;
    \item $\gamma_M = \textbf{1}_{L} \frac{1}{2L}$, with $\textbf{1}_d$ is the vector of length $d$ containing only ones;
    \item $\gamma_X = \frac{\textbf{1}_{K}}{K^2}$;
    \item $\gamma_T=1.2$;
    \item $\gamma_{MT}=\gamma_M$ if we introduce a non-linearity in the outcome model, and the null vector otherwise.
\end{itemize}

In the rest of the article, we refer to misspecification for the outcome or mediator models when we introduce a non-linearity, i.e. when $\gamma_{MT}$ or $\beta_{TX}$ are different from zero respectively.
The notion of misspecification is to understand with respect to the linear parametric specification of the nuisance parameters, used by default for most estimators, as described in Section~\ref{implementation}.

The coefficient $\omega_M$ is used to modulate the mediated proportion, without influencing the respect of the positivity or overlap assumption~\ref{a:seq-positivity}.
Increasing the $\omega_T$ coefficient also increases the indirect effect (and hence the mediated proportion), but can also generate violations to the positivity assumption~\ref{a:seq-positivity}.
We have manually defined three sets of values for the pair $(\omega_T, \omega_M)$ for each of the mediator types (different values as we consider the joint effect, regardless of the dimension of the mediator) to cover the following situations
\begin{itemize}
\item low $\omega_M$ and $\omega_T$: low mediated proportion and no overlap violation,
\item high $\omega_M$ and medium $\omega_T$: high mediated proportion and no overlap violation,
\item medium $\omega_M$ and high $\omega_T$: high mediated proportion and strong overlap violation.
\end{itemize} 
Supplementary Figure~\ref{sim_setting_overlap} illustrates the respect of the overlap assumption in simulated settings representative of those three situations.

The true effects of interest are defined as
\begin{align*}
\theta(t) &= \begin{cases}
\gamma_T + \gamma_{MT} \mathbb{E}_X\left[\mbox{expit}(\beta_0 + X^T\beta_X + (tX)^T\beta_{TX} + \omega_T\beta_T t)\right] & \text{if } M \text{ is binary}\\
\gamma_T + \gamma_{MT} \mathbb{E}_X\left[\beta_0 + X^T\beta_X + (tX)^T\beta_{TX} +\omega_T\beta_T t\right] & \text{if } M \text{ is continuous}
\end{cases} \\
\delta(t) &=
\begin{cases}
\mathbb{E}_X\left[\mathbb{E}\left[(\mbox{expit}(\beta_0 + X^T\beta_X + X^T\beta_{TX} + \omega_T\beta_T) -  \mbox{expit}(\beta_0 + X^T\beta_X)) (\omega_M\gamma_M  + t \gamma_{MT})\right]\right] & \text{if } M \text{ is binary}\\
\mathbb{E}_X\left[\mathbb{E}\left[\left((\beta_0 + X^T\beta_X + (X)^T\beta_{TX} +\omega_T\beta_T) -  (\beta_0 + X^T\beta_X)\right) (\omega_M\gamma_M  + t \gamma_{MT})\right]\right] & \text{if } M \text{ is continuous}.
\end{cases}
\end{align*}

We computed the empirical expectations on our simulated data to obtain the true effect values using extremely large samples ($10^5$).
The full detail of the computation can be found in Supplementary Paragraph~\ref{theoretical_effects}.

We generated 200 repetitions for each of the 26 combinations of parameters, with 3 dataset sizes, making a total of 21,600 simulated datasets.

\subsection{Uncertainty quantification}
Some of the implemented estimators have theoretical results regarding the quantification uncertainty, \emph{i.e.} provide asymptotic convergence results to obtain a confidence interval.
To compare all estimators on an equal ground, and to simplify the implementation work, we use bootstrap to obtain confidence intervals, as it has been shown to be reliable~\citep{hayes2013relative}.
To make a mindful usage of computing resources, we have attempted to assess a reasonable number of bootstrap repetitions.
Indeed, before bootstrap, we have more than $21,600\times20=432,000$ unique runs for all the datasets with all the methods (accounting for the fact that a few methods do not run in all the simulated mediator types), which then will be multiplied by the number of bootstrap repetitions.
The complete benchmark without bootstrap repetitions has a total time slightly above 8.5 CPU days.

We have selected a subset of 100 simulated datasets, encompassing all the configurations and all the dataset sizes, and generated 1,000 bootstrap samples.
We have then analyzed the stability of 95\%-percentile-confidence intervals for an increasing number of bootstrap repetitions, and selected 100 repetitions, as it seems that most methods in most settings have then reached a stable estimate, which already represents 867 CPU days.
We present a few representative illustrations of the confidence interval behavior in Supplementary Figure~\ref{bootstrap}.

\subsection{Results}
We have applied the five representative mediation estimators (with a total of 26 variants listed in Supplementary Table~\ref{list_estim}) on a family of simulated datasets with varying types of mediator, and degree of non-linearity of the outcome and mediator models. We refer to settings with non-linearity of the outcome or mediator models as \emph{misspecification}, in the sense that when estimators use parametric models for nuisance parameters estimation, this parametric model is linear, and hence misspecified to account for non-linearities. 
We compare the relative error for the total, direct and indirect effect, defined as $\frac{\hat{\tau}-\tau}{\tau}$, for example for the total effect $\tau$.

\subsubsection{General trends}
\label{general}

\begin{figure}[h]
\centering
\includegraphics[width=\textwidth]{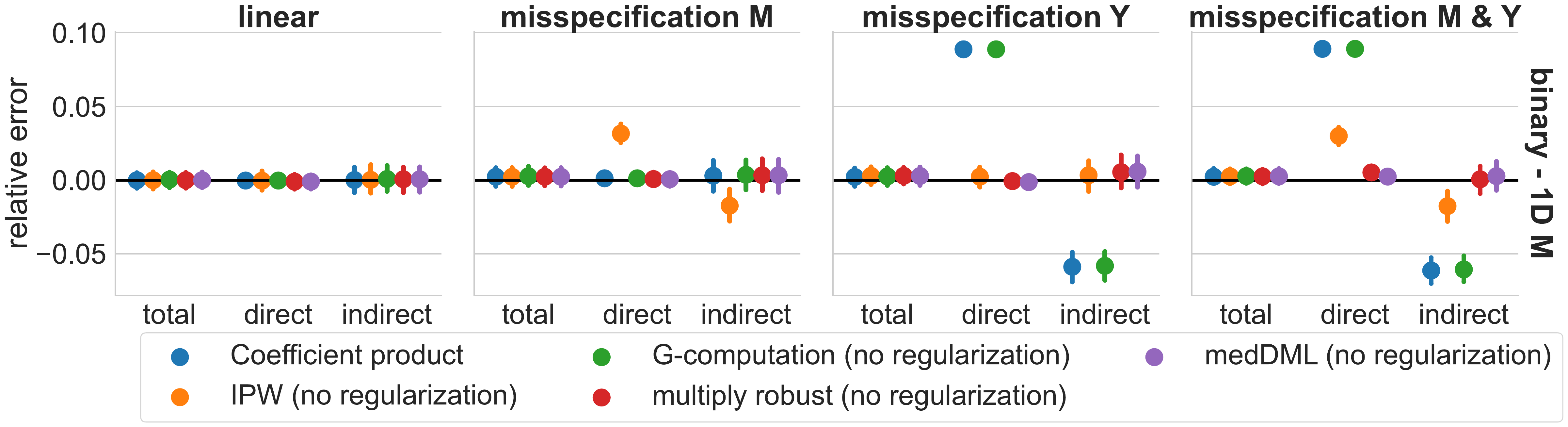}
\caption{
\textbf{Total and natural direct and indirect effects obtained with unregularized estimators for a binary mediator.}
We show results for four scenarios of generative model specification, violating or not the parametric linear nuisance models of some estimators.
Each column corresponds to a distinct specification of simulated models.
Each dot represents the average relative error (i.e. $\frac{\hat{\tau}-\tau}{\tau}$, for example for the total effect $\tau$) over 200 repetitions, and the error bars are the asymptotic normal 95\% confidence intervals from the distribution over the 200 repetitions.
All simulations are in the \emph{high mediated proportion without overlap violation} framework with $n=1,000$ observations, in the case of a binary mediator.
The total effect is generally well estimated in all situations, but not the direct and indirect effects.
The indirect and direct effects have different value, leading to a lack of symmetry in their relative error values, even if their absolute values are equal, leading to no error for the total effect estimation.
Model misspecifications lead to estimation errors for most estimators, especially when the outcome model is misspecified.
The results for all types of mediators are shown in Supplementary Figure~\ref{result_overview_all}.
}
\label{result_overview}
\end{figure}

We first compare the estimators using the simplest setting (non-regularized generalized linear models in Fig.~\ref{result_overview}), and then assess the contribution of more complex estimation approaches.
In Fig.~\ref{result_overview}, we present the relative errors for the total, direct and indirect effects with the \emph{high mediated proportion without overlap violation} framework and $n=1,000$ observations, for a binary mediator.
Complete results for all estimator types are shown in Supplementary Figure~\ref{result_overview_all}
The total effect is almost always well estimated, but not the direct and indirect effects that exhibit opposite errors.

For the simplest simulations with a one-dimensional binary mediator, the different estimators behave mostly as expected depending the theoretical properties.
All estimators achieve a very small or even negligible error in the \emph{linear} setting where all models are well specified.
The misspecification of the mediator model induces an error in the IPW estimator that relies solely on the estimation of the propensity of the treatment given the covariates, and on the propensity of the treatment given the covariates and the mediator; estimation of the latter can be impaired by the introduction of the interaction between the covariates and the mediator.
The misspecification of the outcome model leads to a high estimation error for the estimators that rely the most on regression approaches of the outcome given the treatment, covariates and mediator, namely the coefficient product and the G-computation.
Interestingly, estimators that require both the mediator and the outcome models to be correctly specified (\emph{ie} the coefficient product and the G-estimation estimators) are more robust to a misspecification of the mediator model.
This might be due to our simulation setting where the binary mediator has a more modest range of possible error, while the outcome is a continuous variable.
The multiply robust estimator and the double machine learning estimator have no error in the case where only the mediator or the outcome model, which illustrates their theoretical robustness property.
Finally, when both the mediator and the outcome model are misspecified, all the non-robust estimators exhibit an erroneous estimation of the direct and indirect effects.
This result is surprising as two misspecified models exceed the theoretical robustness property of the multiply robust and the double machine learning estimators, which might be related to the "mildness" of the mediator model misspecification mentioned above.

We observe a similar pattern for the results with a continuous one-dimension mediator.
In that setting, only the coefficient product, the IPW and the double-machine-learning estimator provide a result.
All estimators have a very small or no error when all models are linear; misspecification of the mediator but not the outcome model leads to an error for the IPW estimator, and the opposite for the coefficient product.
The double machine learning estimator~\ref{medDML}, with a simple unregularized linear nuisance estimation, is robust to misspecification, even of both the outcome and the mediator models.

For a mediator, we observe a similar behavior of the IPW and the coefficient product estimators.
The G-computation estimator is also robust to the misspecification of the outcome model, which can be explained by a slightly different implementation for the implicit integration case that is used for continuous mediators, which allows more robustness to interactions, which is the type of misspecification used in this simulated dataset.
In the case of a five-dimensional mediator, the IPW estimator exhibits a smaller error when the mediator model is misspecified, which might be linked to an intrinsic flexibility allowed by more variables and more parameters, that could compensate for the restriction of the linear parametric form~\cite{kunzel2019metalearners}.

Beyond the bias of the estimator, we have also explored the uncertainty of the estimates obtained with the different estimators, to go beyond the uncertainty visible so far that is due to the 200 repeated simulations of each situation.
To that end, we generate 100 bootstrap samples in each case, and obtain a 95\%-confidence intervals based on the 2.5 and 97.5 percentiles of the empirical bootstrap distribution.
Supplementary Figure~\ref{bootstrap_coverage} presents the coverage of the confidence intervals \emph{ie} the proportion of simulations where the true value is inside the confidence interval.
By construction, this proportion should be 95\%.
We observe results that are very similar to what is obtained for the estimation error.
We also consider the confidence interval width, represented in Supplementary Figure~\ref{bootstrap_width}.
A valuable insight is that the IPW estimator systematically has wider confidence intervals compared to the other methods, and that the doubly-robust estimators has slightly narrower intervals than the G-computation approach.
This can constitute a discriminative criterion for the choice of the estimator as both estimators have very similar results for the estimation error.

\subsubsection{Effect of the mediated proportion and overlap assumption}
This first part of the results focuses on the assumptions about the parametric form of the nuisance parameter models, but the identification assumptions are also key to the validity of a causal mediation analysis.
We now explore the impact of the size of the effect to estimate, and the validity of the overlap assumption.
\begin{figure}[h]
\includegraphics[width=\textwidth]{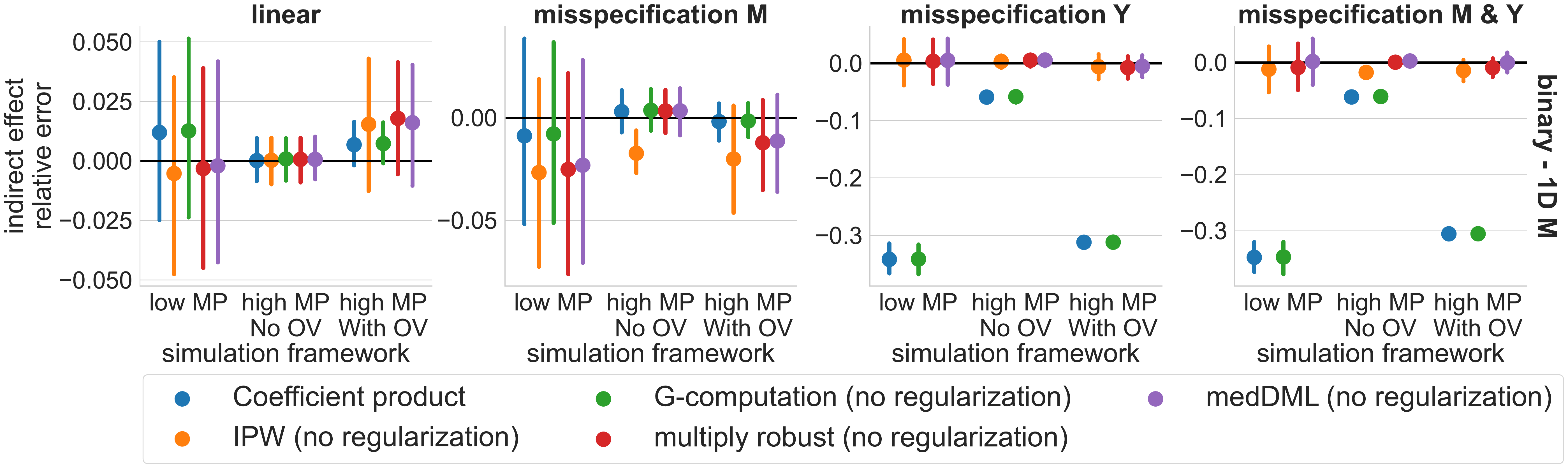}
\caption{
\textbf{Indirect effect estimation across different simulation frameworks with a binary mediator.}
We show results for four scenarios of generative model specification, violating or not the parametric linear nuisance models of some estimators.
Each column corresponds to a distinct specification of simulated models.
The three sets of dots in each panel correspond to the three simulation frameworks, in that order: \emph{low mediated proportion without overlap violation}, \emph{high mediated proportion without overlap violation} and \emph{high mediated proportion with strong overlap violation}.
"MP" stands for "mediated proportion" and "OV" for "overlap violation".
%relative
Each dot represents the average relative error (i.e. $\frac{\hat{\tau}-\tau}{\tau}$, for example for the total effect $\tau$) over 200 repetitions, and the error bars are the asymptotic normal 95\% confidence intervals from the distribution over the 200 repetitions.
All simulations are done with $n=1,000$ observations.
The estimation error for most estimators is greater when there is a violation of the overlap assumption, including in robust estimators.
The absolute value of the indirect effect in the \emph{low mediated proportion without overlap violation} is often very low, leading to very high relative errors compared to other settings.
We found however that the representation of the relative error is informative.
To enable the reader to have a thorough overview of the results, we have included the same figure with the absolute error as Supplementary Figure~\ref{result_framework_abs}, as well as for other types of mediators.
}
\label{result_framework}
\end{figure}

In Figure~\ref{result_framework}, we compare the relative error in the estimation of the indirect effect in the three different simulation frameworks defined in Section~\ref{simulation_generation}:~"low mediated proportion without overlap violation", "high mediated proportion without overlap violation" and "high mediated proportion with strong overlap violation".
For consistency with other graphical representations, and to allow comparison between the different simulation frameworks, the estimation error in Figure~\ref{result_framework} is presented as a relative error.
However, the indirect effect in the "low mediated proportion without overlap violation" is very low, so even a small error leads to a large relative error, which can be misleading, so we also represent the absolute errors in Supplementary Figure~\ref{result_framework_abs}.
In both representations, we observe again that the violation of the overlap or positivity assumption leads to a higher error for almost all estimators, compared to the same order of magnitude of the mediated proportion without violation of this assumption.
In particular, it breaks the multiply robust estimator and the double machine learning estimator that would otherwise exhibit very low error, even with model misspecification.
This is particularly obvious when considering the coverage shown in Supplementary Figure~\ref{result_framework_coverage} and the width of the 95\% bootstrap confidence intervals, shown in Supplementary Figure~\ref{result_framework_width}.

For the interpretation of results for the "low mediated proportion without overlap violation" framework, we refer to the Supplementary Figure~\ref{result_framework_abs}, with the absolute error.
We observe estimation errors similar to the "high mediated proportion without overlap violation" framework largely commented in Section~\ref{general}.
The apparently paradoxical very high relative error when the models are not misspecified observed for this framework in Figure~\ref{result_framework} is only due to the relative error representation.

\subsubsection{Effect of the number of observations}

\begin{figure}[h]
\includegraphics[width=\textwidth]{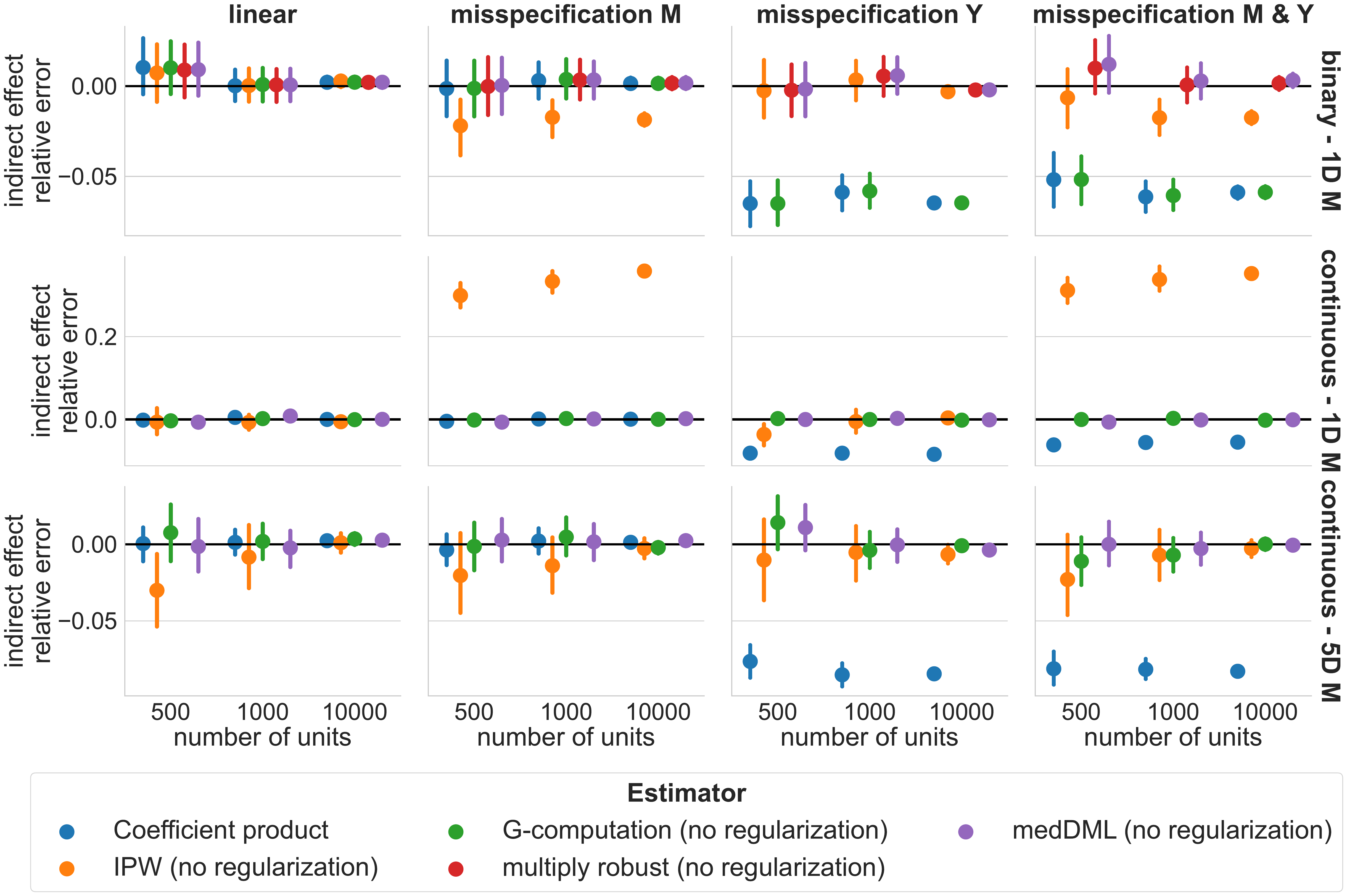}

\caption{
\textbf{Relative indirect  effect for different number of observations.}
We show results for four scenarios of generative model specification, violating or not the parametric linear nuisance models of some estimators.
Each column corresponds to a distinct specification of simulated models.
The rows correspond to different mediator types, labeled on the right.
The three sets of dots in each panel correspond to estimations with an increasing number of observations from left to right ($n=500$, $n=1,000$ and $n=10,000$ on each panel).
Each dot represents the average relative error (i.e. $\frac{\hat{\tau}-\tau}{\tau}$, for example for the total effect $\tau$) over 200 repetitions, and the error bars are the asymptotic normal 95\% confidence intervals from the distribution over the 200 repetitions.
All simulations are in the "high mediated proportion without overlap violation" framework.
The multiply-robust estimator only handles binary one-dimensional mediators, and has no results for the second and third rows.
Overall, the position of each dot is quite steady over the three possible numbers of observations considered.
Increasing the number of observations decreases the estimation uncertainty.}
\label{result_sample_size}
\end{figure}

In Figure~\ref{result_sample_size}, we compare the relative error in the estimation of the indirect effect for three different numbers of observations.
We observe little effect of the number of observations on the magnitude of the error.
The main impact of increasing the number of available observations is to decrease the variance of the estimation, thus resulting in a more reliable result.
For estimators with non-parametric models for the estimation of nuisance functions (Supplementary Figures~\ref{result_linear_reg} and~\ref{result_linear_forest}), the behavior is different, and the relative error in the estimation of the indirect effect decreases with the increase of the number of observations, hinting at slow estimator convergence.

\subsubsection{Practical recommendations to select an estimator and estimate nuisance parameters}

\begin{figure}
\centering
\includegraphics[width=\textwidth]{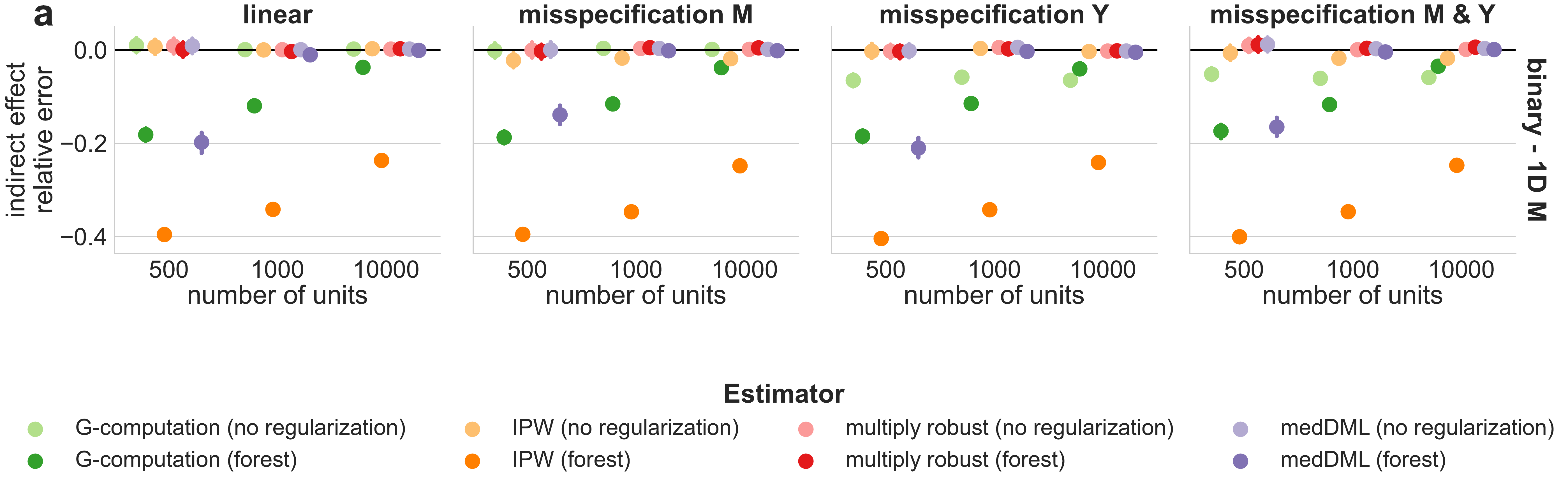}
\includegraphics[width=\textwidth]{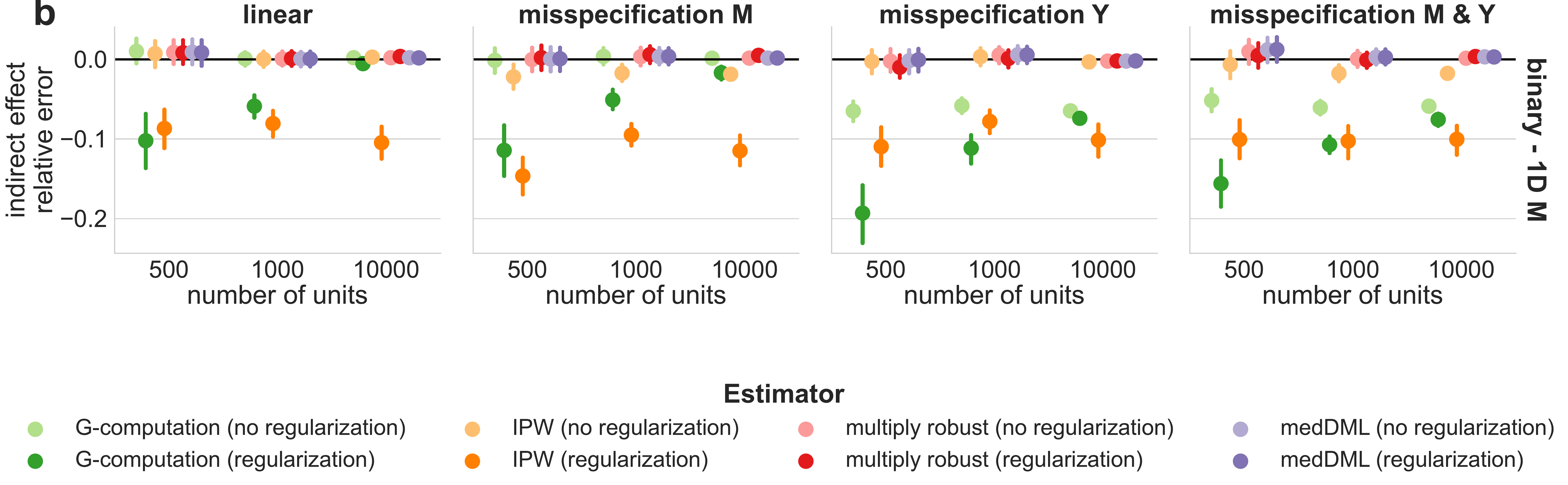}
\includegraphics[width=\textwidth]{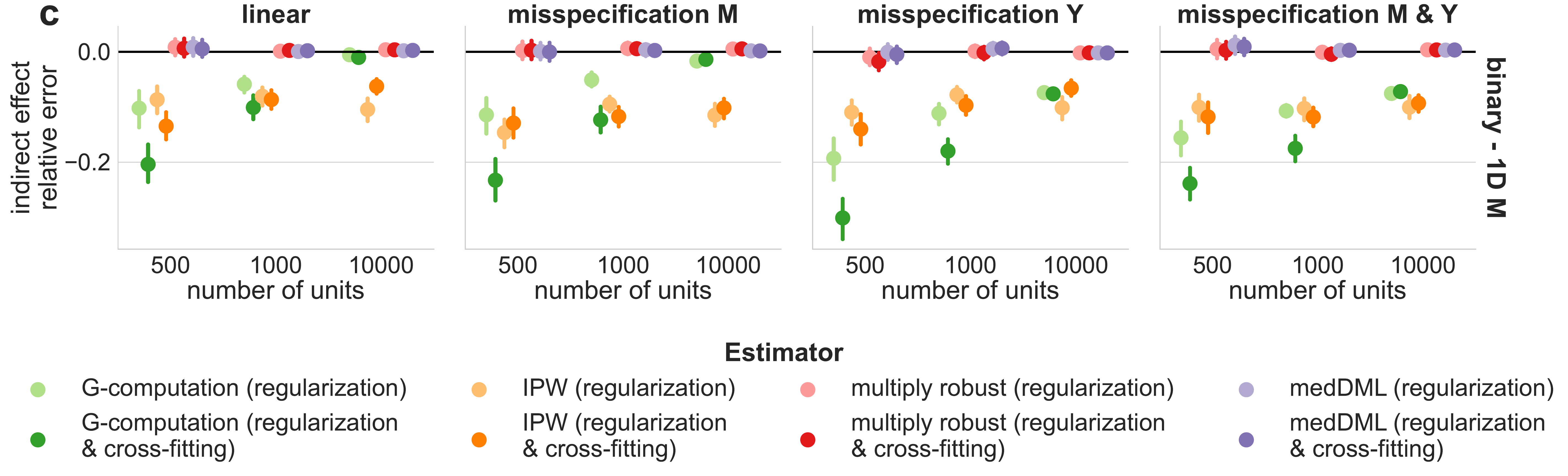}
\includegraphics[width=\textwidth]{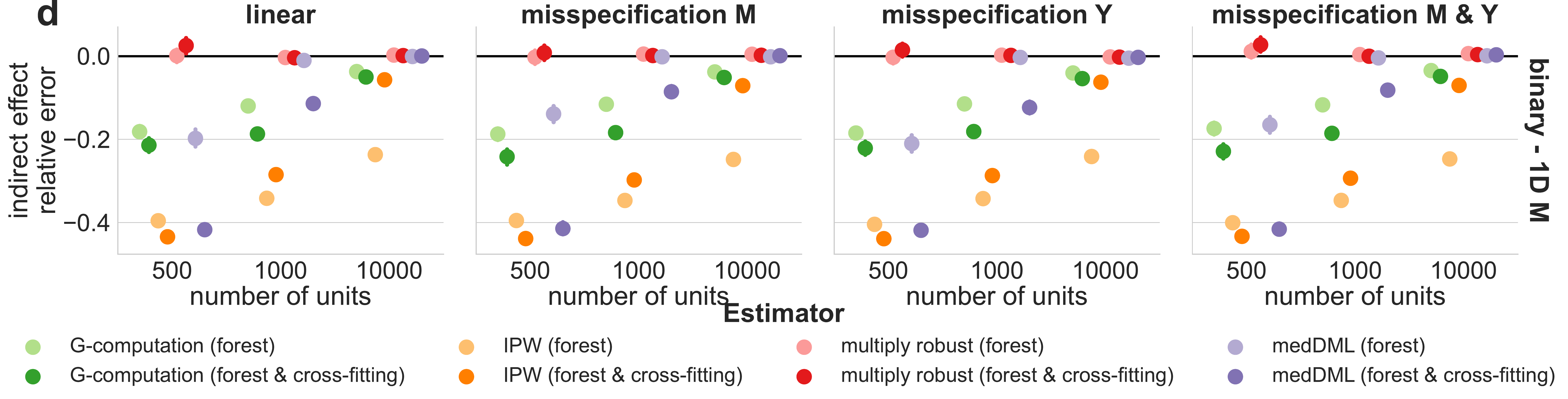}
\caption{
\textbf{Relative indirect  effect for different plug-in nuisance parameters estimation strategies.}
We show results for four scenarios of generative model specification, violating or not the parametric linear nuisance models of some estimators.
Each column corresponds to a distinct specification of simulated models.
Each panel compares two strategies for plug-in nuisance parameters estimation: \textbf{(a)} Random forest versus linear models, \textbf{(b)} linear models with or without L2-regularization, \textbf{(c)} regularized linear models with or without cross-fitting, \textbf{(d)} Random forests with or without cross-fitting.
The three sets of dots in each panel correspond to estimations with an increasing number of observations from left to right ($n=500$, $n=1,000$ and $n=10,000$ on each panel).
Each dot represents the average relative error (i.e. $\frac{\hat{\tau}-\tau}{\tau}$, for example for the total effect $\tau$) over 200 repetitions, and the error bars are the asymptotic normal 95\% confidence intervals from the distribution over the 200 repetitions.
All simulations are in the "high mediated proportion without overlap violation" framework.
Results are shown only for the binary one-dimensional mediators setting, full results are available in Supplementary Figures~\ref{result_linear_reg}) to~\ref{result_forest_cf}).
Overall, alternative estimation strategies show very limited improvements or even performance deteriorations compared to the unregularized linear models, probably due to the very regular and simple simulation setting, and the slower convergence rate.}
\label{summary}
\end{figure}

We further explore the effect of some implementation variations of nuisance parameters estimation, namely the use of non-parametric models such as random forests~\citep{breiman2001random} (Figure~\ref{summary}-a, Supplementary Figure~\ref{result_linear_forest}), the regularization of machine learning models (Figure~\ref{summary}-b, Supplementary Figure~\ref{result_linear_reg}), the use of cross-fitting for a faster convergence~\citep{chernozhukov2018double} (Figure~\ref{summary}-c-d, Supplementary Figures~\ref{result_linear_reg_cf} and~\ref{result_forest_cf}).

Using machine learning models more complex than plain unregularized linear models, such as regularized linear models (Supplementary Figure~\ref{result_linear_reg}) or random forests (Supplementary Figure~\ref{result_linear_forest}) generally yields a larger estimation error for the G-computation and IPW estimators. For the multiply robust and the double machine learning estimators, the results remain almost unbiased for all situations, except with random forests for nuisance estimation with continuous mediators.
Cross-fitting (Figures~\ref{result_linear_reg_cf} and~\ref{result_forest_cf}) does not improve the performances.

Overall, our results show that the statistical robustness properties of some estimators are a better choice than the use of flexible non-parametric machine learning models in the case of model misspecification.
However, the variance of the robust double machine learning or multiply-robust estimator increases importantly in the case of overlap violation.
This failure can be a good way to practically detect this violation, and that the effects of interest are actually not identified.

When overlap is insufficient, interpolating effects from regions with overlap to those without can introduce bias. 
In the case where the user trusts that the effects are not too variant with the value of the covariates (low effect heterogeneity) interpolation of the effects from the regions of the feature space with overlap to the regions that lack overlap.
In such cases, employing G-computation can mitigate this issue by estimating causal effects across the entire feature space.

Alternatively, enhancing the parametric models for both the mediator and outcome variables can improve estimation accuracy. Utilizing variants of the coefficient product estimator, which multiplies coefficients from different model components, can further refine the analysis~\citep{vanderweele2015explanation}.
Those more complex models include for example the use of a proper model depending on the type of the mediator (logistic regression for a binary mediator instead of a standard regression), or parametric regressions with interactions or engineered non-linear transformations of the covariates.
These approaches ensure that causal inferences remain robust and reliable, even in the presence of limited overlap.
We strongly recommend to use different estimators, and the comparison of the obtained results can help diagnose whether the main issue is model misspecification or lack of overlap.

%
% \begin{figure}[h]
% \includegraphics[width=\textwidth]{figures/direct_effect_forest_impact.pdf}
% \caption{\textbf{Effect of using a parametric or a non-parametric model for plug-in nuisance parameters estimation.}}
% \label{effect_forest}
% \end{figure}

\section{Application to cognitive function in the UK Biobank} \label{application}
The results on simulations constitute a valuable resource to understand and gain intuition on the strengths and limitations oh the different estimators.
The analysis of real data often proves much more complex, and we aim at providing a realistic example of how to conduct a causal mediation analysis on actual data, with the associated difficulties, \emph{eg} important number of variables, missing data.
We also illustrate how the results from the simulations can be used to diagnose the behavior of the estimators on unknown data.
Moreover, beyond the nuances between different estimators, we aim to demonstrate that the validity and robustness of a causal mediation analysis also depend on the entire process.
This includes formulating the question, explicitly stating the identifiability assumptions, making modeling choices, and, most importantly, clearly specifying and explaining all choices and reasoning.
This allows for scientific debate on the obtained results, for example, by following the guidelines established by~\citet{lee2021guideline}.

\subsection{Definition of the causal mediation question}
\begin{figure}
\centering
\includegraphics[width=\textwidth]{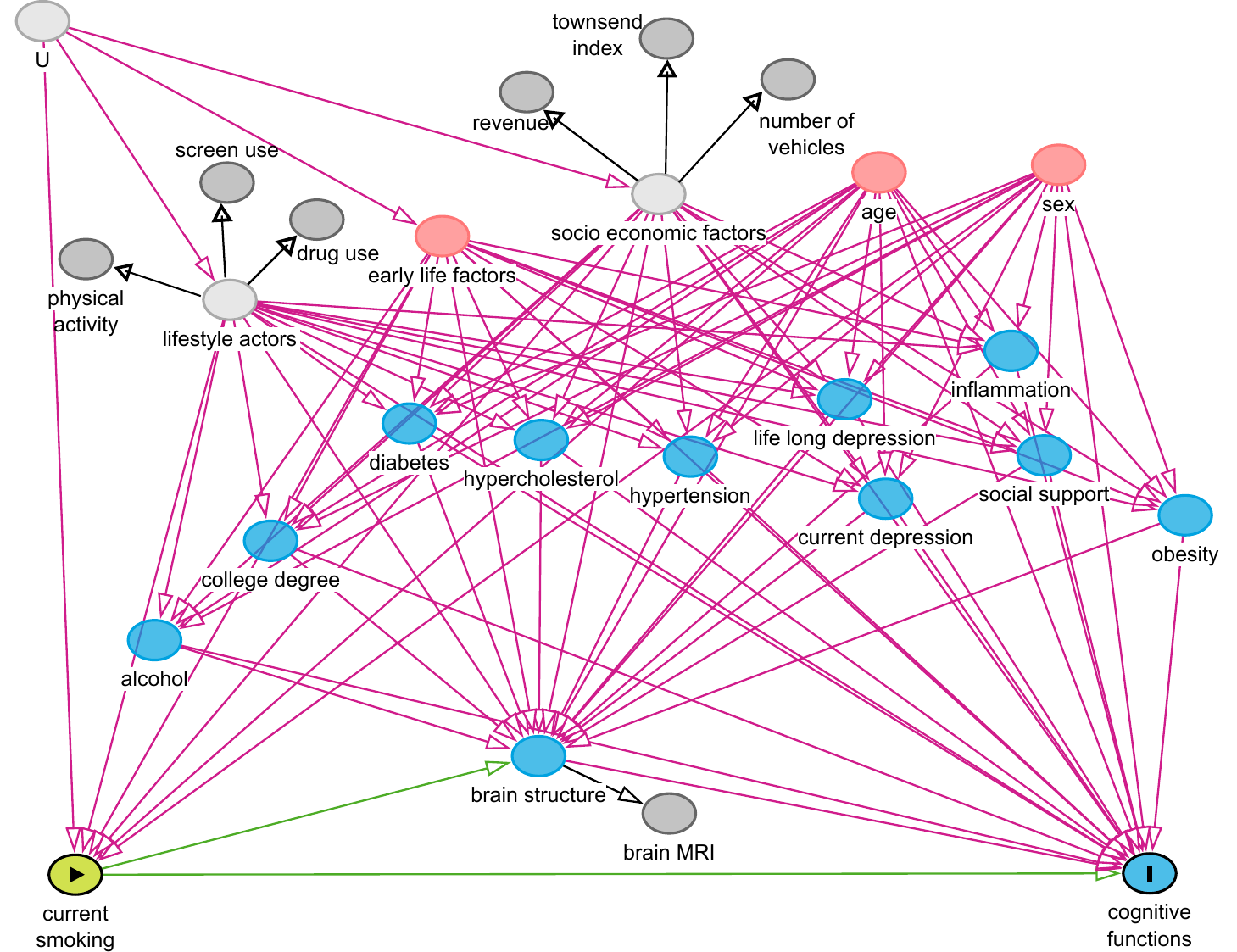}
\caption{\textbf{Simplified causal graph illustrating the potential mediating role of the brain structural characteristics.} To preserve readability, not all variables included in the analysis are represented. Several confounders are not directly observed, are difficult to define and properly assess. Thus they are grouped under the \texttt{U} node (Unobserved), such as genetic factors influencing the behavior or the physiology, environmental factors, developmental factors, major life events, personality traits, propensity to addictions. We assume that observed covariates cover all relevant aspects of those unobserved confounders. Other characteristics are measured and included in the adjustment set, such as lifestyle factors beyond alcohol and tobacco consumption, early life factors and the socio-economic status. We finally adjust for age and sex, as well as the all the alternative exposures, assumed to block the backdoor paths with unobserved nodes. This graph was created with Dagitty~\citep{textor2016robust}, and the source code to reproduce it can be found in Supplementary Code~\ref{dagitty}, with the list of all edges in the graph.}
\label{causal_graph_ukbb}
\end{figure}

The UK Biobank imaging project aims to collect brain scans from 100,000 participants (around 40,000 at the time data was collected for this research).
%
%This constitutes an unprecedented healthy individuals brain imaging cohort size, and makes it possible to use
This study makes it possible to use machine learning to derive objective measures of brain health without requiring in-depth human expert analysis.
Previous studies found that to predict some constructs such as fluid intelligence and neuroticism, the joint use of both image-derived phenotypes (IDPs) and socio-demographic variables brought little to no performance increase compared to socio-demographic alone, although the IDPs alone have a non-zero predictive power~\cite{dadi2021population,cox2019structural}.
This leads us to consider the hypothesis that the potential causal effect of some relevant socio-demographic features on cognitive function is actually mediated by the brain structure, which is captured by brain imaging.
Furthermore, studies have shown that brain MRI captures biomarkers reflecting individuals' overall health status and the pace of their aging~\citep{cole2018brain}.
This supports our mediator hypothesis, which incorporates various types of exposures throughout an individual's life.
This hypothesis is illustrated in the simplified causal graph presented in Figure~\ref{causal_graph_ukbb}.

To explore further this general hypothesis, we consider several exposures (or treatments) that have been shown to influence the brain structure and the cognitive functions.
We have designed the following binary treatment variables from the raw UKBB variables: college or university degree, diabetes, hypercholesterol, hypertension, current depression, lifelong depression, social support, inflammation, current smoking, alcohol consumption, and obesity.
More details about the UKBB variables used, and how they were aggregated to obtain binary exposures can be found in Supplementary Table~\ref{ukbb_exposure}.

The choice of considering several potential treatments with the same mediator allows us to tackle the more general problem of the factors of brain plasticity with respect to complex and intricate life factors rather than making strong claims about the particular mechanism of one exposure in particular.

\subsection{Establishing identifiability: discussion the plausibility of the assumptions}
After the definition of the causal question at hand, and the definition of the hypothesized potential causal relationships between the variables, broadly sketched in the graph of Figure~\ref{causal_graph_ukbb}, one needs to check the identifiability assumptions.
The considered treatments are precisely defined and present no interaction between the subjects, as they are independent participants to the UKBB study, hence SUTVA is satisfied.
%\ref{a:seq-treatment}, \ref{a:seq-mediator}, and \ref{a:seq-positivity}
Assumptions~\ref{a:seq-treatment} and~\ref{a:seq-mediator} focus on the confounding variables of either the treatment and the mediator(s), the mediator(s) and the outcome, and of course between the treatment and the outcome for the identifiability of the total potential causal effect, as well as the direct and indirect effects.
We have included a wide diversity of confounding variables to cover as much as possible the potential behavioral, physiological and societal determinants of the exposures, mediators, and outcomes, as well as sources of technical artifacts such as the evaluation center, or the position of the head in the imaging device, or known brain diseases~\citep{alfaro2021confound,newby2022understanding,schurz2021variability,topiwala2022alcohol}.
The positivity assumption~\ref{a:seq-positivity} can be empirically verified by comparison of the distributions of each confounder among the treated and untreated, relying on the nuisance parameter that predicts the probability of the treated versus untreated, conditionally on the covariates and mediators.

A second aspect of establishing identifiability is the preparation of variables to be included in the estimation models.
Beyond some grouping rules to construct aggregate variables for treatment, like depressive syndromes detailed in the previous subsection, some groups of variables have been combined together as described below.

The first transformed variable is the outcome where several cognitive tests assessing different functions (e.g. working memory, fluid intelligence, vocabulary) into a single score measuring the general cognitive ability ("g-factor"), obtained as the first component of a Principal Component Analysis (PCA)~\citep{fawns2020reliability}.
To avoid missing values, we restricted the construction of the g-factor to four tests at the imaging visit: pairs matching (field 398), the fluid intelligence score (field 20016), the prospective memory result (field 20018), and the reaction time (field 20023).
As expected, the g-factor we obtained with PCA with standard scaling (package Scikit-Learn~\citep{scikit-learn}) was well correlated with the score of each test.

The second transformed group of variables represents imaging data.
To account for a thorough joint mediation analysis by brain imaging data, we collected all the preprocessed variables (IDPs - imaging-derived phenotypes)~\citep{miller2016multimodal,alfaro2018image}, namely derived from the dMRI skeleton (Category 134, 432 variables), Subcortical volumes (Category 1102, 14 variables), Regional gray matter volumes (Category 1101, 139 variables), T1 structural brain MRI (Category 110, 26 variables).
It should be noted that these variables only describe structural information on the brain, and not function. 
We reduced the dimension using PCA after standard scaling, and retained 6 components by visual inspection of the explained variance barplot for each component~(Supplementary Figure~\ref{mri_pca}).

Finally, we aggregated the variables describing the lifestyle of participants, for two reasons: first reduce their dimensionality to facilitate the convergence of nuisance estimation models, and second to account for missing data due to the structure of the questionnaires where some questions are only presented to a participant depending on their answer to a previous question.
We selected the relevant variables, transformed the categorical variables using one-hot-encoding, and only retained the variables with more than 10,000 non-zero values to eliminate the least used modalities.
We then imputed the missing values and reduced the dimension of the data with the R packages missMDA and FactoMineR~\citep{josse2016missmda,le2008factominer}.
The analysis was then restricted on participants with brain MRI data.
We performed the whole analysis with 2 or 5 components, with little influence on the final results for causal effects estimation, as shown in Supplementary Figures~\ref{diff_ncp_hyper} and~\ref{diff_ncp_smoke} for the effect of hypertension and of smoking.
We present in the main text the results with 2 components for imputation.
% \bt{IIUC this is a reduction of X ? It is actually a rather important step, that may affect  the results. Is it standard practice ?}
% \bt{I think it would be good to have somewhere the description of the models proves: what is in X, M, T, Y for each one ? }
% \ja{the models with X, M, T and Y are the tables actually... not so easy to read, but at least it is complete, to enhance reproducibility?}
% \bt{You're not mentioning the bootstrap procedure if I'm not mistaken}

%\ja{actually, there are missing values because of the structure of the questionnaire in UKBB, if you say you are a former smoker, you are asked when you started, how many cigarettes etc, and nothing if you say you have never smoked. Same for mental health questionnaire. And there are a lot of questions. After one hot encoding + filtering for modalities with N>10,000, I had 182 variables for lifestyle, and 239 for mental health}
% \ja{so this was a way to reduce the dimension and impute the missing values automatically rather than by hand for dozens of variables.}
%
We constructed two groupings of such variables: one for mental health variables (Categories Social support (100061), Mental health (100060), Addictions (141), Alcohol use (142), Anxiety (140), Depression (138), Happiness and subjective well-being (147), Unusual and psychotic experiences (144), Self-harm behaviors (146), Traumatic events (145), Mania (139), Mental distress (137)); and the other for lifestyle variables other than alcohol and tobacco use that were adjusted for separately (Categories Physical activity (100054), Electronic device use (100053), Sleep (100057), Cannabis use (143)).
Adjustment on the derived principal components for mental health summary variables was done for all potential treatments except current and lifelong depressions, and for lifestyle summary variables (see Supplementary Tables~\ref{table1_education} to~\ref{table1_obesity}).

\subsection{Estimation: computation of the direct and indirect effects with different estimators}
\begin{figure}
\centering
\includegraphics[width=\textwidth]{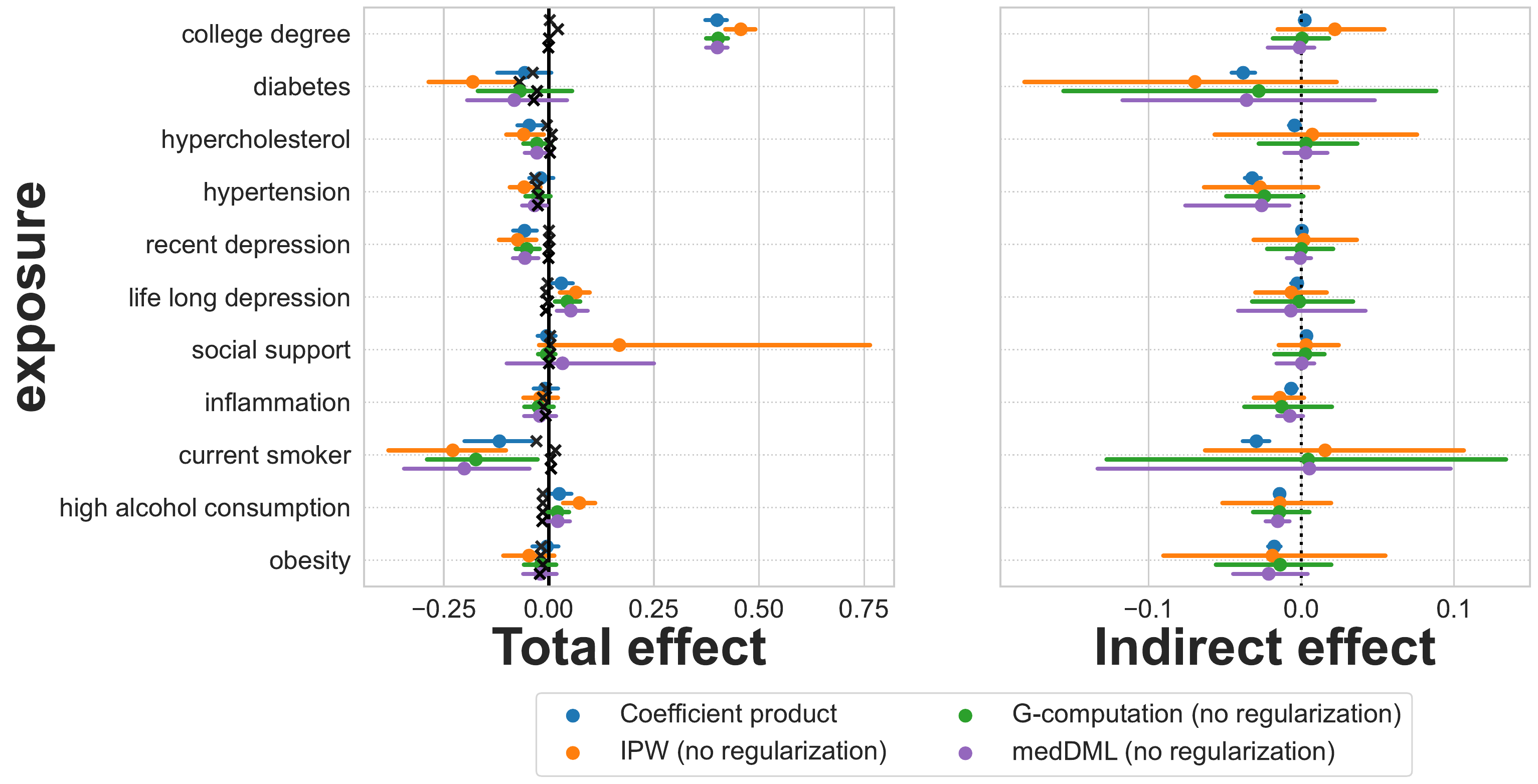}
\caption{\textbf{Total and indirect effect mediated by the brain structure for 11 exposures, estimated by the coefficient product, the double machine learning, the G-computation and the IPW estimators.}
The left panel represents the total effect as a point over 100 bootstrapped samples.
Each point represents the mean of the distribution.
The error bar represent the 95\% percentile-interval, estimated from the empirical bootstrap distribution
We superimpose "x" crosses to represent the indirect effect on the left panel, where each point represents the average of the indirect effect over 100 bootstrapped samples.
The right panel represents a focus on the indirect effects which are of smaller amplitude.
The indirect effect for each exposure is represented by a pointplot, with the same representation rules as for the left panel.
We can consider an effect to be significantly distinct from zero if the error bars do not cross the zero vertical line.}
\label{all_exposures}
\end{figure}

After the selection and the preparation of the different variables to include into the causal mediation analysis (either as an exposure, a confounder, a mediator or an outcome, as detailed in Supplementary Tables~\ref{table1_education} to~\ref{table1_obesity}), we selected participants with brain imaging data, and no missing values for the other considered variables, resulting in a dataset of $N=32,498$ independent participants.
To estimate the total, direct and indirect potential causal effects for each treatment, mediated by brain structure, measured by brain MRI, we applied the estimators able to handle multidimensional mediators: the coefficient product estimator, and several variants of implementation of the double machine learning and IPW estimators.
To assess uncertainty in the estimation, we applied the estimators to 100 bootstrapped samples for each exposure.
We present general results for the three estimators for all considered exposures in Figure~\ref{all_exposures}.
Overall, the results depended weakly on the choice of the estimation method, nor the choice of the dimension for missing values imputation, as shown for two representative exposures: hypertension and current smoking in Supplementary Figures~\ref{diff_ncp_hyper} and~\ref{diff_ncp_smoke}.
However, we notice an important difference in the variance between the different estimators, with a very small variance for the coefficient product, and a much larger one for the double machine learning, the G-computation and IPW estimators.
This indicates that there is no strong non-linearity in the nuisance parameters models, but that there might be overlap issues.

The interpretation of the estimated effects has to be conducted carefully, as those effects are only causal if all the identification assumptions are valid, otherwise those effects are actually associational.
To make it clearer, we will refer to those effects as potentially causal.
The exposures that have the potential larger impact on cognitive functions are the initial training obtained degree, as having obtained a college degree or equivalent ensures a much greater g-factor value.
On the contrary, being currently smoking ($N=605$) or suffering from diabetes is quite deleterious and results in a lower g-factor score.
Several physiological disorders, such as diabetes, hyper-cholesterol, hypertension, inflammation and obesity tend to have a negative total potentially causal effect on cognitive functions, though not always found significant in our study.
Finally psychological exposures (recent or lifelong depression and social support) have conflicting directions that are difficult to interpret.

Regarding the potential indirect effect estimates, they are systematically of much weaker amplitude than the total effect, except for obesity, hypertension, diabetes and alcohol consumption, indicating a strong contribution of the structural modifications to the total estimated causal effect.
We found a significant negative potential indirect effect for exposure to hypertension, inflammation, alcohol consumption, diabetes and obesity, a rather negative but not significant indirect effect for current smoking, and diabetes perhaps due to a high variance associated with the low number of exposed participants.
We finally report a null effect for college degree, hypercholesterol, recent or lifelong depression, and social support, indicating other mechanisms are involved in the variation of potential total effect of those exposures on cognitive function.
Considering the simulated results, and the agreement between the coefficient product and the other methods, for almost all exposures, it could be reasonable to consider this estimator's confidence interval for the exposures where all methods agree, yielding a much narrower confidence interval.

\section{Discussion and Conclusion} \label{discussion_conclusion}

In this work, we have proposed a complete overview of causal mediation analysis.
We first described the framing of the problem in the potential outcomes framework and identifiability assumptions.
We then presented the estimators for the causal effects of interest (the total effect, and the decomposition in the direct and indirect effect through the mediator(s)) ranging from the most classical to more recent approaches.
And finally we detailed an illustrated example with all the analysis steps to grasp the range of possible results: total effect versus path-specific effects, and different proportions of the indirect effect compared to the total effect.
The different exposures are also different with respect to the positivity assumption respect.

In simulations, we illustrated the limitations of the purely parametric models, in particular the widely used method of coefficient product, as such methods can heavily fail when the true relation between variables is not linear.
We have included more recent approaches, as well as a Python implementation that was not available so far, and illustrate their very satisfying performances in a wide variety of simulated settings.
Our experiments on the complex UKBB dataset shows they can be successfully applied to large-scale datasets.

Following previous recommendations for causal mediation analysis~\citep{rohrer2022sa,stuart2021assumptions,lee2021guideline}, we strongly emphasize the importance of the plausibility, but mostly of the clear statement of the underlying identifying assumptions.
A potential limit of our application to the UKBB data is that this is cross-sectional data, where a clear temporal order is not explicitly established by the way the data is collected.
However, we have selected exposures that consist in explicit past events (college education) or chronic habits or conditions that should be relatively stable over time and be less sensitive to the collection date.
Nonetheless, important assumptions regarding the causal relationships between the variables are made to ensure identifiability, as illustrated in the causal graph.
Regarding the confounding variables, we have made the strong assumption that the observed variables are a sufficient adjustment set.
An important confounder would be the brain structure (or brain imaging) at an earlier age, to ensure that some exposures like depression, are not caused by some early brain anomaly.
The exclusion of patients with known brain disease slightly mitigates this risk, but repeated measures would have significantly strengthened the causal claim.
An important underlying assumption is not explicitly stated, but can be deduced from the identifiability in the case of several mediators.
If there are other mediators that are causally related to the mediator of interest, what we compute as the indirect effect might not entirely cover the full indirect effect.
We may think for example to a mediation of obtaining a college degree by the type of professional occupation (intellectual or manual occupation, potentially with odd hours) - node $M_2$ in Figure~\ref{multi_med}, which might in turn affect the brain structure - node $M_1$ in Figure~\ref{multi_med}.
If that is the case, as we do not account for $M_2$, we are actually computing $\delta^{M_{1_p}}$ instead of $\delta^{M_1}$ (with notations introduced in section \ref{sec:several}).
This is a general limit of mediation analysis as one never considers all possible mediators and their complex relationships.
Moreover, we have used a limited repertoire of brain variables (structural only) and reduce them quite strongly with PCA, which may lead to the underestimation of the indirect effect.

The assumptions that support the causal claim for mediation analysis are extremely strong, and stronger than for standard causal inference from observational data, and might be deemed implausible in all situations.
Nonetheless, working in the causal framework encourages to a very rigorous and a very explicit framing of the underlying hypotheses, which can then be discussed through an open scientific discussion.
Even with debatable validity, this constitutes a much more rigorous framework than other fields of data analysis where the line between interpretation of a predictive and causality is very fuzzy~\citep{kumar2020problems,loftus2024causal}.

Overall, this work provides clear guidelines to conduct a causal mediation analysis, although the results rely on strong assumptions, and should always be interpreted carefully.
It becomes increasingly clear that mediation analysis helps formulating in a causal language the many correlations observed in large-scale medical datasets, bridging the gap from basic statistics to causal understanding.

\section{Data availability statement}
The UK Biobank data is publicly available through a procedure described at \url{http://www.ukbiobank.ac.uk/using-the-resource}.
This research has been conducted using the UK Biobank Resource under Application Number 49314.

\section{Code availability}
The code to reproduce the simulations and numerical analyses presented in this work is available at \url{https://github.com/judithabk6/benchmark_mediation_analysis}.
The implementation of the causal mediation estimators can be found at \url{https://github.com/judithabk6/med_bench}.

\section{Acknowledgments}
This work was supported by the SC1-DTH-07-2018 H2020 VirtualBrainCloud Project under grant agreement No. 826421,  the KARAIB AI chair (ANR-20-CHIA-0025-01),  the H2020 Research Infrastructures Grant EBRAIN-Health 101058516, and the ANR under the MUSE grant (ANR-16-IDEX-0006).
This work has benefited of the support of the Hi! PARIS Engineering Team.

\FloatBarrier
\newpage
\bibliography{../biblio}

\begin{thebibliography}{71}
\providecommand{\natexlab}[1]{#1}
\providecommand{\url}[1]{\texttt{#1}}
\expandafter\ifx\csname urlstyle\endcsname\relax
  \providecommand{\doi}[1]{doi: #1}\else
  \providecommand{\doi}{doi: \begingroup \urlstyle{rm}\Url}\fi

\bibitem[Albert and Nelson(2011)]{albert2011generalized}
J.~M. Albert and S.~Nelson.
\newblock Generalized causal mediation analysis.
\newblock \emph{Biometrics}, 67\penalty0 (3):\penalty0 1028--1038, 2011.

\bibitem[Alfaro-Almagro et~al.(2018)Alfaro-Almagro, Jenkinson, Bangerter,
  Andersson, Griffanti, Douaud, Sotiropoulos, Jbabdi, Hernandez-Fernandez,
  Vallee, et~al.]{alfaro2018image}
F.~Alfaro-Almagro, M.~Jenkinson, N.~K. Bangerter, J.~L. Andersson,
  L.~Griffanti, G.~Douaud, S.~N. Sotiropoulos, S.~Jbabdi,
  M.~Hernandez-Fernandez, E.~Vallee, et~al.
\newblock Image processing and quality control for the first 10,000 brain
  imaging datasets from uk biobank.
\newblock \emph{Neuroimage}, 166:\penalty0 400--424, 2018.

\bibitem[Alfaro-Almagro et~al.(2021)Alfaro-Almagro, McCarthy, Afyouni,
  Andersson, Bastiani, Miller, Nichols, and Smith]{alfaro2021confound}
F.~Alfaro-Almagro, P.~McCarthy, S.~Afyouni, J.~L. Andersson, M.~Bastiani, K.~L.
  Miller, T.~E. Nichols, and S.~M. Smith.
\newblock Confound modelling in uk biobank brain imaging.
\newblock \emph{NeuroImage}, 224:\penalty0 117002, 2021.

\bibitem[Avin et~al.(2005)Avin, Shpitser, and Pearl]{avin2005identifiability}
C.~Avin, I.~Shpitser, and J.~Pearl.
\newblock Identifiability of path-specific effects.
\newblock 2005.

\bibitem[Baron and Kenny(1986)]{baron1986moderator}
R.~M. Baron and D.~A. Kenny.
\newblock The moderator--mediator variable distinction in social psychological
  research: Conceptual, strategic, and statistical considerations.
\newblock \emph{Journal of personality and social psychology}, 51\penalty0
  (6):\penalty0 1173, 1986.

\bibitem[Bodory and Huber(2018)]{bodory2018causalweight}
H.~Bodory and M.~Huber.
\newblock The causalweight package for causal inference in r.
\newblock Technical report, University of Fribourg, 2018.

\bibitem[Breiman(2001)]{breiman2001random}
L.~Breiman.
\newblock Random forests.
\newblock \emph{Machine learning}, 45\penalty0 (1):\penalty0 5--32, 2001.

\bibitem[Chernozhukov et~al.(2018)Chernozhukov, Chetverikov, Demirer, Duflo,
  Hansen, Newey, and Robins]{chernozhukov2018double}
V.~Chernozhukov, D.~Chetverikov, M.~Demirer, E.~Duflo, C.~Hansen, W.~Newey, and
  J.~Robins.
\newblock Double/debiased machine learning for treatment and structural
  parameters, 2018.

\bibitem[Chén et~al.(2018)Chén, Crainiceanu, Ogburn, Caffo, Wager, and
  Lindquist]{chen_high-dimensional_2018}
O.~Y. Chén, C.~Crainiceanu, E.~L. Ogburn, B.~S. Caffo, T.~D. Wager, and M.~A.
  Lindquist.
\newblock High-dimensional multivariate mediation with application to
  neuroimaging data.
\newblock \emph{Biostatistics}, 19\penalty0 (2):\penalty0 121--136, Apr. 2018.
\newblock ISSN 1465-4644.
\newblock \doi{10.1093/biostatistics/kxx027}.
\newblock URL
  \url{https://academic.oup.com/biostatistics/article/19/2/121/3868977}.
\newblock Publisher: Oxford Academic.

\bibitem[Cochran(1957)]{cochran1957analysis}
W.~G. Cochran.
\newblock Analysis of covariance: its nature and uses.
\newblock \emph{Biometrics}, 13\penalty0 (3):\penalty0 261--281, 1957.

\bibitem[Cole et~al.(2018)Cole, Ritchie, Bastin, Hern{\'a}ndez,
  Mu{\~n}oz~Maniega, Royle, Corley, Pattie, Harris, Zhang,
  et~al.]{cole2018brain}
J.~H. Cole, S.~J. Ritchie, M.~E. Bastin, V.~Hern{\'a}ndez,
  S.~Mu{\~n}oz~Maniega, N.~Royle, J.~Corley, A.~Pattie, S.~E. Harris, Q.~Zhang,
  et~al.
\newblock Brain age predicts mortality.
\newblock \emph{Molecular psychiatry}, 23\penalty0 (5):\penalty0 1385--1392,
  2018.

\bibitem[Cox et~al.(2019)Cox, Ritchie, Fawns-Ritchie, Tucker-Drob, and
  Deary]{cox2019structural}
S.~R. Cox, S.~J. Ritchie, C.~Fawns-Ritchie, E.~M. Tucker-Drob, and I.~J. Deary.
\newblock Structural brain imaging correlates of general intelligence in uk
  biobank.
\newblock \emph{Intelligence}, 76:\penalty0 101376, 2019.

\bibitem[Dadi et~al.(2021)Dadi, Varoquaux, Houenou, Bzdok, Thirion, and
  Engemann]{dadi2021population}
K.~Dadi, G.~Varoquaux, J.~Houenou, D.~Bzdok, B.~Thirion, and D.~Engemann.
\newblock Population modeling with machine learning can enhance measures of
  mental health.
\newblock \emph{Gigascience}, 10\penalty0 (10):\penalty0 giab071, 2021.

\bibitem[Djordjilovi{\'c} et~al.(2019)Djordjilovi{\'c}, Page, Gran, N{\o}st,
  Sandanger, Veier{\o}d, and Thoresen]{djordjilovic2019global}
V.~Djordjilovi{\'c}, C.~M. Page, J.~M. Gran, T.~H. N{\o}st, T.~M. Sandanger,
  M.~B. Veier{\o}d, and M.~Thoresen.
\newblock Global test for high-dimensional mediation: Testing groups of
  potential mediators.
\newblock \emph{Statistics in medicine}, 38\penalty0 (18):\penalty0 3346--3360,
  2019.

\bibitem[Farbmacher et~al.(2022)Farbmacher, Huber, Laff{\'e}rs, Langen, and
  Spindler]{farbmacher2020causal}
H.~Farbmacher, M.~Huber, L.~Laff{\'e}rs, H.~Langen, and M.~Spindler.
\newblock Causal mediation analysis with double machine learning.
\newblock \emph{The Econometrics Journal}, 25\penalty0 (2):\penalty0 277--300,
  2022.

\bibitem[Fawns-Ritchie and Deary(2020)]{fawns2020reliability}
C.~Fawns-Ritchie and I.~J. Deary.
\newblock Reliability and validity of the uk biobank cognitive tests.
\newblock \emph{PloS one}, 15\penalty0 (4):\penalty0 e0231627, 2020.

\bibitem[Hayes and Scharkow(2013)]{hayes2013relative}
A.~F. Hayes and M.~Scharkow.
\newblock The relative trustworthiness of inferential tests of the indirect
  effect in statistical mediation analysis: does method really matter?
\newblock \emph{Psychological science}, 24\penalty0 (10):\penalty0 1918--1927,
  2013.

\bibitem[Hejazi et~al.(2022)Hejazi, Rudolph, and
  D{\'\i}az]{hejazi2022medoutcon}
N.~Hejazi, K.~Rudolph, and I.~D{\'\i}az.
\newblock medoutcon: Nonparametric efficient causal mediation analysis with
  machine learning in r.
\newblock \emph{Journal of Open Source Software}, 7\penalty0 (69), 2022.

\bibitem[Hejazi et~al.(2021)Hejazi, Duncan, McCoy, and {van der
  Laan}]{hejazi2021tmle3mediate-rpkg}
N.~S. Hejazi, J.~Duncan, D.~McCoy, and M.~J. {van der Laan}.
\newblock {tmle3mediate}: Targeted learning for causal mediation analysis,
  2021.
\newblock URL \url{https://github.com/tlverse/tmle3mediate}.
\newblock R package version 0.0.3.

\bibitem[Hines et~al.(2021)Hines, Vansteelandt, and
  Diaz-Ordaz]{hines2021robust}
O.~Hines, S.~Vansteelandt, and K.~Diaz-Ordaz.
\newblock Robust inference for mediated effects in partially linear models.
\newblock \emph{Psychometrika}, pages 1--24, 2021.

\bibitem[Hong(2010)]{hong2010ratio}
G.~Hong.
\newblock Ratio of mediator probability weighting for estimating natural direct
  and indirect effects.
\newblock In \emph{Proceedings of the American Statistical Association,
  Biometrics Section}, pages 2401--2415. American Statistical Association
  Alexandria, VA, 2010.

\bibitem[Horvitz and Thompson(1952)]{horvitz1952generalization}
D.~G. Horvitz and D.~J. Thompson.
\newblock A generalization of sampling without replacement from a finite
  universe.
\newblock \emph{Journal of the American statistical Association}, 47\penalty0
  (260):\penalty0 663--685, 1952.

\bibitem[Huang and Pan(2016)]{huang2016hypothesis}
Y.-T. Huang and W.-C. Pan.
\newblock Hypothesis test of mediation effect in causal mediation model with
  high-dimensional continuous mediators.
\newblock \emph{Biometrics}, 72\penalty0 (2):\penalty0 402--413, 2016.

\bibitem[Huber(2014)]{huber2014identifying}
M.~Huber.
\newblock Identifying causal mechanisms (primarily) based on inverse
  probability weighting.
\newblock \emph{Journal of Applied Econometrics}, 29\penalty0 (6):\penalty0
  920--943, 2014.

\bibitem[Huber(2019)]{huber2019review}
M.~Huber.
\newblock A review of causal mediation analysis for assessing direct and
  indirect treatment effects.
\newblock 2019.

\bibitem[Huber et~al.(2016)Huber, Lechner, and Mellace]{huber2016finite}
M.~Huber, M.~Lechner, and G.~Mellace.
\newblock The finite sample performance of estimators for mediation analysis
  under sequential conditional independence.
\newblock \emph{Journal of Business \& Economic Statistics}, 34\penalty0
  (1):\penalty0 139--160, 2016.

\bibitem[Imai et~al.(2010)Imai, Keele, and Yamamoto]{imai2010identification}
K.~Imai, L.~Keele, and T.~Yamamoto.
\newblock Identification, inference and sensitivity analysis for causal
  mediation effects.
\newblock \emph{Statistical science}, 25\penalty0 (1):\penalty0 51--71, 2010.

\bibitem[Imbens and Rubin(2015)]{imbens2015causal}
G.~W. Imbens and D.~B. Rubin.
\newblock \emph{Causal inference in statistics, social, and biomedical
  sciences}.
\newblock Cambridge University Press, 2015.

\bibitem[J{\'e}rolon et~al.(2020)J{\'e}rolon, Baglietto, Birmel{\'e}, Alarcon,
  and Perduca]{jerolon2020causal}
A.~J{\'e}rolon, L.~Baglietto, E.~Birmel{\'e}, F.~Alarcon, and V.~Perduca.
\newblock Causal mediation analysis in presence of multiple mediators
  uncausally related.
\newblock \emph{The International Journal of Biostatistics}, 2020.

\bibitem[Josse and Husson(2016)]{josse2016missmda}
J.~Josse and F.~Husson.
\newblock missmda: a package for handling missing values in multivariate data
  analysis.
\newblock \emph{Journal of statistical software}, 70:\penalty0 1--31, 2016.

\bibitem[Judd and Kenny(1981)]{judd1981process}
C.~M. Judd and D.~A. Kenny.
\newblock Process analysis: Estimating mediation in treatment evaluations.
\newblock \emph{Evaluation review}, 5\penalty0 (5):\penalty0 602--619, 1981.

\bibitem[Kennedy(2024)]{kennedy2024semiparametric}
E.~H. Kennedy.
\newblock Semiparametric doubly robust targeted double machine learning: a
  review.
\newblock \emph{Handbook of Statistical Methods for Precision Medicine}, pages
  207--236, 2024.

\bibitem[Kumar et~al.(2020)Kumar, Venkatasubramanian, Scheidegger, and
  Friedler]{kumar2020problems}
I.~E. Kumar, S.~Venkatasubramanian, C.~Scheidegger, and S.~Friedler.
\newblock Problems with shapley-value-based explanations as feature importance
  measures.
\newblock In \emph{International conference on machine learning}, pages
  5491--5500. PMLR, 2020.

\bibitem[K{\"u}nzel et~al.(2019)K{\"u}nzel, Sekhon, Bickel, and
  Yu]{kunzel2019metalearners}
S.~R. K{\"u}nzel, J.~S. Sekhon, P.~J. Bickel, and B.~Yu.
\newblock Metalearners for estimating heterogeneous treatment effects using
  machine learning.
\newblock \emph{Proceedings of the national academy of sciences}, 116\penalty0
  (10):\penalty0 4156--4165, 2019.

\bibitem[L{\^e} et~al.(2008)L{\^e}, Josse, and Husson]{le2008factominer}
S.~L{\^e}, J.~Josse, and F.~Husson.
\newblock Factominer: an r package for multivariate analysis.
\newblock \emph{Journal of statistical software}, 25:\penalty0 1--18, 2008.

\bibitem[Lee et~al.(2021)Lee, Cashin, Lamb, Hopewell, Vansteelandt,
  VanderWeele, MacKinnon, Mansell, Collins, Golub, et~al.]{lee2021guideline}
H.~Lee, A.~G. Cashin, S.~E. Lamb, S.~Hopewell, S.~Vansteelandt, T.~J.
  VanderWeele, D.~P. MacKinnon, G.~Mansell, G.~S. Collins, R.~M. Golub, et~al.
\newblock A guideline for reporting mediation analyses of randomized trials and
  observational studies: the agrema statement.
\newblock \emph{Jama}, 326\penalty0 (11):\penalty0 1045--1056, 2021.

\bibitem[Loftus et~al.(2024)Loftus, Bynum, and Hansen]{loftus2024causal}
J.~Loftus, L.~Bynum, and S.~Hansen.
\newblock Causal dependence plots.
\newblock \emph{Advances in Neural Information Processing Systems},
  37:\penalty0 112656--112683, 2024.

\bibitem[MacKinnon(2012)]{mackinnon2012introduction}
D.~P. MacKinnon.
\newblock \emph{Introduction to statistical mediation analysis}.
\newblock Routledge, 2012.

\bibitem[Miller et~al.(2016)Miller, Alfaro-Almagro, Bangerter, Thomas, Yacoub,
  Xu, Bartsch, Jbabdi, Sotiropoulos, Andersson, et~al.]{miller2016multimodal}
K.~L. Miller, F.~Alfaro-Almagro, N.~K. Bangerter, D.~L. Thomas, E.~Yacoub,
  J.~Xu, A.~J. Bartsch, S.~Jbabdi, S.~N. Sotiropoulos, J.~L. Andersson, et~al.
\newblock Multimodal population brain imaging in the uk biobank prospective
  epidemiological study.
\newblock \emph{Nature neuroscience}, 19\penalty0 (11):\penalty0 1523--1536,
  2016.

\bibitem[Newby and Garfield(2022)]{newby2022understanding}
D.~Newby and V.~Garfield.
\newblock Understanding the inter-relationships of type 2 diabetes and
  hypertension with brain and cognitive health: A uk biobank study.
\newblock \emph{Diabetes, Obesity and Metabolism}, 24\penalty0 (5):\penalty0
  938--947, 2022.

\bibitem[Nguyen et~al.(2021)Nguyen, Schmid, and Stuart]{nguyen2021clarifying}
T.~Q. Nguyen, I.~Schmid, and E.~A. Stuart.
\newblock Clarifying causal mediation analysis for the applied researcher:
  Defining effects based on what we want to learn.
\newblock \emph{Psychological methods}, 26\penalty0 (2):\penalty0 255, 2021.

\bibitem[Nguyen et~al.(2023)Nguyen, Ogburn, Schmid, Sarker, Greifer, Koning,
  and Stuart]{nguyen2023causal}
T.~Q. Nguyen, E.~L. Ogburn, I.~Schmid, E.~B. Sarker, N.~Greifer, I.~M. Koning,
  and E.~A. Stuart.
\newblock Causal mediation analysis: From simple to more robust strategies for
  estimation of marginal natural (in) direct effects.
\newblock \emph{Statistics surveys}, 17:\penalty0 1, 2023.

\bibitem[Niculescu-Mizil and Caruana(2005)]{niculescu2005predicting}
A.~Niculescu-Mizil and R.~Caruana.
\newblock Predicting good probabilities with supervised learning.
\newblock In \emph{Proceedings of the 22nd international conference on Machine
  learning}, pages 625--632, 2005.

\bibitem[Pearl(2001)]{pearl2001direct}
J.~Pearl.
\newblock Direct and indirect effects.
\newblock In \emph{Proceedings of the Seventeenth conference on Uncertainty in
  artificial intelligence}, pages 411--420, 2001.

\bibitem[Pearson et~al.(2003)Pearson, Mensah, Alexander, Anderson, Cannon~III,
  Criqui, Fadl, Fortmann, Hong, Myers, et~al.]{pearson2003markers}
T.~A. Pearson, G.~A. Mensah, R.~W. Alexander, J.~L. Anderson, R.~O. Cannon~III,
  M.~Criqui, Y.~Y. Fadl, S.~P. Fortmann, Y.~Hong, G.~L. Myers, et~al.
\newblock Markers of inflammation and cardiovascular disease: application to
  clinical and public health practice: a statement for healthcare professionals
  from the centers for disease control and prevention and the american heart
  association.
\newblock \emph{circulation}, 107\penalty0 (3):\penalty0 499--511, 2003.

\bibitem[Pedregosa et~al.(2011)Pedregosa, Varoquaux, Gramfort, Michel, Thirion,
  Grisel, Blondel, Prettenhofer, Weiss, Dubourg, Vanderplas, Passos,
  Cournapeau, Brucher, Perrot, and Duchesnay]{scikit-learn}
F.~Pedregosa, G.~Varoquaux, A.~Gramfort, V.~Michel, B.~Thirion, O.~Grisel,
  M.~Blondel, P.~Prettenhofer, R.~Weiss, V.~Dubourg, J.~Vanderplas, A.~Passos,
  D.~Cournapeau, M.~Brucher, M.~Perrot, and E.~Duchesnay.
\newblock Scikit-learn: Machine learning in {P}ython.
\newblock \emph{Journal of Machine Learning Research}, 12:\penalty0 2825--2830,
  2011.

\bibitem[Pepys et~al.(2003)Pepys, Hirschfield, et~al.]{pepys2003c}
M.~B. Pepys, G.~M. Hirschfield, et~al.
\newblock C-reactive protein: a critical update.
\newblock \emph{The Journal of clinical investigation}, 111\penalty0
  (12):\penalty0 1805--1812, 2003.

\bibitem[Petersen et~al.(2006)Petersen, Sinisi, and van~der
  Laan]{petersen2006estimation}
M.~L. Petersen, S.~E. Sinisi, and M.~J. van~der Laan.
\newblock Estimation of direct causal effects.
\newblock \emph{Epidemiology}, pages 276--284, 2006.

\bibitem[Rijnhart et~al.(2017)Rijnhart, Twisk, Chinapaw, de~Boer, and
  Heymans]{rijnhart2017comparison}
J.~J. Rijnhart, J.~W. Twisk, M.~J. Chinapaw, M.~R. de~Boer, and M.~W. Heymans.
\newblock Comparison of methods for the analysis of relatively simple mediation
  models.
\newblock \emph{Contemporary clinical trials communications}, 7:\penalty0
  130--135, 2017.

\bibitem[Robins(2003)]{robinssemantics}
J.~M. Robins.
\newblock Semantics of causal dag models and the identification of direct and
  indirect effects.
\newblock In P.~J. Green, S.~Richardson, and Hjort, editors, \emph{In Highly
  Structured Stochastic Systems}, pages 70--81. Oxford University Press, 2003.

\bibitem[Robins and Greenland(1992)]{robins1992identifiability}
J.~M. Robins and S.~Greenland.
\newblock Identifiability and exchangeability for direct and indirect effects.
\newblock \emph{Epidemiology}, pages 143--155, 1992.

\bibitem[Rohrer et~al.(2022)Rohrer, H{\"u}nermund, Arslan, and
  Elson]{rohrer2022sa}
J.~M. Rohrer, P.~H{\"u}nermund, R.~C. Arslan, and M.~Elson.
\newblock That’sa lot to process! pitfalls of popular path models.
\newblock \emph{Advances in Methods and Practices in Psychological Science},
  5\penalty0 (2):\penalty0 25152459221095827, 2022.

\bibitem[Rudolph et~al.(2020)Rudolph, Goin, and Stuart]{rudolph2020peril}
K.~E. Rudolph, D.~E. Goin, and E.~A. Stuart.
\newblock The peril of power: a tutorial on using simulation to better
  understand when and how we can estimate mediating effects.
\newblock \emph{American journal of epidemiology}, 189\penalty0 (12):\penalty0
  1559--1567, 2020.

\bibitem[Schurz et~al.(2021)Schurz, Uddin, Kanske, Lamm, Sallet, Bernhardt,
  Mars, and Bzdok]{schurz2021variability}
M.~Schurz, L.~Q. Uddin, P.~Kanske, C.~Lamm, J.~Sallet, B.~C. Bernhardt, R.~B.
  Mars, and D.~Bzdok.
\newblock Variability in brain structure and function reflects lack of peer
  support.
\newblock \emph{Cerebral Cortex}, 31\penalty0 (10):\penalty0 4612--4627, 2021.

\bibitem[Scott and Sain(2004)]{Scott04multi-dimensionaldensity}
D.~W. Scott and S.~R. Sain.
\newblock Multi-dimensional density estimation, 2004.

\bibitem[Shi et~al.(2021)Shi, Choirat, Coull, VanderWeele, and
  Valeri]{shi2021cmaverse}
B.~Shi, C.~Choirat, B.~A. Coull, T.~J. VanderWeele, and L.~Valeri.
\newblock Cmaverse: a suite of functions for reproducible causal mediation
  analyses.
\newblock \emph{Epidemiology}, 32\penalty0 (5):\penalty0 e20--e22, 2021.

\bibitem[Stolicyn et~al.(2020)Stolicyn, Harris, Shen, Barbu, Adams, Hawkins,
  de~Nooij, Yeung, Murray, Lawrie, et~al.]{stolicyn2020automated}
A.~Stolicyn, M.~A. Harris, X.~Shen, M.~C. Barbu, M.~J. Adams, E.~L. Hawkins,
  L.~de~Nooij, H.~W. Yeung, A.~D. Murray, S.~M. Lawrie, et~al.
\newblock Automated classification of depression from structural brain measures
  across two independent community-based cohorts.
\newblock \emph{Human brain mapping}, 41\penalty0 (14):\penalty0 3922--3937,
  2020.

\bibitem[Stuart et~al.(2021)Stuart, Schmid, Nguyen, Sarker, Pittman, Benke,
  Rudolph, Badillo-Goicoechea, and Leoutsakos]{stuart2021assumptions}
E.~A. Stuart, I.~Schmid, T.~Nguyen, E.~Sarker, A.~Pittman, K.~Benke,
  K.~Rudolph, E.~Badillo-Goicoechea, and J.-M. Leoutsakos.
\newblock Assumptions not often assessed or satisfied in published mediation
  analyses in psychology and psychiatry.
\newblock \emph{Epidemiologic reviews}, 43\penalty0 (1):\penalty0 48--52, 2021.

\bibitem[Sudlow et~al.(2015)Sudlow, Gallacher, Allen, Beral, Burton, Danesh,
  Downey, Elliott, Green, Landray, et~al.]{sudlow2015uk}
C.~Sudlow, J.~Gallacher, N.~Allen, V.~Beral, P.~Burton, J.~Danesh, P.~Downey,
  P.~Elliott, J.~Green, M.~Landray, et~al.
\newblock Uk biobank: an open access resource for identifying the causes of a
  wide range of complex diseases of middle and old age.
\newblock \emph{PLoS medicine}, 12\penalty0 (3):\penalty0 e1001779, 2015.

\bibitem[Tchetgen and Shpitser(2012)]{tchetgen2012semiparametric}
E.~J.~T. Tchetgen and I.~Shpitser.
\newblock Semiparametric theory for causal mediation analysis: efficiency
  bounds, multiple robustness, and sensitivity analysis.
\newblock \emph{Annals of statistics}, 40\penalty0 (3):\penalty0 1816, 2012.

\bibitem[Textor et~al.(2016)Textor, Van~der Zander, Gilthorpe, Li{\'s}kiewicz,
  and Ellison]{textor2016robust}
J.~Textor, B.~Van~der Zander, M.~S. Gilthorpe, M.~Li{\'s}kiewicz, and G.~T.
  Ellison.
\newblock Robust causal inference using directed acyclic graphs: the r package
  ‘dagitty’.
\newblock \emph{International journal of epidemiology}, 45\penalty0
  (6):\penalty0 1887--1894, 2016.

\bibitem[Topiwala et~al.(2022)Topiwala, Ebmeier, Maullin-Sapey, and
  Nichols]{topiwala2022alcohol}
A.~Topiwala, K.~P. Ebmeier, T.~Maullin-Sapey, and T.~E. Nichols.
\newblock Alcohol consumption and mri markers of brain structure and function:
  Cohort study of 25,378 uk biobank participants.
\newblock \emph{NeuroImage: Clinical}, 35:\penalty0 103066, 2022.

\bibitem[Valente et~al.(2020)Valente, Rijnhart, Smyth, Muniz, and
  MacKinnon]{valente2020causal}
M.~J. Valente, J.~J. Rijnhart, H.~L. Smyth, F.~B. Muniz, and D.~P. MacKinnon.
\newblock Causal mediation programs in r, m plus, sas, spss, and stata.
\newblock \emph{Structural equation modeling: a multidisciplinary journal},
  27\penalty0 (6):\penalty0 975--984, 2020.

\bibitem[VanderWeele(2015)]{vanderweele2015explanation}
T.~VanderWeele.
\newblock \emph{Explanation in causal inference: methods for mediation and
  interaction}.
\newblock Oxford University Press, 2015.

\bibitem[VanderWeele and Vansteelandt(2014)]{vanderweele2014mediation}
T.~VanderWeele and S.~Vansteelandt.
\newblock Mediation analysis with multiple mediators.
\newblock \emph{Epidemiologic methods}, 2\penalty0 (1):\penalty0 95--115, 2014.

\bibitem[Vansteelandt and VanderWeele(2012)]{vansteelandt2012natural}
S.~Vansteelandt and T.~J. VanderWeele.
\newblock Natural direct and indirect effects on the exposed: effect
  decomposition under weaker assumptions.
\newblock \emph{Biometrics}, 68\penalty0 (4):\penalty0 1019--1027, 2012.

\bibitem[Zadrozny and Elkan(2002)]{zadrozny2002transforming}
B.~Zadrozny and C.~Elkan.
\newblock Transforming classifier scores into accurate multiclass probability
  estimates.
\newblock In \emph{Proceedings of the eighth ACM SIGKDD international
  conference on Knowledge discovery and data mining}, pages 694--699, 2002.

\bibitem[Zenati et~al.(2025)Zenati, Ab{\'e}cassis, Josse, and
  Thirion]{zenati2025double}
H.~Zenati, J.~Ab{\'e}cassis, J.~Josse, and B.~Thirion.
\newblock Double debiased machine learning for mediation analysis with
  continuous treatments.
\newblock \emph{arXiv preprint arXiv:2503.06156}, 2025.

\bibitem[Zhang et~al.(2021)Zhang, Chen, Feng, Wang, Li, and
  Liu]{zhang2021mediation}
H.~Zhang, J.~Chen, Y.~Feng, C.~Wang, H.~Li, and L.~Liu.
\newblock Mediation effect selection in high-dimensional and compositional
  microbiome data.
\newblock \emph{Statistics in medicine}, 40\penalty0 (4):\penalty0 885--896,
  2021.

\bibitem[Zhao et~al.(2020)Zhao, Lindquist, and Caffo]{zhao2020sparse}
Y.~Zhao, M.~A. Lindquist, and B.~S. Caffo.
\newblock Sparse principal component based high-dimensional mediation analysis.
\newblock \emph{Computational Statistics and Data Analysis}, 142:\penalty0
  106835, 2020.

\bibitem[Zheng and van~der Laan(2012)]{zheng2012targeted}
W.~Zheng and M.~J. van~der Laan.
\newblock Targeted maximum likelihood estimation of natural direct effects.
\newblock \emph{The international journal of biostatistics}, 8\penalty0
  (1):\penalty0 1--40, 2012.

\end{thebibliography}

\newpage
\appendix
\renewcommand\thefigure{S\arabic{figure}}    
\setcounter{figure}{0}
\renewcommand\thetable{S\arabic{table}}    
\setcounter{table}{0}    
\renewcommand*{\thesection}{S\arabic{section}}
\setcounter{section}{0}
\renewcommand*{\thesubsection}{S\arabic{section}.\arabic{subsection}}
\setcounter{subsection}{0}
\renewcommand*{\thesubsubsection}{S\arabic{section}.\arabic{subsection}.\arabic{subsubsection}}
\setcounter{subsubsection}{0}
\renewcommand{\theHtable}{Supplement.\thetable}
\renewcommand{\theHfigure}{Supplement.\thefigure}
\renewcommand{\theHsection}{Supplement.\thesection}
\renewcommand{\theHsubsection}{Supplement.\thesubsection}
\renewcommand{\theHsubsubsection}{Supplement.\thesubsubsection}

\section{Identifiability of the total effect, and the natural direct and indirect effect}
\label{identifiability_proof}
Let us first consider the potential outcomes $Y(t, M(t))$: 

\small{\begin{DispWithArrows}
& \mathbb{E}\left[Y\left(t, M\left(t\right)\right)\right] \notag \\
=& \mathbb{E}\left[\mathbb{E}\left[Y(t, M(t))|X=x\right]\right]  \tiny{\Arrow{sequential ignorability (assumption \ref{a:seq-treatment})}} \notag \\
=& \mathbb{E}\left[\mathbb{E}\left[Y(t, M(t))|T=t, X=x\right]\right] \tiny{\Arrow{SUTVA (assumption \ref{a:sutva_con})}}\notag \\
=& \mathbb{E}\left[\mathbb{E}\left[Y|T=t,X=x\right]\right] \tiny{\Arrow{overlap (assumption~\ref{a:seq-positivity})}}\notag \\
=& \mathbb{E}\left[\mathbb{E}\left[\frac{Y \cdot I\{T=t\}}{\mathbb{P}(T=t | X)}|X=x\right]\right] \notag \\
=& \mathbb{E}\left[\frac{Y \cdot I\{T=t\}}{\mathbb{P}(T=t | X)}\right] \label{total_effect_identifiability}
\end{DispWithArrows}}

And then the cross-world potential outcomes $Y(t, M(t'))$.
\small{\begin{DispWithArrows}
& \mathbb{E}\left[Y\left(t, M\left(t^{\prime}\right)\right)\right] \notag \\
&= \mathbb{E}\left[\mathbb{E}\left[Y\left(t, M\left(t^{\prime}\right)\right)|X=x\right]\right] \notag \\
&= \mathbb{E}\left[\mathbb{E}\left[\mathbb{E}\left[Y\left(t, m\right)|X=x, M(t^{\prime})=m\right]\right]\right] \notag \\
=& \iint \mathbb{E}\left[Y(t, m) | M\left(t^{\prime}\right)=m, X=x\right] f_{M\left(t^{\prime}\right)=m | X=x} d m f_{X=x} d x \tiny{\Arrow{ignorability (assumption \ref{a:seq-treatment})\\and SUTVA (assumption \ref{a:sutva_con})}}\notag \\
=& \iint \mathbb{E}[Y | T=t, M=m, X=x] f_{M\left(t^{\prime}\right)=m | X=x} f_{X=x} d m d x \tiny{\Arrow{ignorability (assumption \ref{a:seq-mediator})}}\notag \\
=& \iint \mathbb{E}[Y | T=t, M=m, X=x] f_{M=m | T=t^{\prime}, X=x} f_{X=x} d m d x \label{mediation_formula_origin} \\
=& \iint \mathbb{E}[Y | T=t, M=m, X=x] \cdot \frac{\mathbb{P}\left(T=t^{\prime} | M=m, X=x\right)}{\mathbb{P}\left(T=t^{\prime} | X=x\right)} f_{M=m | X=x} d m f_{X=x} d x \notag \\
=& \mathbb{E}\left[\mathbb{E}\left[\mathbb{E}\left[\frac{Y \cdot I\{T=t\}}{\mathbb{P}(T=t | M, X)} | M, X\right] \cdot \frac{\mathbb{P}\left(T=t^{\prime} | M, X\right)}{\mathbb{P}\left(T=t^{\prime} | X\right)} | X\right]\right]  \notag \\
=& \mathbb{E}\left[\frac{Y \cdot I\{T=t\}}{\mathbb{P}(T=t | M, X)} \cdot \frac{\mathbb{P}\left(T=t^{\prime} | M, X\right)}{\mathbb{P}\left(T=t^{\prime} | X\right)}\right] \label{propensity_weighting}\\
=& \mathbb{E}\left[\frac{Y \cdot I\{T=t\}}{\mathbb{P}(T=t | X)} \cdot \frac{f\left(M=m | T=t^{\prime}, X\right)}{f(M=m | T=t, X)}\right] \label{density_weighting}
\end{DispWithArrows}}

\section{More details about the causal mediation estimators derivation}
\label{sup:detail_estimator}
\subsection{G-computation without mediator density estimation}
\label{sup:g_comput}

We recall the mediation formula
$$
\begin{aligned}
&\theta(t)=\iint\{\mathbb{E}[Y \mid T=1, M=m, X=x]-\mathbb{E}[Y \mid T=0, M=m, X=x]\} f_{M=m \mid T=t, X=x} d m f_{X=x} d x\\
&\delta(t)=\iint \mathbb{E}[Y \mid T=t, M=m, X=x]\left\{f_{M=m \mid T=1, X=x}-f_{M=m \mid T=0, X=x}\right\} d m f_{X=x} d x.%
\end{aligned}
$$

Instead of doing the double integration explicitly in the estimation, we can integrate over the mediator in the mediation formula above, yielding:
$$
\begin{aligned}
&\theta(t)=\int\left(\mathbb{E}\left[\mu_{Y}(1, m, x)|T=t, X=x\right] - \mathbb{E}\left[\mu_{Y}(0, m, x)|T=t, X=x\right]\right) f_{X=x} d x\\
&\delta(t)=\int \left(\mathbb{E}\left[\mu_{Y}(t, m, x)|T=1, X=x\right] - \mathbb{E}\left[\mu_{Y}(t, m, x)|T=0, X=x\right]\right) f_{X=x} d x.%
\end{aligned}
$$

with $\mu_{Y}(t, m, x) = \mathbb{E}[Y \mid T=t, M=m, X=x]$.
Those terms involving two different values of the treatment are called cross-world conditional mean outcomes, $\omega_Y(t, t', x)=\mathbb{E}\left[\mu_{Y}(t, m, x)|T=t', X=x\right]$, following the notations of~\citep{zenati2025double}, and represent the expected outcome for values of the treatment $t$ in the population that has actually received the treatment $t'$.
It can be estimated by
\begin{enumerate}
  \item fit a regression model to predict $Y$ from $(T, M, X)$ (model $\hat{\mu}$)
  \item fit a second regression model on the predictions $\hat{Y}$ from $\hat{\mu}$ on $X$, in the population where $T=t'$ (model $\hat{\nu}_{t'}(\hat{\mu}(t, m, x), x)$)
  \item take the empirical mean of those predictions from the second regression over the whole sample
\end{enumerate}

This leads to the following for estimation: $\hat{\omega_Y}(t, t', x) = \hat{\nu}_{t'}(\hat{\mu}(t, M_i, X_i), X_i)$

\begin{align}\label{implicit_g_computation}
\begin{split}
    \hat{\theta}(t) &= \frac{1}{n} \sum_{i=1}^n \hat{\omega_Y}(1, t, x) - \hat{\omega_Y}(0, t, x)  \\&= \frac{1}{n} \sum_{i=1}^n   \hat{\nu}_t(\hat{\mu}(1, M_i, X_i), X_i) - \hat{\nu}_t(\hat{\mu}(0, M_i, X_i), X_i) \\
    \hat{\delta}(t) &= \frac{1}{n} \sum_{i=1}^n \hat{\omega_Y}(t, 1, x) - \hat{\omega_Y}(t, 0, x)  \\&=\frac{1}{n} \sum_{i=1}^n  \hat{\nu}_1(\hat{\mu}(t, M_i, X_i), X_i) - \hat{\nu}_0(\hat{\mu}(t, M_i, X_i), X_i) 
    \end{split}
\end{align}

This new formula does not necessitate the estimation of the mediator density, but two nested regression steps.

\section{More details on simulation settings}
\label{sup:simulation_settings}

We describe the main parameters of each considered simulation setting in Table~\ref{sim_setting}.

\begin{table}[h!!]
\resizebox{\textwidth}{!}{%
\begin{tabular}{lrrrrlrrllrrrrrr}
\toprule
  setting\_nb &    n &  dim\_x &  dim\_m &      type\_m &  wt\_list &  wm\_list &  m\_misspec &  y\_misspec &  mediated\_prop &  total &  direct\_1 &  direct\_0 &  indirect\_1 &  indirect\_0 \\
\midrule
\rowcolor{GreenYellow}        1 &  500 &      5 &      1 &      binary &     0.50 &     0.50 &      False &      False &           0.02 &   1.23 &      1.20 &      1.20 &        0.03 &        0.03 \\
\rowcolor{GreenYellow}        2 &  500 &      5 &      1 &      binary &     0.50 &     0.50 &       True &      False &           0.02 &   1.23 &      1.20 &      1.20 &        0.03 &        0.03 \\
\rowcolor{GreenYellow}        3 &  500 &      5 &      1 &      binary &     0.50 &     0.50 &      False &       True &           0.06 &   1.54 &      1.51 &      1.45 &        0.09 &        0.03 \\
\rowcolor{GreenYellow}        4 &  500 &      5 &      1 &      binary &     0.50 &     0.50 &       True &       True &           0.05 &   1.53 &      1.51 &      1.45 &        0.08 &        0.03 \\
\rowcolor{GreenYellow}        5 &  500 &      5 &      1 &  continuous &     0.50 &     0.50 &      False &      False &           0.10 &   1.33 &      1.20 &      1.20 &        0.13 &        0.13 \\
\rowcolor{GreenYellow}        6 &  500 &      5 &      1 &  continuous &     0.50 &     0.50 &       True &      False &           0.09 &   1.32 &      1.20 &      1.20 &        0.12 &        0.12 \\
\rowcolor{GreenYellow}       7 &  500 &      5 &      1 &  continuous &     0.50 &     0.50 &      False &       True &           0.24 &   1.58 &      1.45 &      1.20 &        0.38 &        0.13 \\
\rowcolor{GreenYellow}       8 &  500 &      5 &      1 &  continuous &     0.50 &     0.50 &       True &       True &           0.25 &   1.60 &      1.46 &      1.20 &        0.40 &        0.13 \\
\rowcolor{GreenYellow}      9 &  500 &      5 &      5 &  continuous &     0.20 &     0.20 &      False &      False &           0.02 &   1.22 &      1.20 &      1.20 &        0.02 &        0.02 \\
\rowcolor{GreenYellow}    10 &  500 &      5 &      5 &  continuous &     0.20 &     0.20 &       True &      False &           0.02 &   1.22 &      1.20 &      1.20 &        0.02 &        0.02 \\
\rowcolor{GreenYellow}     11 &  500 &      5 &      5 &  continuous &     0.20 &     0.20 &      False &       True &           0.08 &   1.31 &      1.29 &      1.19 &        0.11 &        0.02 \\
\rowcolor{GreenYellow}     12 &  500 &      5 &      5 &  continuous &     0.20 &     0.20 &       True &       True &           0.09 &   1.32 &      1.30 &      1.20 &        0.11 &        0.02 \\
\rowcolor{Apricot}       13 &  500 &      5 &      1 &      binary &     2.00 &    10.00 &      False &      False &           0.61 &   3.05 &      1.20 &      1.20 &        1.85 &        1.85 \\
\rowcolor{Apricot}       14 &  500 &      5 &      1 &      binary &     2.00 &    10.00 &       True &      False &           0.60 &   3.00 &      1.20 &      1.20 &        1.80 &        1.80 \\
\rowcolor{Apricot}     15 &  500 &      5 &      1 &      binary &     2.00 &    10.00 &      False &       True &           0.59 &   3.52 &      1.64 &      1.45 &        2.07 &        1.88 \\
\rowcolor{Apricot}         16 &  500 &      5 &      1 &      binary &     2.00 &    10.00 &       True &       True &           0.58 &   3.43 &      1.63 &      1.45 &        1.98 &        1.80 \\
\rowcolor{Apricot}      17 &  500 &      5 &      1 &  continuous &     0.80 &     5.00 &      False &      False &           0.63 &   3.27 &      1.20 &      1.20 &        2.07 &        2.07 \\
\rowcolor{Apricot}      18 &  500 &      5 &      1 &  continuous &     0.80 &     5.00 &       True &      False &           0.59 &   2.95 &      1.20 &      1.20 &        1.75 &        1.75 \\
\rowcolor{Apricot}      19 &  500 &      5 &      1 &  continuous &     0.80 &     5.00 &      False &       True &           0.62 &   3.24 &      1.57 &      1.17 &        2.01 &        1.67 \\
\rowcolor{Apricot}      20 &  500 &      5 &      1 &  continuous &     0.80 &     5.00 &       True &       True &           0.68 &   3.75 &      1.61 &      1.21 &        2.57 &        2.14 \\
\rowcolor{Apricot}      21 &  500 &      5 &      5 &  continuous &     0.30 &     5.00 &      False &      False &           0.37 &   1.90 &      1.20 &      1.20 &        0.70 &        0.70 \\
\rowcolor{Apricot}      22 &  500 &      5 &      5 &  continuous &     0.30 &     5.00 &       True &      False &           0.39 &   1.97 &      1.20 &      1.20 &        0.77 &        0.77 \\
\rowcolor{Apricot}      23 &  500 &      5 &      5 &  continuous &     0.30 &     5.00 &      False &       True &           0.41 &   2.03 &      1.34 &      1.19 &        0.82 &        0.69 \\
\rowcolor{Apricot}     24 &  500 &      5 &      5 &  continuous &     0.30 &     5.00 &       True &       True &           0.44 &   2.13 &      1.35 &      1.20 &        0.94 &        0.78 \\
\rowcolor{lightgray}        25 &  500 &      5 &      1 &      binary &     4.00 &     2.00 &      False &      False &           0.28 &   1.67 &      1.20 &      1.20 &        0.47 &        0.47 \\
\rowcolor{lightgray}       26 &  500 &      5 &      1 &      binary &     4.00 &     2.00 &       True &      False &           0.29 &   1.68 &      1.20 &      1.20 &        0.48 &        0.48 \\
\rowcolor{lightgray}       27 &  500 &      5 &      1 &      binary &     4.00 &     2.00 &      False &       True &           0.33 &   2.17 &      1.69 &      1.45 &        0.72 &        0.48 \\
\rowcolor{lightgray}     28 &  500 &      5 &      1 &      binary &     4.00 &     2.00 &       True &       True &           0.33 &   2.16 &      1.69 &      1.45 &        0.71 &        0.47 \\
\rowcolor{lightgray}       29 &  500 &      5 &      1 &  continuous &     2.00 &     1.00 &      False &      False &           0.47 &   2.26 &      1.20 &      1.20 &        1.06 &        1.06 \\
\rowcolor{lightgray}      30 &  500 &      5 &      1 &  continuous &     2.00 &     1.00 &       True &      False &           0.45 &   2.18 &      1.20 &      1.20 &        0.98 &        0.98 \\
\rowcolor{lightgray}      31 &  500 &      5 &      1 &  continuous &     2.00 &     1.00 &      False &       True &           0.63 &   3.20 &      2.20 &      1.20 &        2.01 &        1.00 \\
\rowcolor{lightgray}     32 &  500 &      5 &      1 &  continuous &     2.00 &     1.00 &       True &       True &           0.62 &   3.20 &      2.20 &      1.20 &        2.00 &        1.00 \\
\rowcolor{lightgray}     33 &  500 &      5 &      5 &  continuous &     1.00 &     1.00 &      False &      False &           0.29 &   1.69 &      1.20 &      1.20 &        0.49 &        0.49 \\
\rowcolor{lightgray}      34 &  500 &      5 &      5 &  continuous &     1.00 &     1.00 &       True &      False &           0.30 &   1.71 &      1.20 &      1.20 &        0.51 &        0.51 \\
\rowcolor{lightgray}      35 &  500 &      5 &      5 &  continuous &     1.00 &     1.00 &      False &       True &           0.46 &   2.22 &      1.71 &      1.21 &        1.02 &        0.51 \\
\rowcolor{lightgray}     36 &  500 &      5 &      5 &  continuous &     1.00 &     1.00 &       True &       True &           0.45 &   2.20 &      1.70 &      1.20 &        1.00 &        0.50 \\
\bottomrule
\end{tabular}
}
\caption{\textbf{parameter settings for simulations.} This table presents the different sets of parameters chosen for the simulations. The different color blocks represent modulations of the mediated proportion, with a low proportion (green), a high proportion but no overlap violation (orange) or a high proportion with a strong overlap violation (gray). Those proportions are adjusted either through $\omega_M$ which sets the strength of the relation between $M$ and $Y$, or through $\omega_T$ which sets the strength of the relation between $T$ and $M$. A value of $\omega_T$ too high can lead to the violation of the overlap assumption.}
\label{sim_setting}
\end{table}

\begin{figure}
 \centering
     \begin{subfigure}[b]{0.32\textwidth} % "0.45" donne ici la largeur de l'image
        \centering \includegraphics[width=\textwidth]{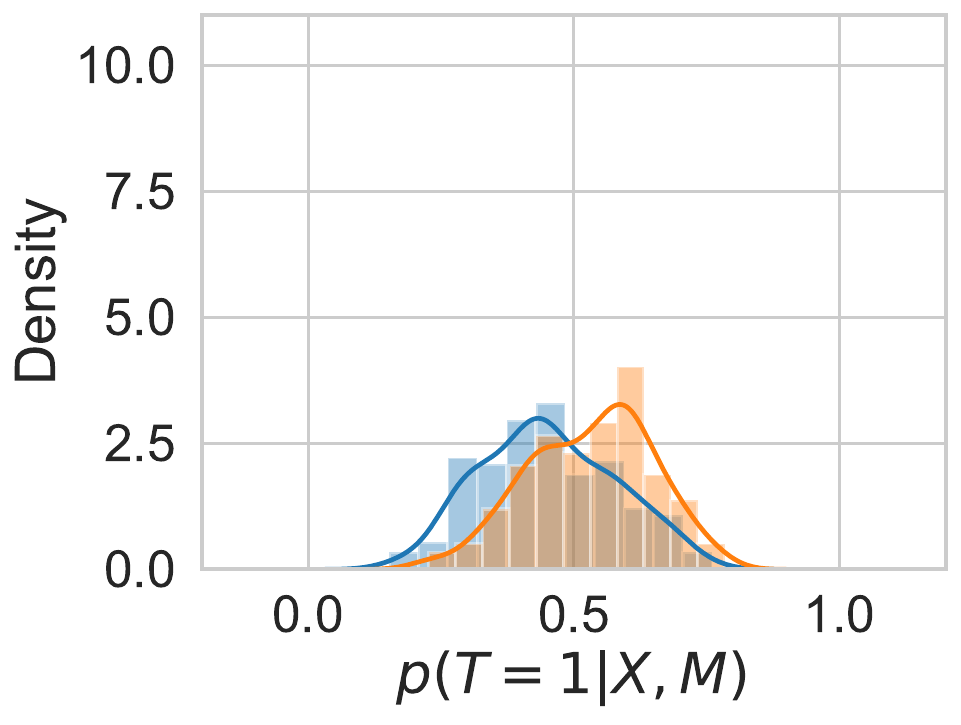}
        \caption{setting 3}\label{fig3:set2}
    \end{subfigure}
    \begin{subfigure}[b]{0.32\textwidth} % "0.45" donne ici la largeur de l'image
        \centering \includegraphics[width=\textwidth]{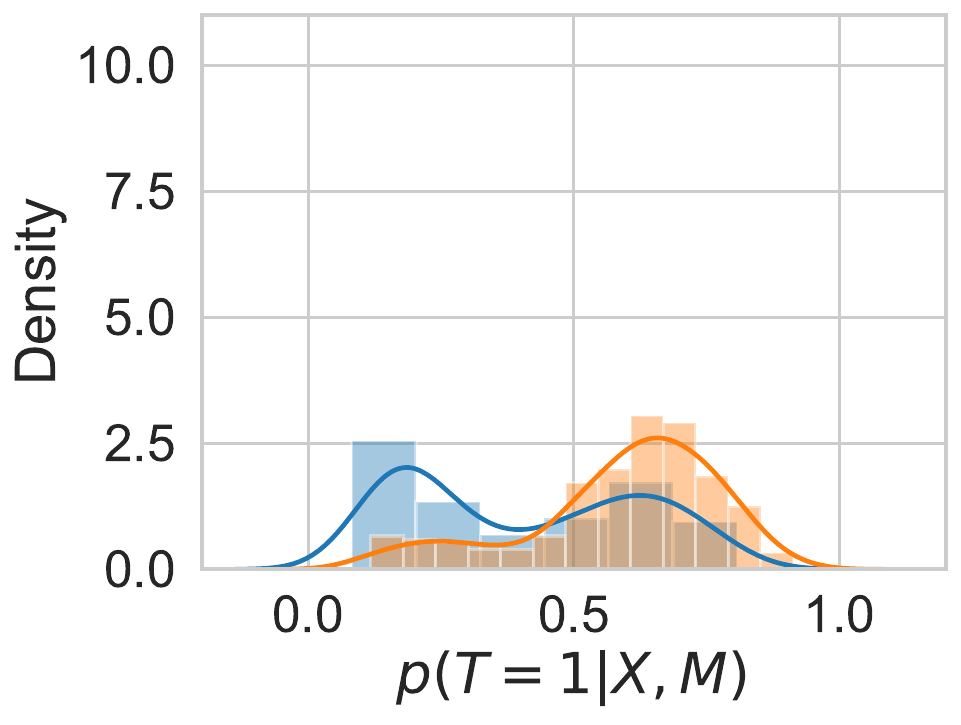}
        \caption{setting 15}\label{fig3:set15}
    \end{subfigure}
    \begin{subfigure}[b]{0.32\textwidth} % "0.45" donne ici la largeur de l'image
        \centering \includegraphics[width=\textwidth]{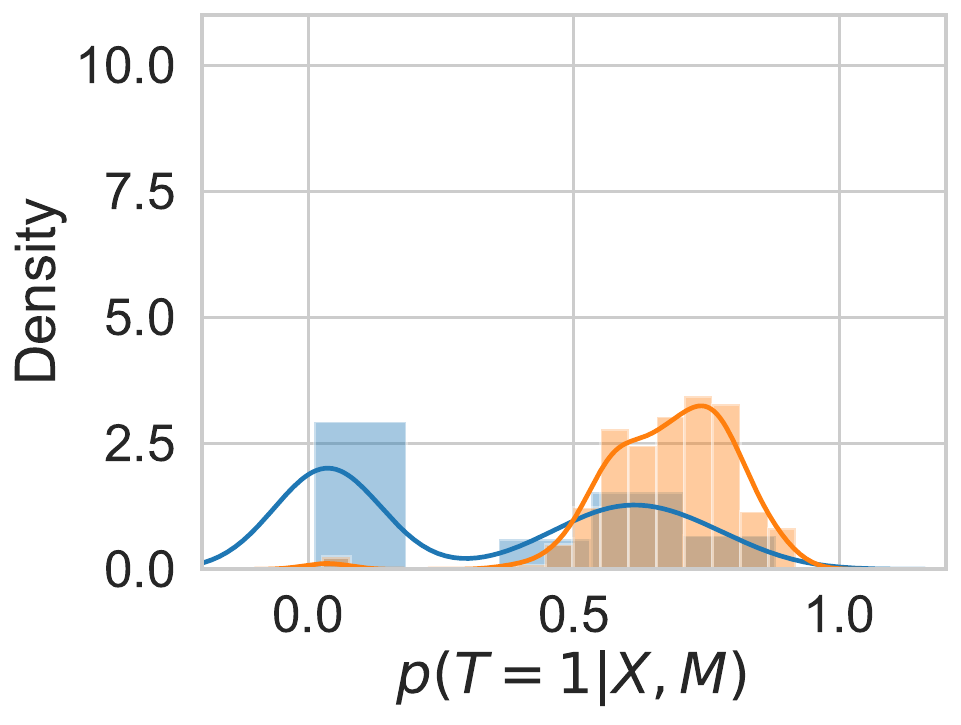}
        \caption{setting 27}\label{fig3:set27}
    \end{subfigure}

         \begin{subfigure}[b]{0.32\textwidth} % "0.45" donne ici la largeur de l'image
        \centering \includegraphics[width=\textwidth]{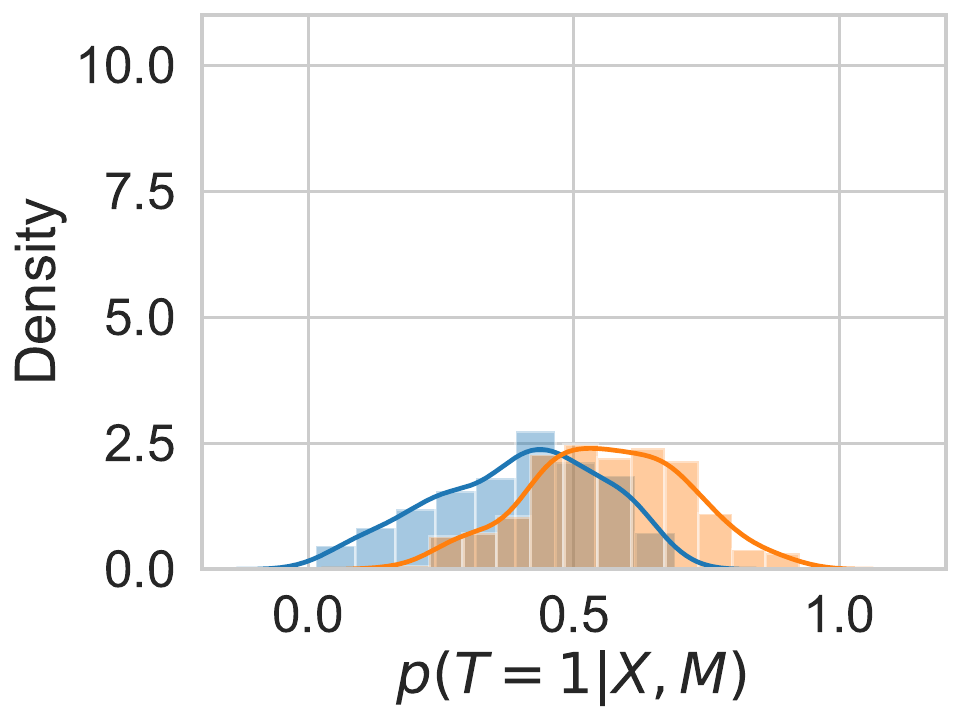}
        \caption{setting 6}\label{fig3:set6}
    \end{subfigure}
    \begin{subfigure}[b]{0.32\textwidth} % "0.45" donne ici la largeur de l'image
        \centering \includegraphics[width=\textwidth]{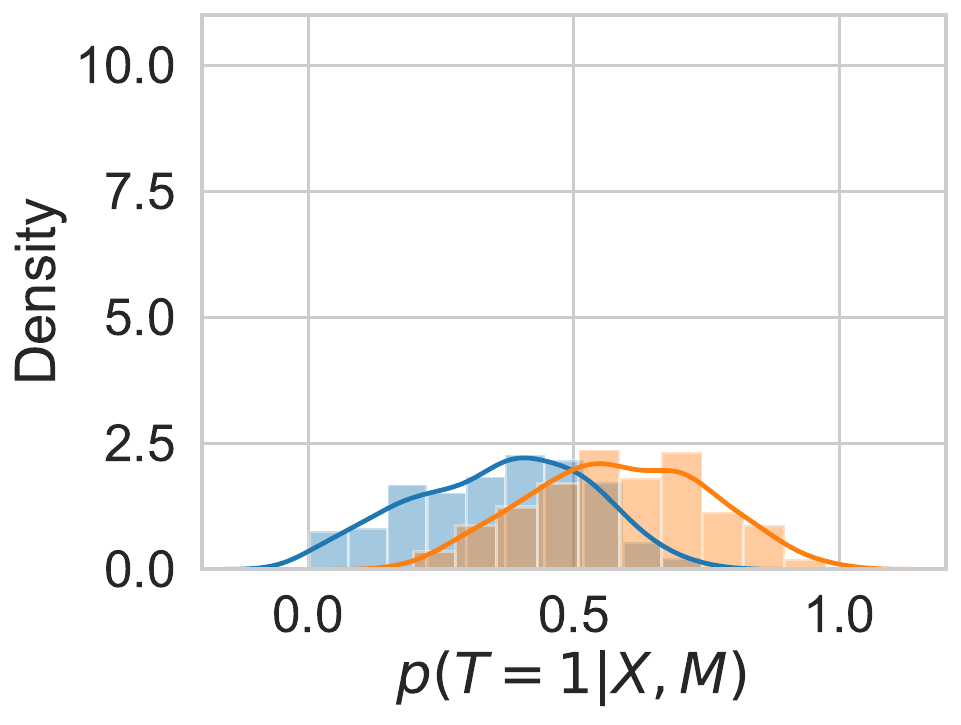}
        \caption{setting 18}\label{fig3:set18}
    \end{subfigure}
    \begin{subfigure}[b]{0.32\textwidth} % "0.45" donne ici la largeur de l'image
        \centering \includegraphics[width=\textwidth]{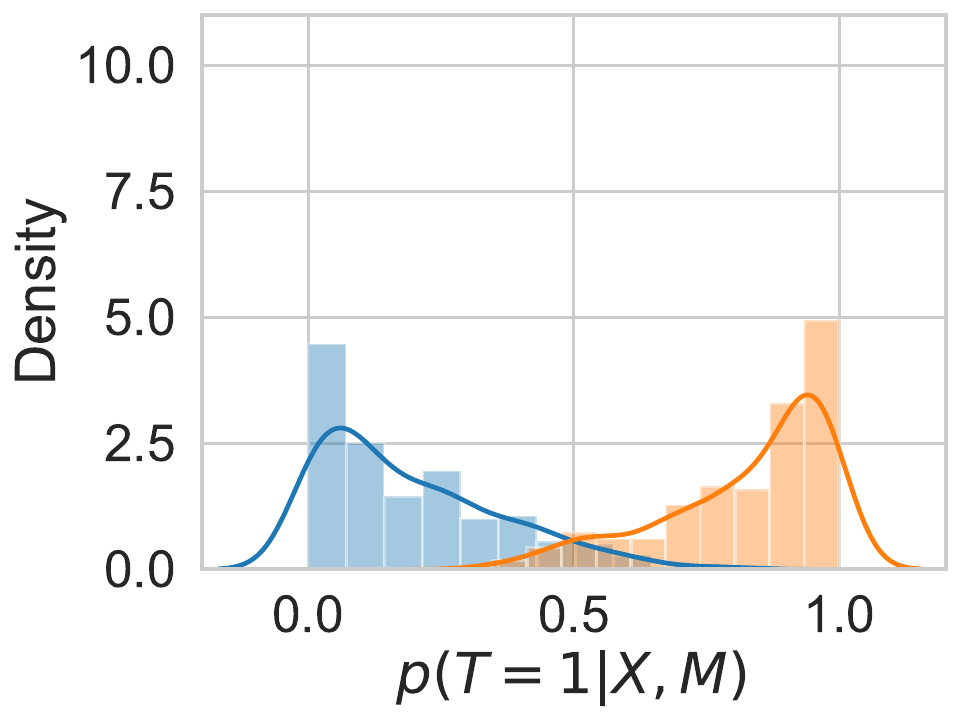}
        \caption{setting 30}\label{fig3:set30}
    \end{subfigure}

             \begin{subfigure}[b]{0.32\textwidth} % "0.45" donne ici la largeur de l'image
        \centering \includegraphics[width=\textwidth]{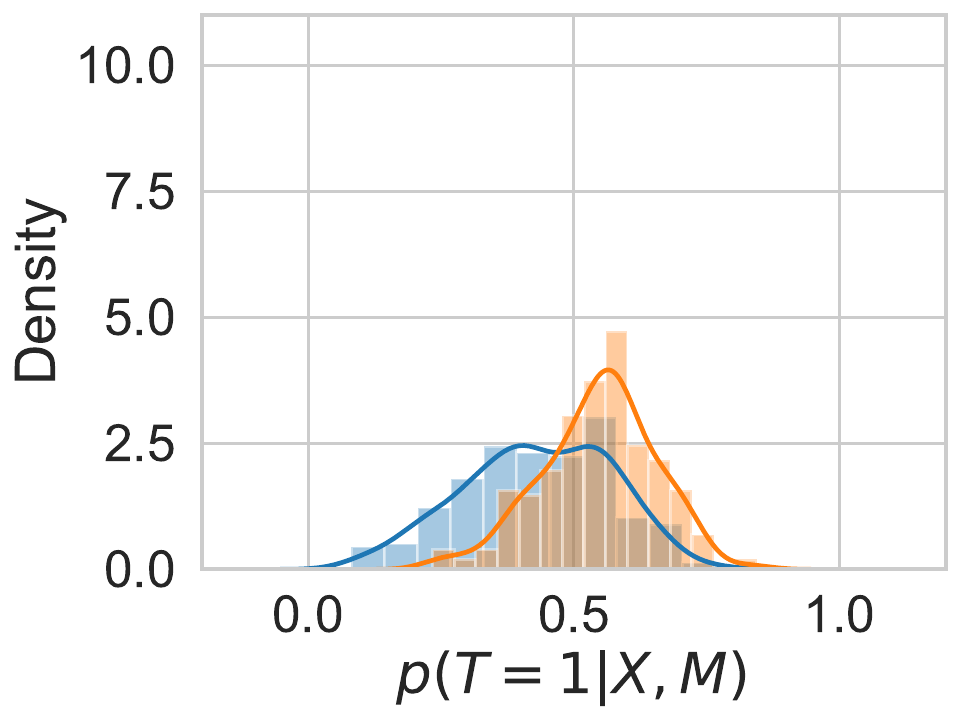}
        \caption{setting 12}\label{fig3:set12}
    \end{subfigure}
    \begin{subfigure}[b]{0.32\textwidth} % "0.45" donne ici la largeur de l'image
        \centering \includegraphics[width=\textwidth]{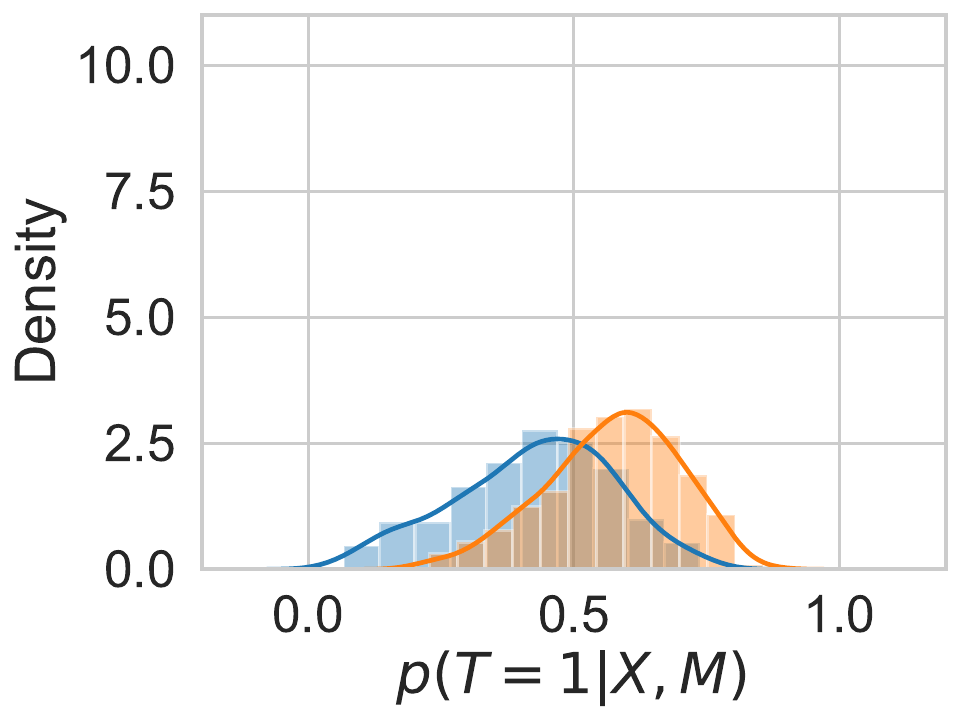}
        \caption{setting 24}\label{fig3:set24}
    \end{subfigure}
    \begin{subfigure}[b]{0.32\textwidth} % "0.45" donne ici la largeur de l'image
        \centering \includegraphics[width=\textwidth]{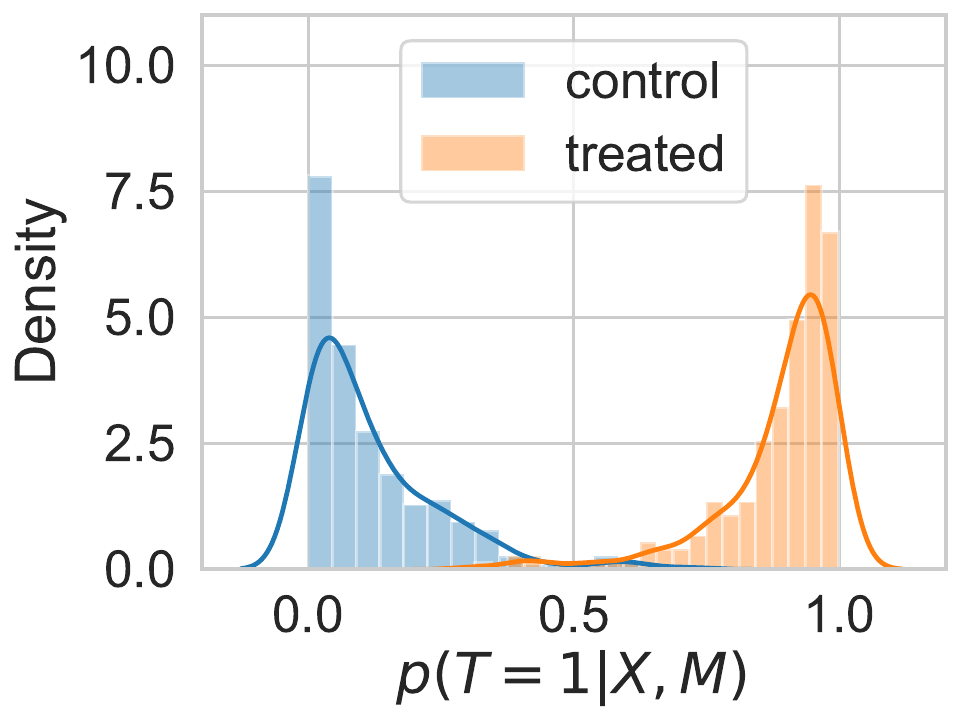}
        \caption{setting 36}\label{fig3:set36}
    \end{subfigure}
    \caption{Propensity score for different simulation settings, with low mediated proportion (left column), high mediated proportion and no overlap violation (middle column) or high mediated proportion and strong overlap violation (right column), for a binary one-dimension mediator (first row), a continuous one-dimension mediator (second row) or a continuous five-dimension mediator (last row). As expected, we see the strongest overlap violations for the right column, with a high mediated proportion but driven by a stronger link between $T$ and $M$ (high $\omega_T$ coefficient).}
    \label{sim_setting_overlap}
    \end{figure}

In Figure~\ref{sim_setting_overlap}, we illustrate the respect of the overlap assumption~\ref{a:seq-positivity} by representing the conditional probability of treatment given covariates and mediator(s), $\rho(M, X)=\mathbb{P}(T=1 \mid X, M)$ for the treated and the untreated in representative simulation settings for each of the three $(\omega_T, \omega_M)$ value sets. 

\subsection{Theoretical effects}
\label{theoretical_effects}
We are interested in the natural direct effect, $\theta$, and in the natural indirect effect, $\delta$. We provide here a computation of the theoretical value of those effects.

\subsubsection{with a binary mediator}
\begin{align*}
    \theta(t) &= \mathbb{E}_X\left[\mathbb{E}\left[Y(1, M(t)) - Y(0, M(t)) | X=x\right]\right] \\
    &= \mathbb{E}_X\left[\mathbb{E}\left[\gamma_0 + x^T \gamma_X + \gamma_T + \omega_M\gamma_M M(t) + \gamma_{MT} M(t) \right.\right.\\ & \left.\left.- (\gamma_0 + x^T \gamma_X + \omega_M\gamma_M M(t)) | X=x\right]\right] \\
    &=\mathbb{E}_X\left[\mathbb{E}\left[\gamma_T + \gamma_{MT} M(t)  | X=x\right]\right] \\
    &= \gamma_T + \gamma_{MT} \mathbb{E}_X\left[\mbox{expit}(\beta_0 + X^T\beta_X + (tX)^T\beta_{TX} + \omega_T\beta_T t)\right]
\end{align*}

\begin{align*}
    \delta(t) &= \mathbb{E}_X\left[\mathbb{E}\left[Y(t, M(1)) - Y(t, M(0)) | X=x\right]\right] \\
    &= \mathbb{E}_X\left[\mathbb{E}\left[\gamma_0 + X^T \gamma_X + \gamma_T t + \omega_M\gamma_M M(1) + \gamma_{MT} M(1)t \right.\right.\\ & \left.\left.- (\gamma_0 + X^T \gamma_X + \gamma_T t + \omega_M\gamma_M M(0) + \gamma_{MT} M(0)t)\right]\right] \\
    &=  \mathbb{E}_X\left[\mathbb{E}\left[(M(1) - M(0)) (\omega_M\gamma_M + t \gamma_{MT})\right]\right] \\
    &= \mathbb{E}_X\left[\mathbb{E}\left[(\mbox{expit}(\beta_0 + X^T\beta_X + X^T\beta_{TX} + \omega_T\beta_T) - \right.\right.\\ & \left.\left. \mbox{expit}(\beta_0 + X^T\beta_X)) (\omega_M\gamma_M  + t \gamma_{MT})\right]\right]
\end{align*}

\subsubsection{with a continuous mediator}
\begin{align*}
    \theta(t) &= \mathbb{E}_X\left[\mathbb{E}\left[Y(1, M(t)) - Y(0, M(t)) | X=x\right]\right] \\
    &= \mathbb{E}_X\left[\mathbb{E}\left[\gamma_0 + x^T \gamma_X + \gamma_T + \omega_M\gamma_M M(t) + \gamma_{MT} M(t) \right.\right.\\ & \left.\left.- (\gamma_0 + x^T \gamma_X + \omega_M\gamma_M M(t)) | X=x\right]\right] \\
    &=\mathbb{E}_X\left[\mathbb{E}\left[\gamma_T + \gamma_{MT} M(t)  | X=x\right]\right] \\
    &= \gamma_T + \gamma_{MT} \mathbb{E}_X\left[\beta_0 + X^T\beta_X + (tX)^T\beta_{TX} +\omega_T\beta_T t\right]
\end{align*}

\begin{align*}
    \delta(t) &= \mathbb{E}_X\left[\mathbb{E}\left[Y(t, M(1)) - Y(t, M(0)) | X=x\right]\right] \\
    &= \mathbb{E}_X\left[\mathbb{E}\left[\gamma_0 + X^T \gamma_X + \gamma_T t + \omega_M\gamma_M M(1) + \gamma_{MT} M(1)t \right.\right.\\ & \left.\left.- (\gamma_0 + X^T \gamma_X + \gamma_T t + \omega_M\gamma_M M(0) + \gamma_{MT} M(0)t)\right]\right] \\
    &=  \mathbb{E}_X\left[\mathbb{E}\left[(M(1) - M(0)) (\omega_M\gamma_M + t \gamma_{MT})\right]\right] \\
    &= \mathbb{E}_X\left[\mathbb{E}\left[\left((\beta_0 + X^T\beta_X + (X)^T\beta_{TX} +\omega_T\beta_T) - \right.\right.\right.\\ & \left.\left.\left. (\beta_0 + X^T\beta_X)\right) (\omega_M\gamma_M  + t \gamma_{MT})\right]\right]
\end{align*}

\section{How many bootstrap repetitions are required to obtain decent percentile confidence intervals?}

To assess a reasonable number of bootstrap repetitions to assess the characteristics of estimators' convergence, we have conducted 1,000 bootstrap repetitions on 100 varied simulated datasets.
We represent the mean, and the 2.5 and 97.5 percentiles for different number of bootstrap samples ranging from 1 to 1,000 in Supplementary Figure~\ref{bootstrap}.

\begin{figure}[!!!h]
\centering
\includegraphics[height=0.8\textheight]{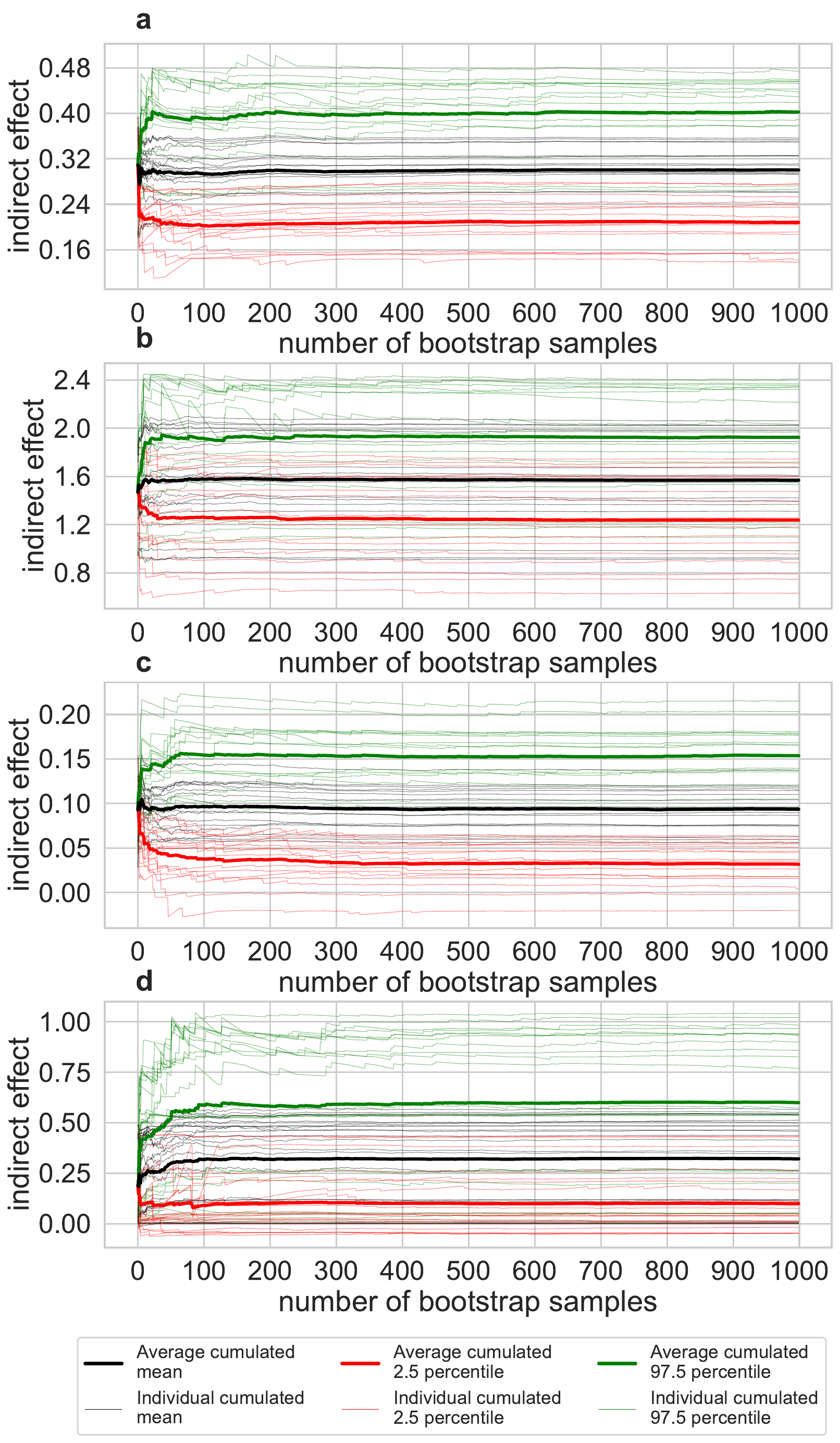}
\caption{\textbf{Exploration of an adequate number of bootstrap samples for uncertainty quantification} We show the value of the bounds of the 95\%-percentile confidence interval for different numbers of bootstrap samples. The bold curves represent the average over the 26 individual estimators, and the thin lines are the results for each estimator.
Each panel represents a different simulation configuration: \textbf{a} 500 observations, with a one-dimensional continuous mediator, in the setting of a low mediated proportion without overlap violation, and a misspecification of the outcome model; \textbf{b} 500 observations, with a one-dimensional continuous mediator, in the setting of a high mediated proportion with overlap violation, and a misspecification of both the outcome and the mediator models; \textbf{c} 1,000 observations, with a five-dimensional continuous mediator, in the setting of a low mediated proportion without overlap violation, and a misspecification of the outcome model; \textbf{d} 1,000 observations, with a binary mediator, in the setting of a high mediated proportion with overlap violation, and a misspecification of the mediator model.
This experiment shows that most estimators have reached a stable point with 100 bootstrap samples.
}
\label{bootstrap}
\end{figure}

\section{Extended results on simulations}
In this section, we report more details on the simulation results, including results for the total and indirect effects, and the effect of calibration and cross-fitting for all algorithms for nuisance parameters estimation.

\begin{table}
\begin{tabular}{l}
Coefficient product\\
G-computation (forest)\\
G-computation (forest \& cross-fitting)\\
G-computation (no regularization)\\
G-computation (regularization)\\
G-computation (regularization \& cross-fitting)\\
G-computation (forest)\\
G-computation (forest \& cross-fitting)\\
G-computation (no regularization)\\
G-computation (regularization)\\
G-computation (regularization \& cross-fitting)\\
IPW (forest)\\
IPW (forest \& cross-fitting)\\
IPW (no regularization)\\
IPW (regularization)\\
IPW (regularization \& cross-fitting)\\
multiply robust (forest)\\
multiply robust (forest \& cross-fitting)\\
multiply robust (no regularization)\\
multiply robust (regularization)\\
multiply robust (regularization \& cross-fitting)\\
medDML (no regularization)\\
medDML (regularization)\\
medDML (regularization \& cross-fitting)\\
medDML (forest)\\
medDML (forest \& cross-fitting)\\
\end{tabular}
\caption{\textbf{List of considered estimators and their implementation variants}}
\label{list_estim}
\end{table}

We consider the variant estimators listed in Table~\ref{list_estim}.

\begin{figure}[h]
\includegraphics[width=\textwidth]{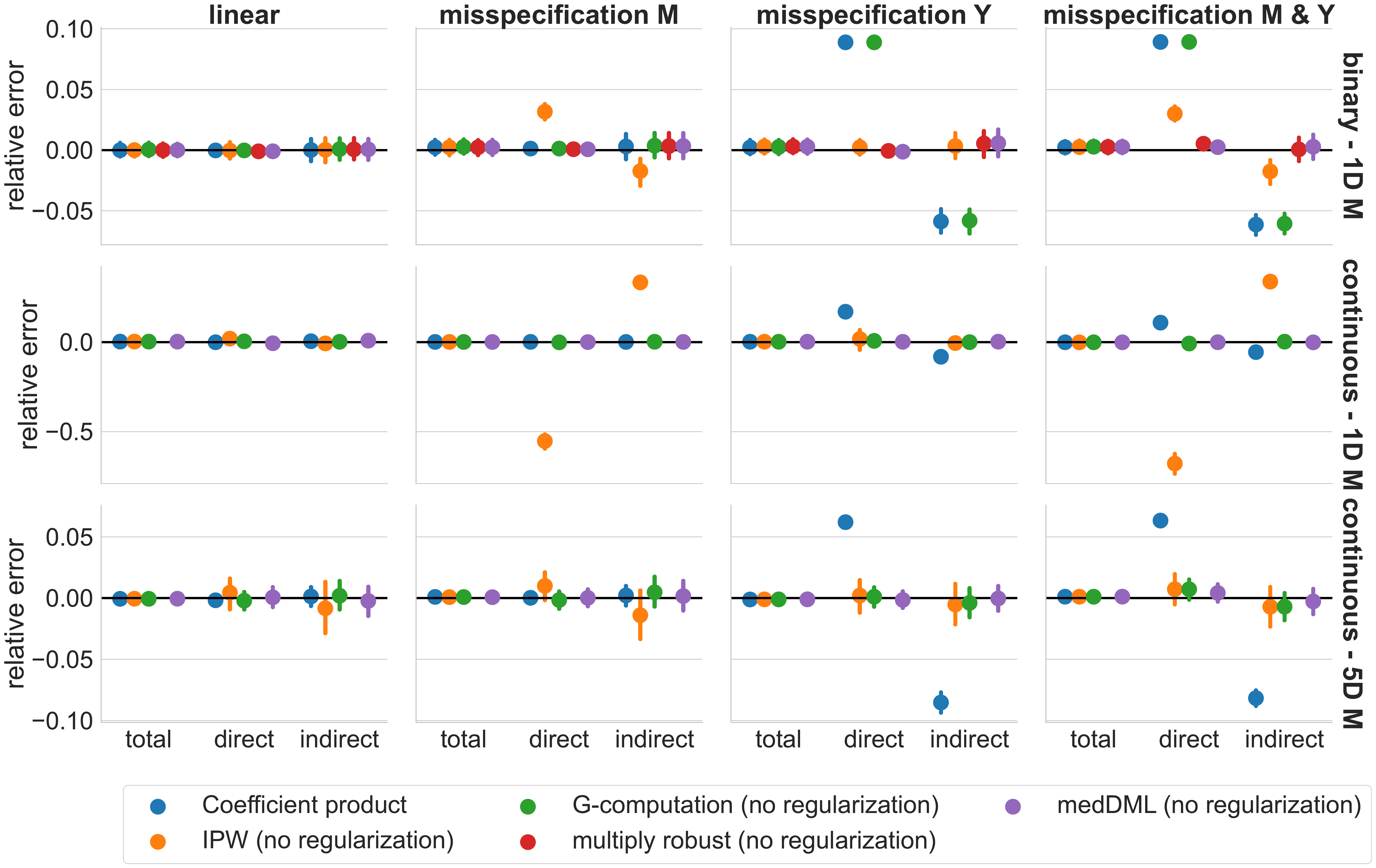}
\caption{
\textbf{Total and natural direct and indirect effects obtained with unregularized estimators for all mediator types.}
We show results for four scenarios of generative model specification, violating or not the parametric linear nuisance models of some estimators.
Each column corresponds to a distinct specification of simulated models.
The rows correspond to different mediator types, labeled on the right.
Each dot represents the average relative error (i.e. $\frac{\hat{\tau}-\tau}{\tau}$, for example for the total effect $\tau$) over 200 repetitions, and the error bars are the asymptotic normal 95\% confidence intervals from the distribution over the 200 repetitions.
All simulations are in the \emph{high mediated proportion without overlap violation} framework with $n=1,000$ observations.
The multiply-robust estimator only handles binary one-dimensional mediators, and has no results for the second and third rows.
The total effect is generally well estimated in all situations, but not the direct and indirect effects.
The indirect and direct effects have different value, leading to a lack of symmetry in their relative error values, even if their absolute values are equal, leading to no error for the total effect estimation.
Model misspecifications lead to estimation errors for most estimators, especially when the outcome model is misspecified.
}
\label{result_overview_all}
\end{figure}

\begin{figure}[h]
\includegraphics[width=\textwidth]{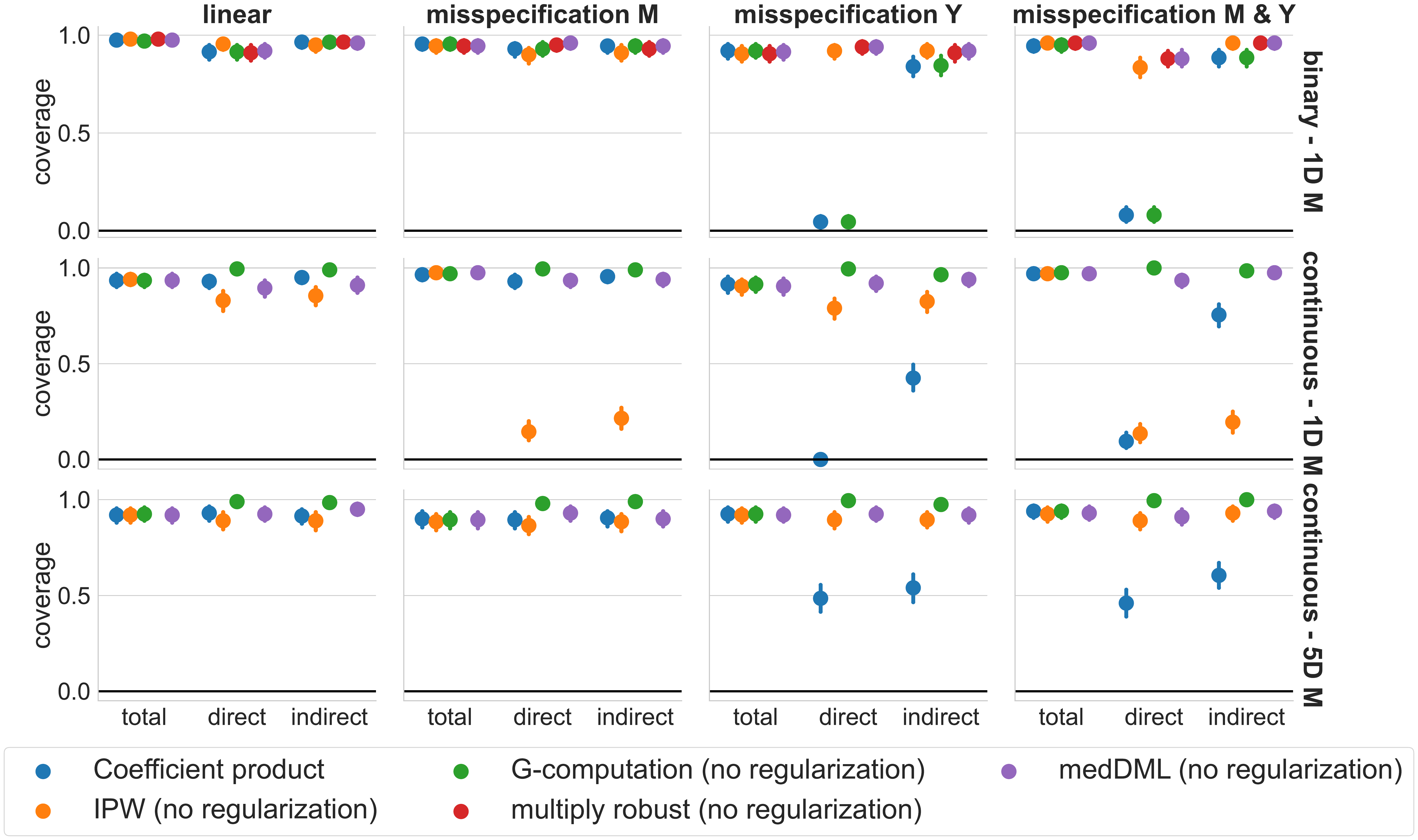}
\caption{
\textbf{95\%-confidence-interval coverage for the total and natural direct and indirect effects obtained with unregularized estimators for all mediator types.}
We show results for four scenarios of generative model specification, violating or not the parametric linear nuisance models of some estimators.
Each column corresponds to a distinct specification of simulated models.
The rows correspond to different mediator types, labeled on the right.
Each dot represents the average coverage obtained with 100 bootstrap samples, for example for the total effect $\tau$) over 200 repetitions, and the error bars are the asymptotic normal 95\% confidence intervals from the distribution over the 200 repetitions.
All simulations are in the \emph{high mediated proportion without overlap violation} framework with $n=1,000$ observations.
}
\label{bootstrap_coverage}
\end{figure}

\begin{figure}[h]
\includegraphics[width=\textwidth]{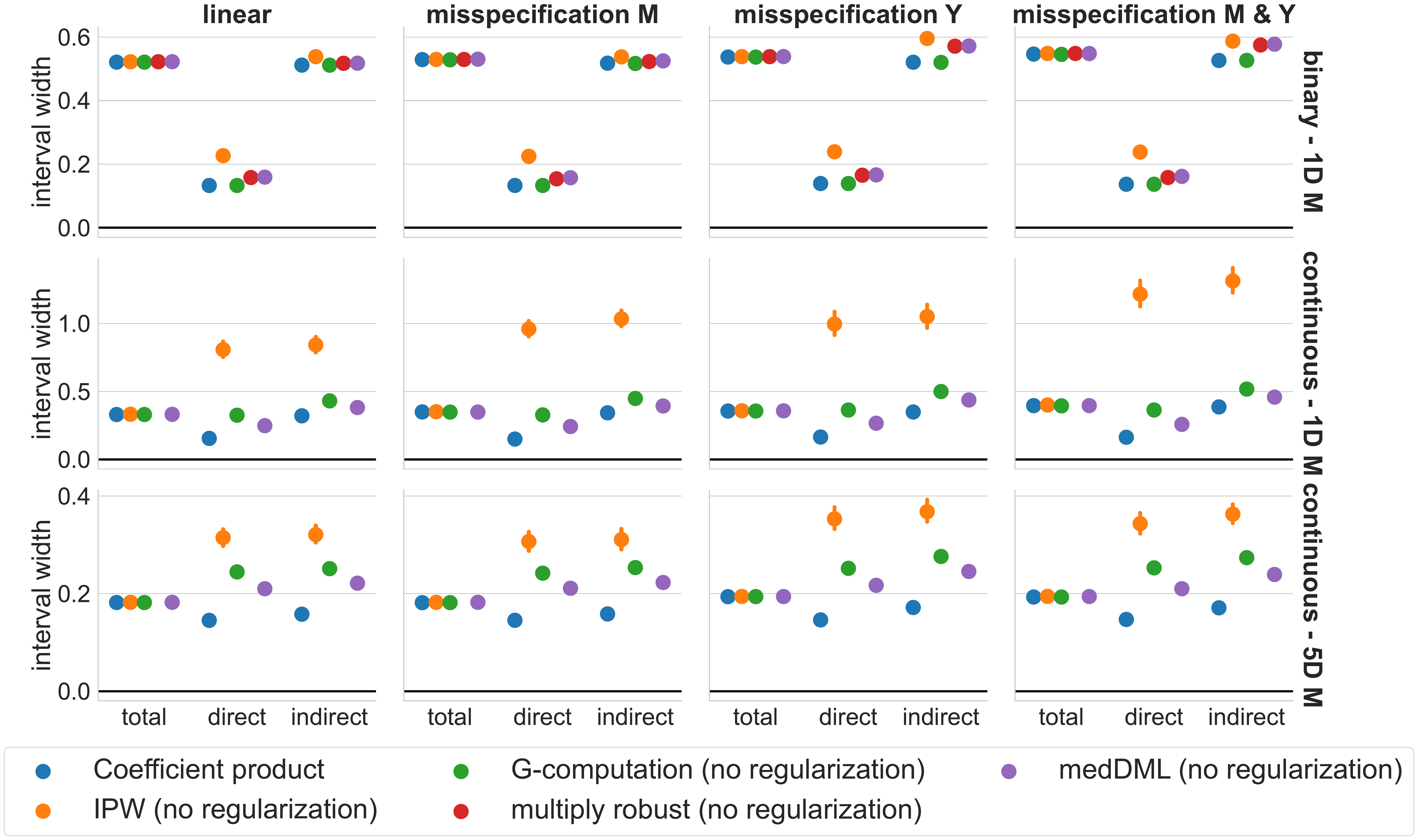}
\caption{
\textbf{95\%-confidence-interval width for the total and natural direct and indirect effects obtained with unregularized estimators for all mediator types.}
We show results for four scenarios of generative model specification, violating or not the parametric linear nuisance models of some estimators.
Each column corresponds to a distinct specification of simulated models.
The rows correspond to different mediator types, labeled on the right.
Each dot represents the average interval width obtained with 100 bootstrap samples, for example for the total effect $\tau$) over 200 repetitions, and the error bars are the asymptotic normal 95\% confidence intervals from the distribution over the 200 repetitions.
All simulations are in the \emph{high mediated proportion without overlap violation} framework with $n=1,000$ observations.
}
\label{bootstrap_width}
\end{figure}

Supplementary Figure~\ref{bootstrap_coverage} represents the coverage, and Supplementary Figure~\ref{bootstrap_width} the width of 95\% percentile bootstrap confidence intervals for the estimation of the different effects.
We observe results that are very similar to what is obtained for the estimation error.
A valuable insight is that the IPW estimator systematically has wider confidence intervals compared to the other methods, and that the doubly-robust estimators has slightly narrower intervals than the G-computation approach.
This can constitute a discriminative criterion for the choice of the estimator as both estimators have very similar results for the estimation error.

In Supplementary Figure~\ref{result_overview_all}, we present the same results as Figure~\ref{result_overview} with the continuous one-dimensional mediator (second row) and the multi-dimensional mediator (third row).
The pattern of bias is similar to the pattern observed for the binary mediator, with an important bias of the IPW estimator when the mediator model is misspecified, and for the coefficient product when the outcome model is misspecified.
In the case of a multi-dimensional mediator, the bias is less important for the IPW estimator when the mediator model is misspecified.
The higher number of mediators may provide additional intrinsic flexibility to the propensity score model, as the mediators are used solely to predict the binary treatment and are not explicitly modeled themselves.
The G-computation estimator does not exhibit a bias in the estimation when the mediator is continuous.
This is related to our implementation for the case of the implicit integration for the mediator density involves two distinct regressions for the outcome model, one regression for the treated units, and one for the untreated units.
This pattern is similar to the difference between the S-learner and the T-learner estimators in standard causal inference without mediation, and is known to be able to model properly interactions between the treatment and the covariates~\citep{kunzel2019metalearners}.
This property explains why the G-computation in this implementation does not suffer from the outcome model misspecification.
However, in theory, it would handle badly more complex non-linearities of the covariates.

\begin{figure}[h]
\includegraphics[width=\textwidth]{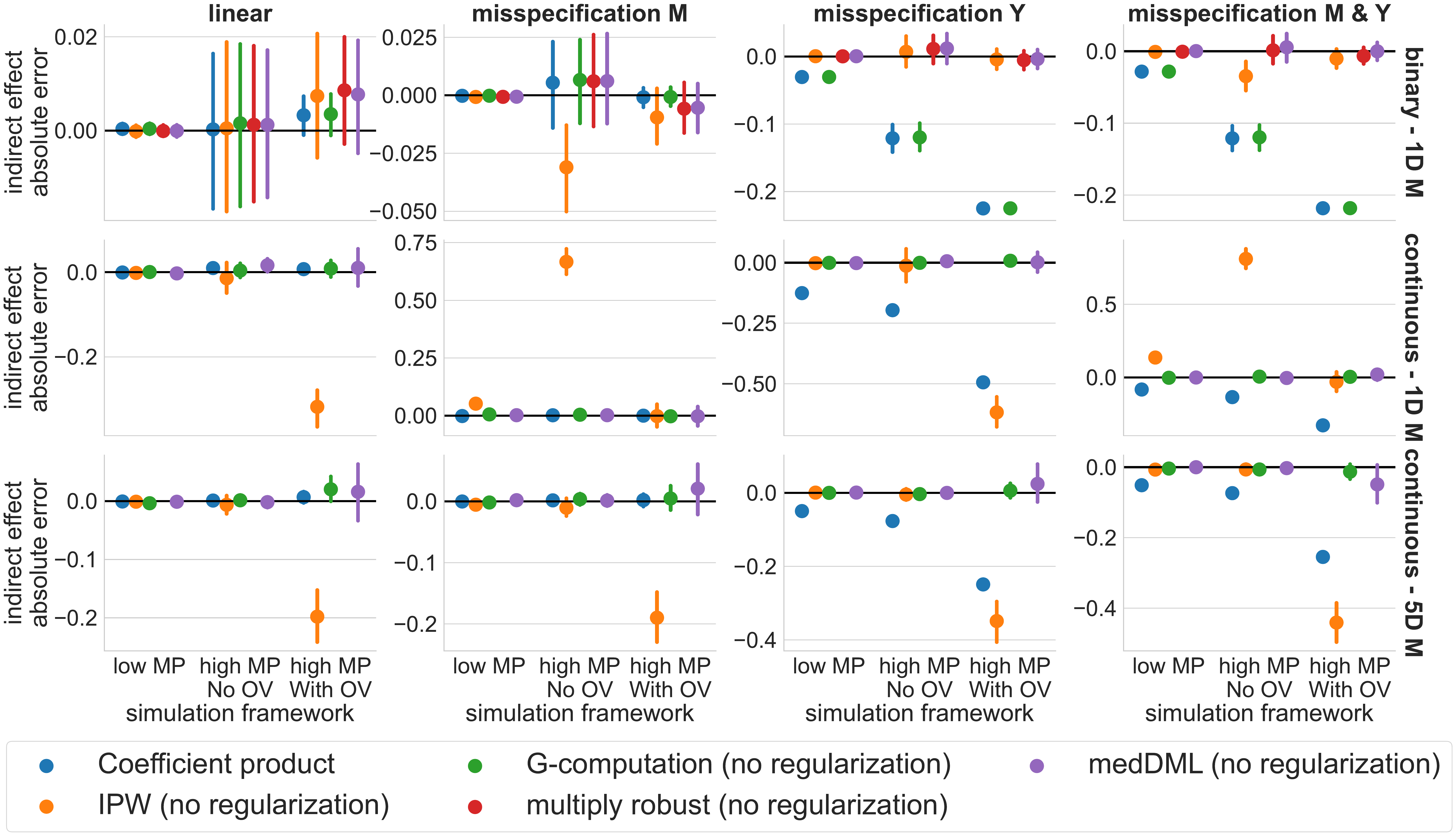}
\caption{
  \textbf{Indirect absolute effect for different simulation frameworks.}
We show results for four scenarios of generative model specification, violating or not the parametric linear nuisance models of some estimators.
Each column corresponds to a distinct specification of simulated models.
The rows correspond to different mediator types, labeled on the right.
The three sets of dots in each panel correspond to the three simulation frameworks, in that order:  "low mediated proportion without overlap violation", "high mediated proportion without overlap violation" and "high mediated proportion with strong overlap violation".
"MP" stands for "mediated proportion" and "OV" for "overlap violation".
Each dot represents the average relative error (i.e. $\frac{\hat{\tau}-\tau}{\tau}$, for example for the total effect $\tau$) over 200 repetitions, and the error bars are the asymptotic normal 95\% confidence intervals from the distribution over the 200 repetitions.
All simulations are in the case where $n=1,000$ observations.
The multiply-robust estimator only handles binary one-dimensional mediators, and has no results for the second and third rows.
The estimation error for most estimators is greater when there is a violation of the overlap assumption, including in robust estimators.
The absolute value of the indirect effect in the "low mediated proportion without overlap violation" is often very low, leading to very high relative errors compared to other settings.
We found however that the representation of the relative error in Figure~\ref{result_framework} is informative.
We represent here the same results as Figure~\ref{result_framework} with the absolute error (and not relative).
}
\label{result_framework_abs}
\end{figure}

In Supplementary Figure~\ref{result_framework_abs}, we present the same results as Figure~\ref{result_framework} with the absolute error, so that the reader can fully interpret the simulation errors, and include results for the continuous mediators.
The IPW estimators exhibits an important error rate for continuous mediators and overlap violation even in the absence of misspecification of the mediator model.

\begin{figure}[h]
\includegraphics[width=\textwidth]{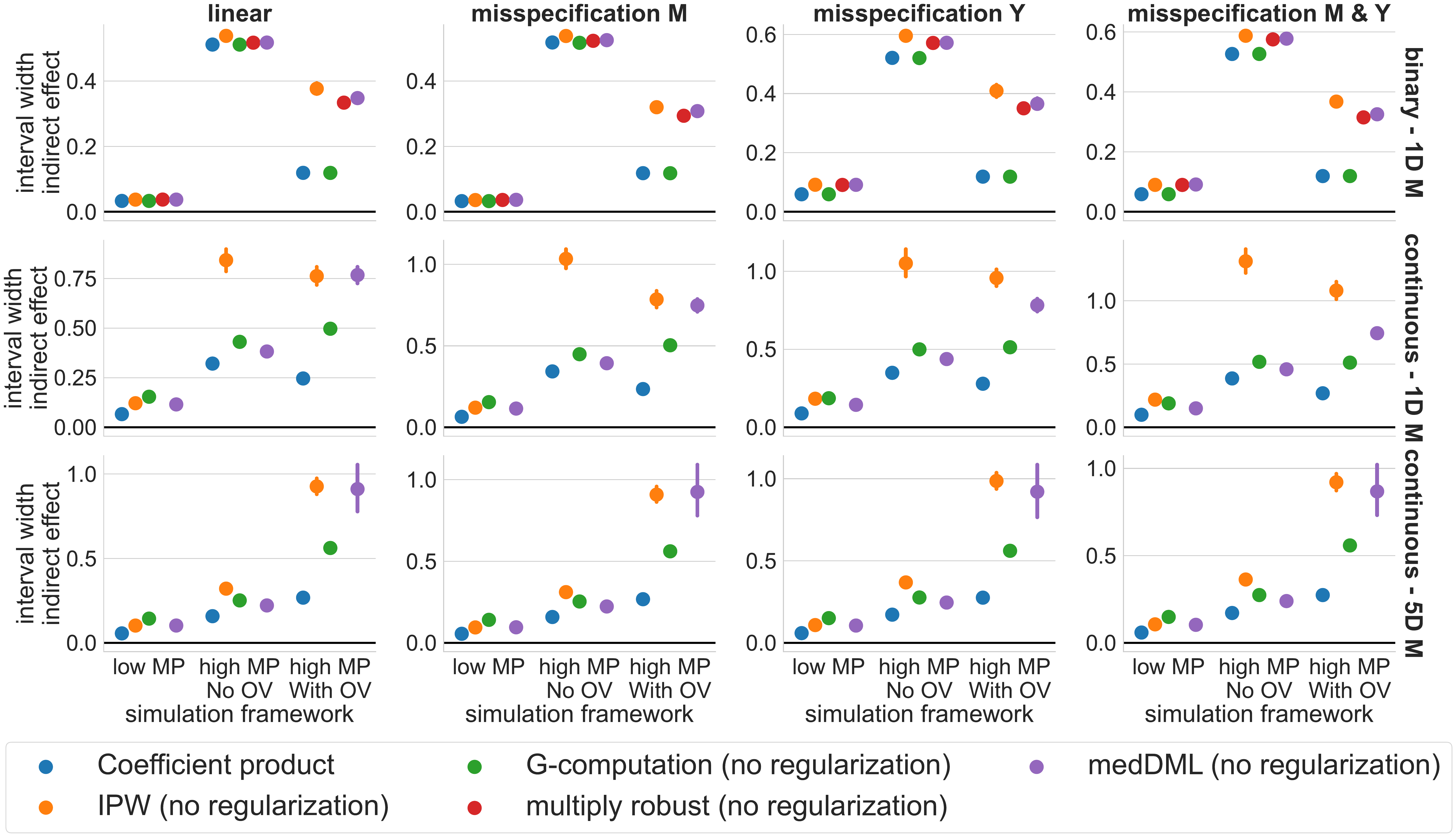}
\caption{
  \textbf{Confidence interval width for the indirect absolute effect for different simulation frameworks.}
We show results for four scenarios of generative model specification, violating or not the parametric linear nuisance models of some estimators.
Each column corresponds to a distinct specification of simulated models.
The rows correspond to different mediator types, labeled on the right.
The three sets of dots in each panel correspond to the three simulation frameworks, in that order:  "low mediated proportion without overlap violation", "high mediated proportion without overlap violation" and "high mediated proportion with strong overlap violation".
"MP" stands for "mediated proportion" and "OV" for "overlap violation".
Each dot represents the average interval width over 200 repetitions, and the error bars are the asymptotic normal 95\% confidence intervals from the distribution over the 200 repetitions.
All simulations are in the case where $n=1,000$ observations.
The multiply-robust estimator only handles binary one-dimensional mediators, and has no results for the second and third rows.
}
\label{result_framework_width}
\end{figure}

We also consider the behavior of the confidence interval coverage and width over those different configurations in Supplementary Figures~\ref{result_framework_coverage} and~\ref{result_framework_width}.
In this case, the behavior of those additional performance metrics is very different from the estimation error.
Indeed, the double machine learning estimator with multi-dimensional mediators exhibits a very important drop in coverage when there is an overlap violation, but only a modest estimation error, that does not appear statistically significant.
For the estimators involving inverse-propensity-weighting (IPW, G-computation and ) continuous mediators, we observe an important increase 

\begin{figure}[h]
\includegraphics[width=\textwidth]{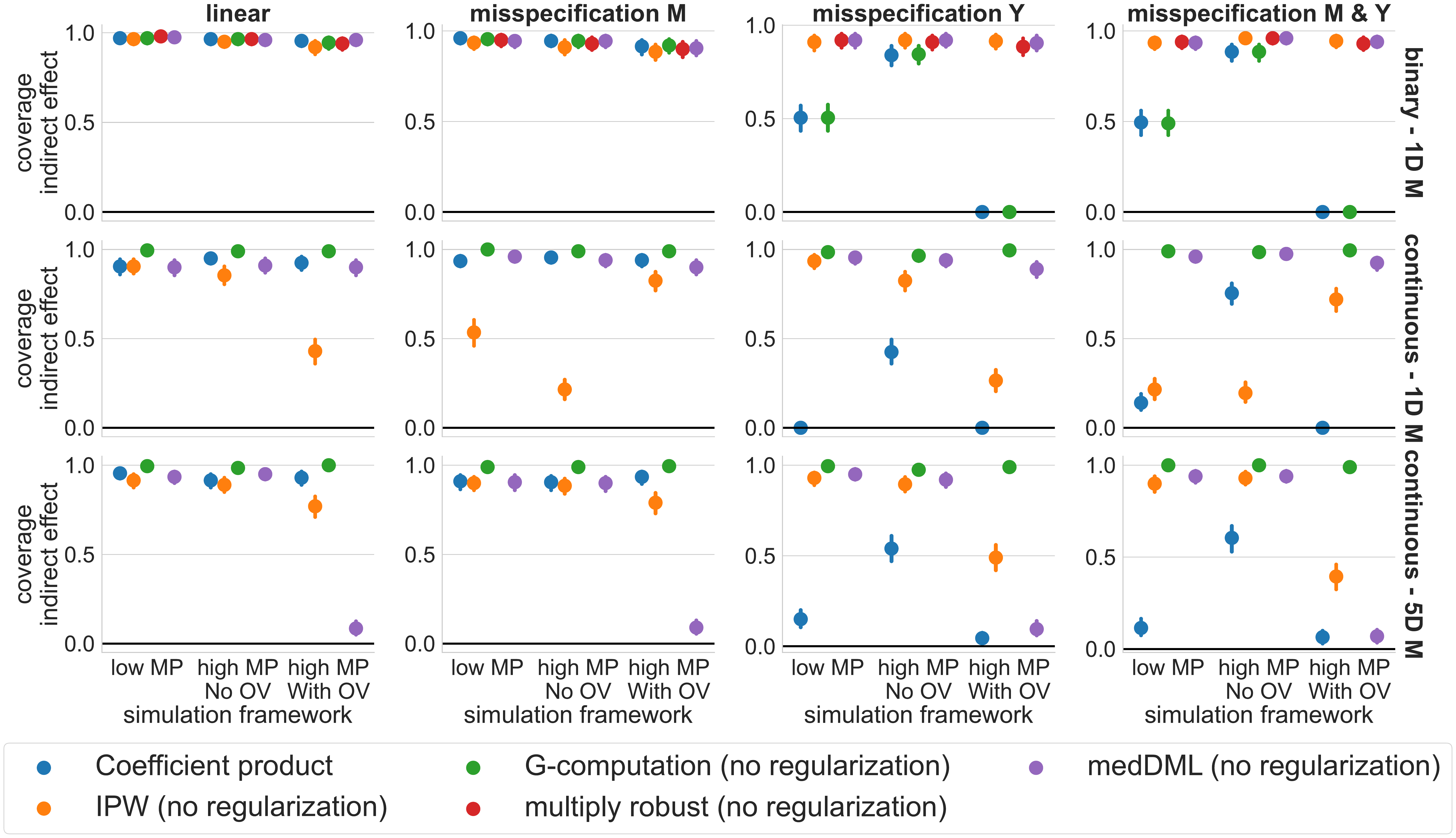}
\caption{
  \textbf{Coverage for the indirect absolute effect for different simulation frameworks.}
We show results for four scenarios of generative model specification, violating or not the parametric linear nuisance models of some estimators.
Each column corresponds to a distinct specification of simulated models.
The rows correspond to different mediator types, labeled on the right.
The three sets of dots in each panel correspond to the three simulation frameworks, in that order:  "low mediated proportion without overlap violation", "high mediated proportion without overlap violation" and "high mediated proportion with strong overlap violation".
"MP" stands for "mediated proportion" and "OV" for "overlap violation".
Each dot represents the average coverage over 200 repetitions, and the error bars are the asymptotic normal 95\% confidence intervals from the distribution over the 200 repetitions.
All simulations are in the case where $n=1,000$ observations.
The multiply-robust estimator only handles binary one-dimensional mediators, and has no results for the second and third rows.
}
\label{result_framework_coverage}
\end{figure}

\begin{figure}[h]
\includegraphics[width=\textwidth]{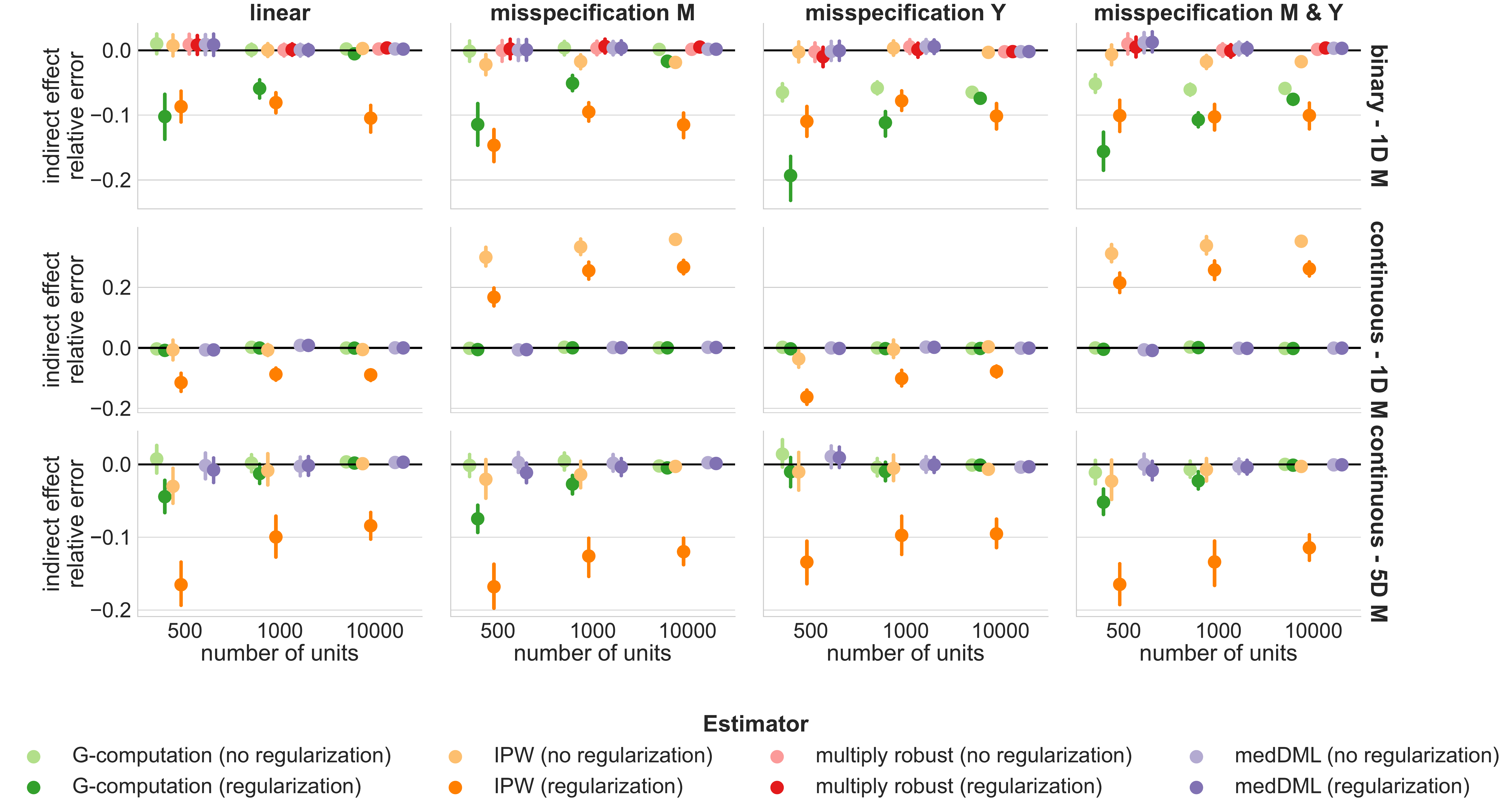}
\caption{
\textbf{Indirect effect relative error with regularized and unregularized linear models for plug-in nuisance parameters estimation.}
We show results for four scenarios of generative model specification, violating or not the parametric linear nuisance models of some estimators.
Each column corresponds to a distinct specification of simulated models.
The rows correspond to different mediator types, labeled on the right.
The four sets of dots in each panel correspond to the different number of observations tested in our set of simulations.
Each dot represents the average relative error (i.e. $\frac{\hat{\tau}-\tau}{\tau}$, for example for the total effect $\tau$) over 200 repetitions, and the error bars are the asymptotic normal 95\% confidence intervals from the distribution over the 200 repetitions.
All simulations are in the "high mediated proportion without overlap violation" framework.
We focus on four estimators that were re-implemented in Python for this study, and assess the impact of using regularized models for the nuisance functions.
In all simulated cases, it appears preferable to use unregularized models.
}
\label{result_linear_reg}
\end{figure}

In the following experiments, we focus on four estimators that we have re-implemented in Python for this study, and assess the impact of using machine learning for the nuisance functions, instead of plain parametric linear models.
We consider the IPW estimator, the G-computation estimator, the multiply-robust estimator and the double-machine learning estimator.
A first improvement we can apply is regularization, by imposing a penalty on the size of the coefficients.
In our implementation, the hyperparameter setting the amount of shrinkage is set by cross-validation.
Supplementary Figure~\ref{result_linear_reg} presents the results for the unregularized and the regularized versions of those three estimators in the "high mediated proportion without overlap violation" framework.
Overall, the regularized estimators underperform their unregularized equivalent.
Contrary to what we observed with the unregularized estimators (Figure~\ref{result_sample_size}, and Supplementary Figure~\ref{result_linear_reg}), we see that the estimation error decreases with the number of samples, indicating a slow convergence.
However, even with a large sample, the error is still much larger for the regularized estimators.
This could be explained by the fact that this is a well-posed regime, with not-so-high dimensional data and a lot of samples.
Regularization becomes futile and may bias estimators when we compare different parameter settings.
%\bt{Maybe we can say a bit more here: we are in a well-posed regime, with not-so-high dimensional data and a lot of samples. Regularization becomes futile and may bias estimators when we compare different parameter settings.}
% \ja{dire quelque chose des données UKBB!!!!}

\begin{figure}[h]
\includegraphics[width=\textwidth]{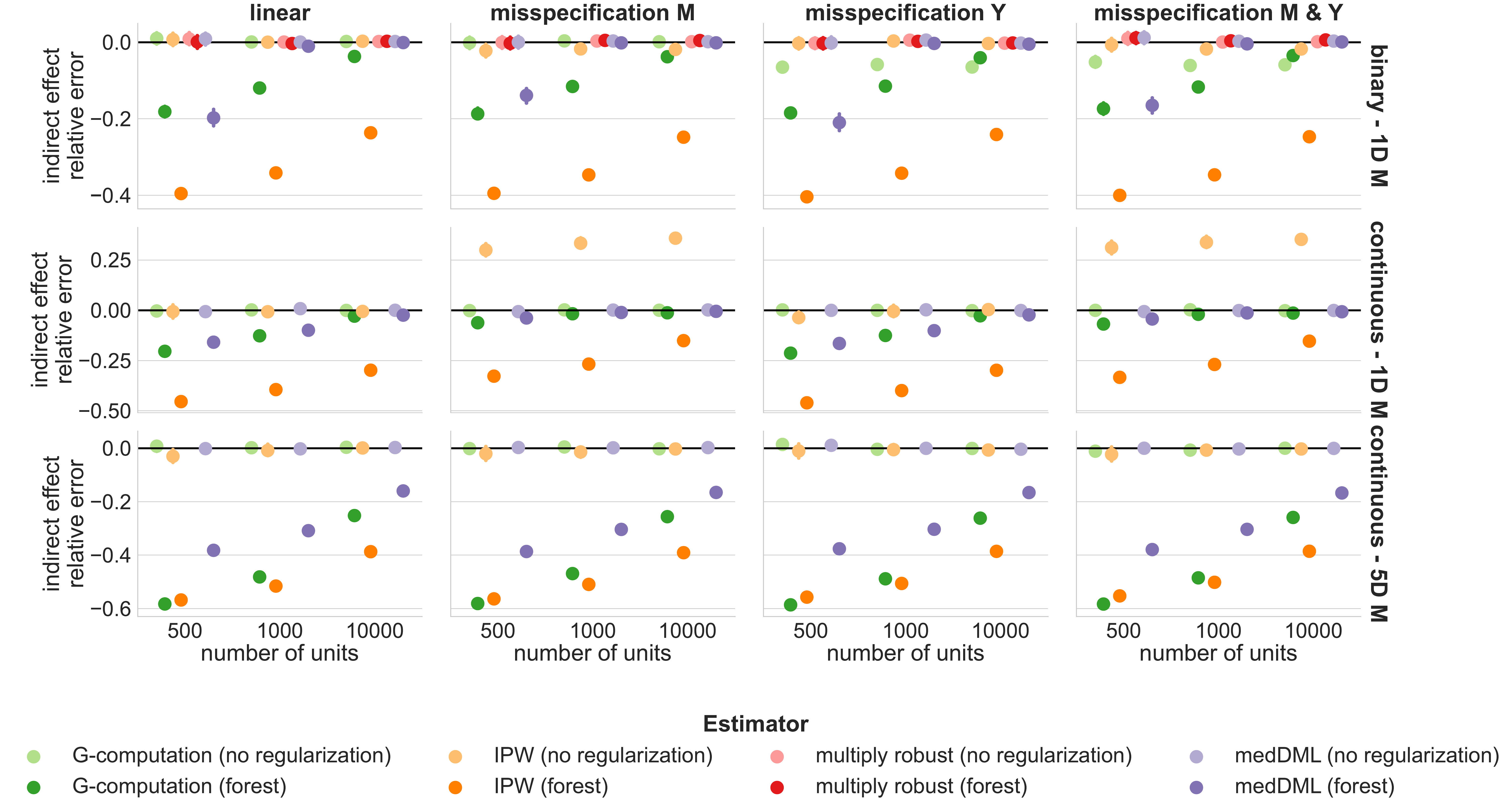}
\caption{
\textbf{Indirect effect relative error with random forests for plug-in nuisance parameters estimation.}
We show results for four scenarios of generative model specification, violating or not the parametric linear nuisance models of some estimators.
Each column corresponds to a distinct specification of simulated models.
The rows correspond to different mediator types, labeled on the right.
The three sets of dots in each panel correspond to the different number of observations tested in our set of simulations.
Each dot represents the average relative error (i.e. $\frac{\hat{\tau}-\tau}{\tau}$, for example for the total effect $\tau$) over 200 repetitions, and the error bars are the asymptotic normal 95\% confidence intervals from the distribution over the 200 repetitions.
All simulations are in the "high mediated proportion without overlap violation" framework.
}
\label{result_linear_forest}
\end{figure}

We observe the same trend, and no improvement on the estimation accuracy when we replace the nuisance functions models by random forests~\citep{breiman2001random} (Supplementary Figure~\ref{result_linear_forest}, even when the mediator or outcome models are misspecified.
In the case of the G-computation, with a very large number of observations, the performance of the estimator with forest nuisance models seems to catch up with the linear model, indicating that it could eventually converge to an unbiased estimation for the one-dimensional binary or continuous mediator, however, the convergence is really slow for the multi-dimensional case.
For the double machine learning estimator, the robustness property with misspecified nuisance function seems to be the best solution.

In Supplementary Figures~\ref{result_linear_reg_cf} and~\ref{result_forest_cf}, we explore the potential improvement obtained by using cross-fitting for the nuisance models.
This is generally presented as a solution to the slow convergence of machine learning models for the nuisance parameters~\cite{chernozhukov2018double}.
However, we observe rather a slightly slower convergence, probably due to the smaller sample size.

\begin{figure}[h]
\includegraphics[width=\textwidth]{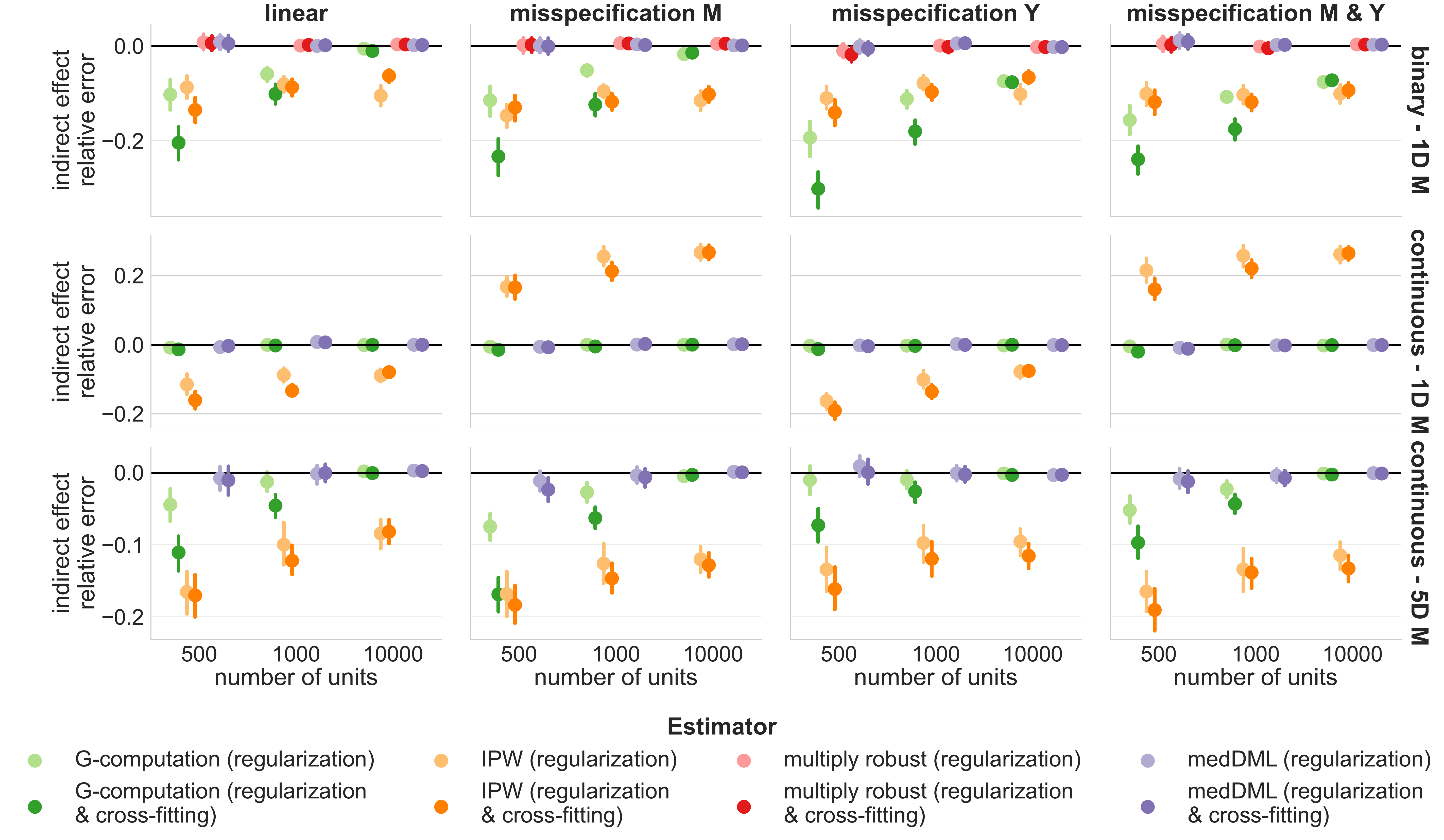}
\caption{
\textbf{Indirect effect relative error with linear models with and without cross-fitting for plug-in nuisance parameters estimation.}
We show results for four scenarios of generative model specification, violating or not the parametric linear nuisance models of some estimators.
Each column corresponds to a distinct specification of simulated models.
The rows correspond to different mediator types, labeled on the right.
The four sets of dots in each panel correspond to the different number of observations tested in our set of simulations.
Each dot represents the average relative error (i.e. $\frac{\hat{\tau}-\tau}{\tau}$, for example for the total effect $\tau$) over 200 repetitions, and the error bars are the asymptotic normal 95\% confidence intervals from the distribution over the 200 repetitions.
All simulations are in the "high mediated proportion without overlap violation" framework.
}
\label{result_linear_reg_cf}
\end{figure}

\begin{figure}[h]
\includegraphics[width=\textwidth]{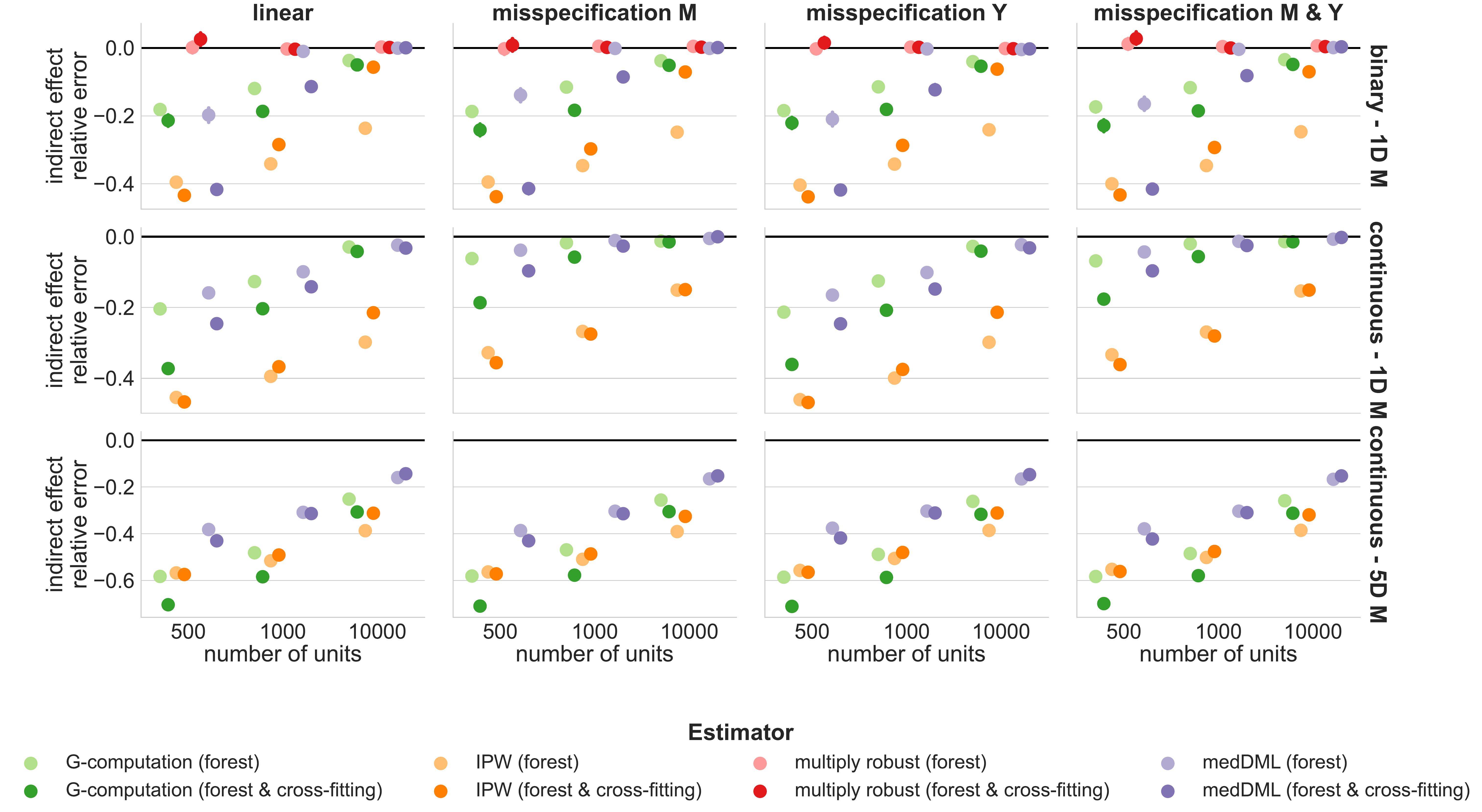}
\caption{
\textbf{Indirect effect relative error with random forests models with and without cross-fitting for plug-in nuisance parameters estimation.}
We show results for four scenarios of generative model specification, violating or not the parametric linear nuisance models of some estimators.
Each column corresponds to a distinct specification of simulated models.
The rows correspond to different mediator types, labeled on the right.
The four sets of dots in each panel correspond to the different number of observations tested in our set of simulations.
Each dot represents the average relative error (i.e. $\frac{\hat{\tau}-\tau}{\tau}$, for example for the total effect $\tau$) over 200 repetitions, and the error bars are the asymptotic normal 95\% confidence intervals from the distribution over the 200 repetitions.
All simulations are in the "high mediated proportion without overlap violation" framework.
}
\label{result_forest_cf}
\end{figure}

\FloatBarrier
\newpage

\begin{figure}
\centering
\includegraphics[width=\textwidth]{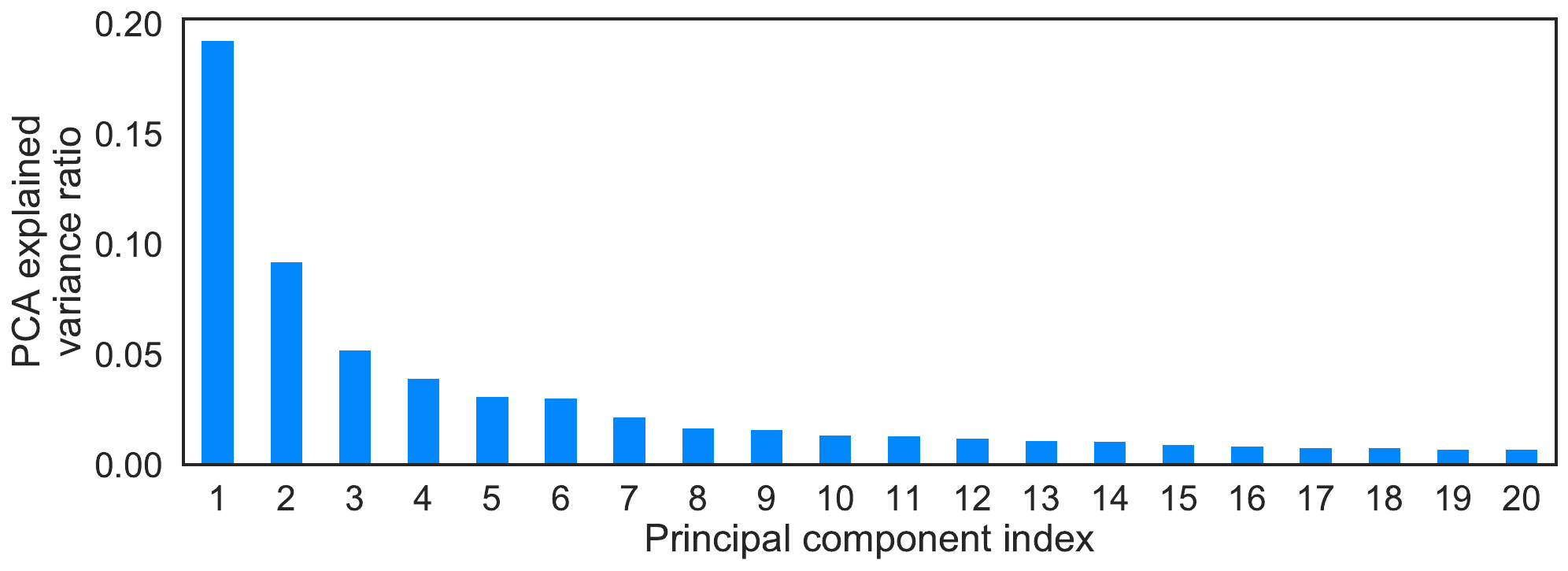}
\caption[Explained variance of PCA components of sMRI and dMRI IDPs (Image Derived Phenotypes)]{\textbf{Explained variance of PCA components of sMRI and dMRI IDPs (Image Derived Phenotypes).} This barplot shows the percentage of the dataset variation that is explained by each component of a PCA applied to the 311 retained IDPs variables from the raw brain MRI data. We retained 6 components after visual inspection.}
\label{mri_pca}
\end{figure}

\FloatBarrier

\begin{longtable}{|p{2cm}|p{8cm}|p{5cm}|}
\caption{\textbf{Description of considered exposures.} This table describes how several fields from the raw UKBB data are combined together to define binary exposures.}
\label{ukbb_exposure} \\ \hline

\textbf{exposure} & \textbf{UBKK fields} & \textbf{aggregation rationale} \\ \hline
\endfirsthead
college or university degree & \textbf{6138} (Qualifications) & equal to 'College or University degree' \\ \hline
diabetic & fields \textbf{20002} (Non-cancer illness code, self-reported) and \textbf{6177} (Medication for cholesterol, blood pressure or diabetes) and \textbf{6153} (Medication for cholesterol, blood pressure, diabetes, or take exogenous hormones) at the imaging visit & self-declared diabetes or medication for diabetes, inspired from~\citet{newby2022understanding} \\ \hline
hypercholesterol & fields \textbf{20002} (Non-cancer illness code, self-reported) and \textbf{6177} (Medication for cholesterol, blood pressure or diabetes) and \textbf{6153} (Medication for cholesterol, blood pressure, diabetes, or take exogenous hormones) at the imaging visit & self-declared hypercholesterol or medication for hypercholesterol, inspired from~\citet{newby2022understanding} \\ \hline
hypertension & fields \textbf{20002} (Non-cancer illness code, self-reported) and \textbf{6177} (Medication for cholesterol, blood pressure or diabetes) and \textbf{6153} (Medication for cholesterol, blood pressure, diabetes, or take exogenous hormones) or \texttt{6150} (Vascular/heart problems diagnosed by doctor) at the imaging visit & self-declared hypertension or medication for hypertension, or diagnosed by a medical doctor, inspired from~\citet{newby2022understanding} \\
\bottomrule
\end{longtable}

\newgeometry{left=1.5cm,bottom=1.5cm,right=1.5cm,top=1.5cm}
\begin{scriptsize}
\begin{landscape}
\setlength{\tabcolsep}{2pt}

\FloatBarrier
\newpage

\begin{longtable}{|p{2.5cm}|p{4cm}|p{3cm}|p{2cm}|p{2cm}|p{2cm}|p{2cm}|p{2cm}|p{2cm}|}
\caption[Baseline characteristics of for the included participants, according to their higher education degree]{\textbf{Baseline characteristics of for the included participants, according to their higher education degree}. Adequate statistical tests depending on the variable type assess whether each variable has a distinct distributions between the "exposed" and the "unexposed" group (with exposure being graduating a College or University degree). We also report the Standardized mean difference (SMD) for each variable.}
\label{table1_education}\\ \hline 
\hline

variable role & variable name   &      variable modalities    & Overall        & No College or University degree              & College or University degree              & P-Value   & Test                                      & SMD (0,1)   \\ \hline
\hline
\endfirsthead
\multicolumn{8}{@{}l}{\ldots continued}\\\hline
variable role & variable name   &      variable modalities     & Overall        & No College or University degree              & College or University degree              & P-Value   & Test                                      & SMD (0,1)   \\ \hline
\endhead

Sample size & n                                                                            &                                               & 32498          & 16295          & 16203          &           &                                           &             \\
Outcome & G factor, mean (SD)                                                          &                                               & -0.0 (1.1)     & 0.2 (1.2)      & -0.3 (1.1)     & \ensuremath{<}0.001    & Two Sample T-test                         & -0.470      \\
\multirow[t]{16}{*}{Mediators} & Dmri pc 0, mean (SD)                                                         &                                               & -0.3 (10.2)    & 0.1 (10.3)     & -0.7 (10.1)    & \ensuremath{<}0.001    & Two Sample T-test                         & -0.085      \\
 & Dmri pc 1, mean (SD)                                                         &                                               & 0.1 (5.2)      & 0.3 (5.2)      & -0.2 (5.2)     & \ensuremath{<}0.001    & Two Sample T-test                         & -0.091      \\
 & Dmri pc 2, mean (SD)                                                         &                                               & -0.1 (5.0)     & 0.2 (5.1)      & -0.4 (4.9)     & \ensuremath{<}0.001    & Two Sample T-test                         & -0.111      \\
 & Dmri pc 3, mean (SD)                                                         &                                               & 0.0 (4.4)      & 0.2 (4.4)      & -0.2 (4.4)     & \ensuremath{<}0.001    & Two Sample T-test                         & -0.096      \\
 & Dmri pc 4, mean (SD)                                                         &                                               & 0.0 (4.2)      & 0.0 (4.3)      & 0.0 (4.1)      & 0.880     & Two Sample T-test                         & 0.002       \\
\bottomrule
\end{longtable}

\begin{longtable}{|p{2.5cm}|p{4cm}|p{3cm}|p{2cm}|p{2cm}|p{2cm}|p{2cm}|p{2cm}|p{2cm}|}
\caption[Baseline characteristics of for the included participants, according to their diabetes status]{\textbf{Baseline characteristics of for the included participants, according to their higher diabetes status}. Adequate statistical tests depending on the variable type assess whether each variable has a distinct distributions between the "exposed" and the "unexposed" group (with exposure being diagnosed with diabetes). We also report the Standardized mean difference (SMD) for each variable.}
\label{table1_diabetes}\\ \hline 
\hline

variable role & variable name   &      variable modalities    & Overall        & No diabetes              & Diabetes              & P-Value   & Test                                      & SMD (0,1)   \\ \hline
\hline
\endfirsthead
\multicolumn{8}{@{}l}{\ldots continued}\\\hline
variable role & variable name   &      variable modalities     & Overall        & No diabetes              & Diabetes              & P-Value   & Test                                      & SMD (0,1)   \\ \hline
\endhead

Sample size &  n                                                                            &                                                                 & 32498          & 31154          & 1344           &           &                                           &             \\
Outcome &  G factor, mean (SD)                                                          &                                                                 & -0.0 (1.1)     & -0.0 (1.1)     & 0.2 (1.2)      & \ensuremath{<}0.001    & Two Sample T-test                         & 0.202       \\
\multirow[t]{16}{*}{Mediators} &  Dmri pc 0, mean (SD)                                                         &                                                                 & -0.3 (10.2)    & -0.4 (10.1)    & 3.4 (11.7)     & \ensuremath{<}0.001    & Two Sample T-test                         & 0.356       \\
 &  Dmri pc 1, mean (SD)                                                         &                                                                 & 0.1 (5.2)      & 0.1 (5.2)      & -1.2 (6.2)     & \ensuremath{<}0.001    & Two Sample T-test                         & -0.231      \\
 &  Dmri pc 2, mean (SD)                                                         &                                                                 & -0.1 (5.0)     & -0.2 (4.8)     & 1.9 (7.9)      & \ensuremath{<}0.001    & Two Sample T-test                         & 0.330       \\
 &  Dmri pc 3, mean (SD)                                                         &                                                                 & 0.0 (4.4)      & -0.0 (4.4)     & 0.6 (4.9)      & \ensuremath{<}0.001    & Two Sample T-test                         & 0.139       \\
\bottomrule
\end{longtable} 

\begin{longtable}{|p{2.5cm}|p{4cm}|p{3cm}|p{2cm}|p{2cm}|p{2cm}|p{2cm}|p{2cm}|p{2cm}|}
\caption[Baseline characteristics of for the included participants, according to their hypercholesterol status]{\textbf{Baseline characteristics of for the included participants, according to their higher hypercholesterol status}. Adequate statistical tests depending on the variable type assess whether each variable has a distinct distributions between the "exposed" and the "unexposed" group (with exposure having hypercholesterol). We also report the Standardized mean difference (SMD) for each variable.}
\label{table1_hypercholesterol}\\ \hline 
\hline

variable role & variable name   &      variable modalities    & Overall        & No hypercholesterol              & Hypercholesterol              & P-Value   & Test                                      & SMD (0,1)   \\ \hline
\hline
\endfirsthead
\multicolumn{8}{@{}l}{\ldots continued}\\\hline
variable role & variable name   &      variable modalities     & Overall        & No hypercholesterol              & Hypercholesterol              & P-Value   & Test                                      & SMD (0,1)   \\ \hline
\endhead

Sample size &  n                                                                            &                                                                 & 32498          & 24538          & 7960           &           &                                           &             \\
Outcome &  G factor, mean (SD)                                                          &                                                                 & -0.0 (1.1)     & -0.1 (1.1)     & 0.2 (1.2)      & \ensuremath{<}0.001    & Two Sample T-test                         & 0.230       \\
\multirow[t]{16}{*}{Mediators}  &  Dmri pc 0, mean (SD)                                                         &                                                                 & -0.3 (10.2)    & -1.4 (9.6)     & 3.0 (11.2)     & \ensuremath{<}0.001    & Two Sample T-test                         & 0.418       \\
 &  Dmri pc 1, mean (SD)                                                         &                                                                 & 0.1 (5.2)      & 0.5 (5.0)      & -1.2 (5.5)     & \ensuremath{<}0.001    & Two Sample T-test                         & -0.308      \\
 &  Dmri pc 2, mean (SD)                                                         &                                                                 & -0.1 (5.0)     & -0.5 (4.7)     & 1.0 (5.7)      & \ensuremath{<}0.001    & Two Sample T-test                         & 0.288       \\
 &  Dmri pc 3, mean (SD)                                                         &                                                                 & 0.0 (4.4)      & -0.2 (4.3)     & 0.5 (4.6)      & \ensuremath{<}0.001    & Two Sample T-test                         & 0.159       \\
\bottomrule
\end{longtable} 

\begin{longtable}{|p{2.5cm}|p{4cm}|p{3cm}|p{2cm}|p{2cm}|p{2cm}|p{2cm}|p{2cm}|p{2cm}|}
\caption[Baseline characteristics of for the included participants, according to their hypertensive status]{\textbf{Baseline characteristics of for the included participants, according to their higher hypertensive status}. Adequate statistical tests depending on the variable type assess whether each variable has a distinct distributions between the "exposed" and the "unexposed" group (with exposure having hypertension). We also report the Standardized mean difference (SMD) for each variable.}
\label{table1_hypertensive}\\ \hline 
\hline

variable role & variable name   &      variable modalities    & Overall        & No hypertension              & Hypertension              & P-Value   & Test                                      & SMD (0,1)   \\ \hline
\hline
\endfirsthead
\multicolumn{8}{@{}l}{\ldots continued}\\\hline
variable role & variable name   &      variable modalities     & Overall        & No hypertension              &  Hypertension             & P-Value   & Test                                      & SMD (0,1)   \\ \hline
\endhead
Sample size & n                                                                            &                                                                & 32498          & 23962          & 8536           &           &                                           &             \\
Outcome & G factor, mean (SD)                                                          &                                                                & -0.0 (1.1)     & -0.1 (1.1)     & 0.1 (1.2)      & \ensuremath{<}0.001    & Two Sample T-test                         & 0.169       \\
\multirow[t]{16}{*}{Mediators}  & Dmri pc 0, mean (SD)                                                         &                                                                & -0.3 (10.2)    & -1.6 (9.3)     & 3.4 (11.5)     & \ensuremath{<}0.001    & Two Sample T-test                         & 0.476       \\
 & Dmri pc 1, mean (SD)                                                         &                                                                & 0.1 (5.2)      & 0.4 (5.1)      & -0.8 (5.5)     & \ensuremath{<}0.001    & Two Sample T-test                         & -0.225      \\
 & Dmri pc 2, mean (SD)                                                         &                                                                & -0.1 (5.0)     & -0.4 (4.9)     & 0.8 (5.2)      & \ensuremath{<}0.001    & Two Sample T-test                         & 0.241       \\
 & Dmri pc 3, mean (SD)                                                         &                                                                & 0.0 (4.4)      & -0.3 (4.3)     & 0.7 (4.5)      & \ensuremath{<}0.001    & Two Sample T-test                         & 0.226       \\
\bottomrule
\end{longtable} 

\begin{longtable}{|p{2.5cm}|p{4cm}|p{3cm}|p{2cm}|p{2cm}|p{2cm}|p{2cm}|p{2cm}|p{2cm}|}
\caption[Baseline characteristics of for the included participants, according to their probable current major depressive disorder (cMMD) status]{\textbf{Baseline characteristics of for the included participants, according to their probable current major depressive disorder (cMMD)  status}. Adequate statistical tests depending on the variable type assess whether each variable has a distinct distributions between the "exposed" and the "unexposed" group (with exposure having a probable current major depressive disorder). We also report the Standardized mean difference (SMD) for each variable.}
\label{table1_cMMD}\\ \hline 
\hline

variable role & variable name   &      variable modalities    & Overall        & No cMMD              & cMMD              & P-Value   & Test                                      & SMD (0,1)   \\ \hline
\hline
\endfirsthead
\multicolumn{8}{@{}l}{\ldots continued}\\\hline
variable role & variable name   &      variable modalities     & Overall        & No cMMD              & cMMD              & P-Value   & Test                                      & SMD (0,1)   \\ \hline
\endhead
Sample size & n                                                                            &                                                                & 32498          & 23413          & 9085           &           &                                           &                    \\
Outcome & G factor, mean (SD)                                                          &                                                                & -0.0 (1.1)     & -0.0 (1.1)     & -0.0 (1.1)     & 0.064     & Two Sample T-test                         & 0.023              \\
\multirow[t]{16}{*}{Mediators}  & Dmri pc 0, mean (SD)                                                         &                                                                & -0.3 (10.2)    & -0.0 (10.2)    & -1.0 (10.1)    & \ensuremath{<}0.001    & Two Sample T-test                         & -0.096             \\
 & Dmri pc 1, mean (SD)                                                         &                                                                & 0.1 (5.2)      & -0.1 (5.2)     & 0.5 (5.1)      & \ensuremath{<}0.001    & Two Sample T-test                         & 0.118              \\
 & Dmri pc 2, mean (SD)                                                         &                                                                & -0.1 (5.0)     & -0.1 (5.1)     & -0.2 (4.9)     & 0.143     & Two Sample T-test                         & -0.018             \\
 & Dmri pc 3, mean (SD)                                                         &                                                                & 0.0 (4.4)      & 0.0 (4.4)      & -0.0 (4.3)     & 0.519     & Two Sample T-test                         & -0.008             \\
\bottomrule
\end{longtable} 

\begin{longtable}{|p{2.5cm}|p{4cm}|p{3cm}|p{2cm}|p{2cm}|p{2cm}|p{2cm}|p{2cm}|p{2cm}|}
\caption[Baseline characteristics of for the included participants, according to their life long depression status]{\textbf{Baseline characteristics of for the included participants, according to their life long depression status}. Adequate statistical tests depending on the variable type assess whether each variable has a distinct distributions between the "exposed" and the "unexposed" group (with exposure having a lifelong depression). We also report the Standardized mean difference (SMD) for each variable.}
\label{table1_life_long_depression}\\ \hline 
\hline

variable role & variable name   &      variable modalities    & Overall        & No lifelong depression              & Lifelong depression              & P-Value   & Test                                      & SMD (0,1)   \\ \hline
\hline
\endfirsthead
\multicolumn{8}{@{}l}{\ldots continued}\\\hline
variable role & variable name   &      variable modalities     & Overall        & No lifelong depression              & Lifelong depression              & P-Value   & Test                                      & SMD (0,1)   \\ \hline
\endhead
Sample size & n                                                                            &                                                                 & 32498          & 27461          & 5037           &           &                                           &                    \\
Outcome & G factor, mean (SD)                                                          &                                                                 & -0.0 (1.1)     & -0.0 (1.2)     & -0.1 (1.1)     & \ensuremath{<}0.001    & Two Sample T-test                         & -0.104             \\
\multirow[t]{16}{*}{Mediators}  & Dmri pc 0, mean (SD)                                                         &                                                                 & -0.3 (10.2)    & -0.2 (10.3)    & -1.0 (9.9)     & \ensuremath{<}0.001    & Two Sample T-test                         & -0.081             \\
 & Dmri pc 1, mean (SD)                                                         &                                                                 & 0.1 (5.2)      & -0.0 (5.3)     & 0.6 (4.9)      & \ensuremath{<}0.001    & Two Sample T-test                         & 0.119              \\
 & Dmri pc 2, mean (SD)                                                         &                                                                 & -0.1 (5.0)     & -0.1 (5.1)     & -0.3 (4.6)     & \ensuremath{<}0.001    & Two Sample T-test                         & -0.055             \\
 & Dmri pc 3, mean (SD)                                                         &                                                                 & 0.0 (4.4)      & 0.0 (4.4)      & -0.1 (4.3)     & 0.050     & Two Sample T-test                         & -0.030             \\
\bottomrule
\end{longtable} 

\begin{longtable}{|p{2.5cm}|p{4cm}|p{3cm}|p{2cm}|p{2cm}|p{2cm}|p{2cm}|p{2cm}|p{2cm}|}
\caption[Baseline characteristics of for the included participants, according to their social support status]{\textbf{Baseline characteristics of for the included participants, according to their social support status}. Adequate statistical tests depending on the variable type assess whether each variable has a distinct distributions between the "exposed" and the "unexposed" group (with exposure having a good social support). We also report the Standardized mean difference (SMD) for each variable.}
\label{table1_life_social_support}\\ \hline 
\hline

variable role & variable name   &      variable modalities    & Overall        & Lack of social support              & Social support              & P-Value   & Test                                      & SMD (0,1)   \\ \hline
\hline
\endfirsthead
\multicolumn{8}{@{}l}{\ldots continued}\\\hline
variable role & variable name   &      variable modalities     & Overall        & Lack of social support              & Social support              & P-Value   & Test                                      & SMD (0,1)   \\ \hline
\endhead
Sample size & n                                                                            &                                                                & 32498          & 12720          & 19778          &           &                                           &                    \\
Outcome & G factor, mean (SD)                                                          &                                                                & -0.0 (1.1)     & 0.0 (1.2)      & -0.1 (1.1)     & \ensuremath{<}0.001    & Two Sample T-test                         & -0.072             \\
\multirow[t]{16}{*}{Mediators}  & Dmri pc 0, mean (SD)                                                         &                                                                & -0.3 (10.2)    & 0.2 (10.4)     & -0.6 (10.1)    & \ensuremath{<}0.001    & Two Sample T-test                         & -0.084             \\
 & Dmri pc 1, mean (SD)                                                         &                                                                & 0.1 (5.2)      & -0.1 (5.3)     & 0.2 (5.2)      & \ensuremath{<}0.001    & Two Sample T-test                         & 0.068              \\
 & Dmri pc 2, mean (SD)                                                         &                                                                & -0.1 (5.0)     & 0.1 (5.0)      & -0.2 (5.0)     & \ensuremath{<}0.001    & Two Sample T-test                         & -0.063             \\
 & Dmri pc 3, mean (SD)                                                         &                                                                & 0.0 (4.4)      & 0.0 (4.5)      & 0.0 (4.3)      & 0.732     & Two Sample T-test                         & -0.004             \\
\bottomrule
\end{longtable}  

\begin{longtable}{|p{2.5cm}|p{4cm}|p{3cm}|p{2cm}|p{2cm}|p{2cm}|p{2cm}|p{2cm}|p{2cm}|}
\caption[Baseline characteristics of for the included participants, according to their inflammation status]{\textbf{Baseline characteristics of for the included participants, according to their inflammation status}. Adequate statistical tests depending on the variable type assess whether each variable has a distinct distributions between the "exposed" and the "unexposed" group (with exposure being chronic inflammation). We also report the Standardized mean difference (SMD) for each variable.}
\label{table1_inflammation}\\ \hline 
\hline

variable role & variable name   &      variable modalities    & Overall        & No inflammation             & Inflammation              & P-Value   & Test                                      & SMD (0,1)   \\ \hline
\hline
\endfirsthead
\multicolumn{8}{@{}l}{\ldots continued}\\\hline
variable role & variable name   &      variable modalities     & Overall        & No inflammation             & Inflammation              & P-Value   & Test                                      & SMD (0,1)   \\ \hline
\endhead

Sample size & n                                                                            &                                                                 & 32498          & 27267          & 5231           &           &                                           &                    \\
Outcome & G factor, mean (SD)                                                          &                                                                 & -0.0 (1.1)     & -0.1 (1.1)     & 0.1 (1.2)      & \ensuremath{<}0.001    & Two Sample T-test                         & 0.095              \\
\multirow[t]{16}{*}{Mediators}  & Dmri pc 0, mean (SD)                                                         &                                                                 & -0.3 (10.2)    & -0.4 (10.1)    & 0.4 (10.5)     & \ensuremath{<}0.001    & Two Sample T-test                         & 0.083              \\
 & Dmri pc 1, mean (SD)                                                         &                                                                 & 0.1 (5.2)      & 0.1 (5.2)      & -0.1 (5.4)     & 0.009     & Two Sample T-test                         & -0.040             \\
 & Dmri pc 2, mean (SD)                                                         &                                                                 & -0.1 (5.0)     & -0.3 (4.9)     & 0.7 (5.8)      & \ensuremath{<}0.001    & Two Sample T-test                         & 0.190              \\
 & Dmri pc 3, mean (SD)                                                         &                                                                 & 0.0 (4.4)      & -0.1 (4.4)     & 0.6 (4.4)      & \ensuremath{<}0.001    & Two Sample T-test                         & 0.152              \\
\bottomrule
\end{longtable}  

\begin{longtable}{|p{2.5cm}|p{4cm}|p{3cm}|p{2cm}|p{2cm}|p{2cm}|p{2cm}|p{2cm}|p{2cm}|}
\caption[Baseline characteristics of for the included participants, according to their current smoker status]{\textbf{Baseline characteristics of for the included participants, according to their current smoker status}. Adequate statistical tests depending on the variable type assess whether each variable has a distinct distributions between the "exposed" and the "unexposed" group (with exposure being current smoking). We also report the Standardized mean difference (SMD) for each variable.}
\label{table1_current_smoker}\\ \hline 
\hline

variable role & variable name   &      variable modalities    & Overall        & No current smoking              & Current smoking              & P-Value   & Test                                      & SMD (0,1)   \\ \hline
\hline
\endfirsthead
\multicolumn{8}{@{}l}{\ldots continued}\\\hline
variable role & variable name   &      variable modalities     & Overall        & No current smoking              & Current smoking              & P-Value   & Test                                      & SMD (0,1)   \\ \hline
\endhead

Sample size & n                                                                            &                                                                 & 32498          & 31893          & 605            &           &                                           &             \\
Outcome & G factor, mean (SD)                                                          &                                                                 & -0.0 (1.1)     & -0.0 (1.1)     & 0.1 (1.1)      & 0.006     & Two Sample T-test                         & 0.114       \\
\multirow[t]{16}{*}{Mediators}  & Dmri pc 0, mean (SD)                                                         &                                                                 & -0.3 (10.2)    & -0.3 (10.2)    & 0.7 (10.6)     & 0.026     & Two Sample T-test                         & 0.093       \\
 & Dmri pc 1, mean (SD)                                                         &                                                                 & 0.1 (5.2)      & 0.1 (5.2)      & 0.3 (5.1)      & 0.278     & Two Sample T-test                         & 0.044       \\
 & Dmri pc 2, mean (SD)                                                         &                                                                 & -0.1 (5.0)     & -0.1 (5.0)     & -0.1 (4.5)     & 0.890     & Two Sample T-test                         & 0.005       \\
 & Dmri pc 3, mean (SD)                                                         &                                                                 & 0.0 (4.4)      & -0.0 (4.4)     & 0.8 (4.0)      & \ensuremath{<}0.001    & Two Sample T-test                         & 0.195       \\
 & Dmri pc 4, mean (SD)                                                         &                                                                 & 0.0 (4.2)      & 0.0 (4.2)      & 0.2 (4.2)      & 0.189     & Two Sample T-test                         & 0.054       \\
\bottomrule
\end{longtable}

\begin{longtable}{|p{2.5cm}|p{4cm}|p{3cm}|p{2cm}|p{2cm}|p{2cm}|p{2cm}|p{2cm}|p{2cm}|}
\caption[Baseline characteristics of for the included participants, according to their current smoker status versus never smokers]{\textbf{Baseline characteristics of for the included participants, according to their current smoker status versus never smokers}. Adequate statistical tests depending on the variable type assess whether each variable has a distinct distributions between the "exposed" and the "unexposed" group (with exposure being a current smoker, and no treatment being a never smoker). We also report the Standardized mean difference (SMD) for each variable.}
\label{table1_current_never_smoker}\\ \hline 
\hline

variable role & variable name   &      variable modalities    & Overall        & Never smoker              & Current smoker              & P-Value   & Test                                      & SMD (0,1)   \\ \hline
\hline
\endfirsthead
\multicolumn{8}{@{}l}{\ldots continued}\\\hline
variable role & variable name   &      variable modalities     & Overall        & Never smoker              & Current smoker              & P-Value   & Test                                      & SMD (0,1)   \\ \hline
\endhead

Sample size & n                                                                            &                                                                & 21203          & 20598          & 605            &           &                                           &                 \\
Outcome & G factor, mean (SD)                                                          &                                                                & -0.1 (1.1)     & -0.1 (1.1)     & 0.1 (1.1)      & 0.001     & Two Sample T-test                         & 0.132           \\
\multirow[t]{16}{*}{Mediators}  & Dmri pc 0, mean (SD)                                                         &                                                                & -0.8 (10.0)    & -0.8 (10.0)    & 0.7 (10.6)     & 0.001     & Two Sample T-test                         & 0.147           \\
 & Dmri pc 1, mean (SD)                                                         &                                                                & 0.4 (5.2)      & 0.4 (5.2)      & 0.3 (5.1)      & 0.698     & Two Sample T-test                         & -0.016          \\
 & Dmri pc 2, mean (SD)                                                         &                                                                & -0.3 (5.0)     & -0.3 (5.0)     & -0.1 (4.5)     & 0.275     & Two Sample T-test                         & 0.043           \\
 & Dmri pc 3, mean (SD)                                                         &                                                                & -0.1 (4.4)     & -0.1 (4.4)     & 0.8 (4.0)      & \ensuremath{<}0.001    & Two Sample T-test                         & 0.228           \\
 & Dmri pc 4, mean (SD)                                                         &                                                                & -0.0 (4.2)     & -0.0 (4.2)     & 0.2 (4.2)      & 0.139     & Two Sample T-test                         & 0.061           \\
\bottomrule
\end{longtable}

\begin{longtable}{|p{2.5cm}|p{4cm}|p{3cm}|p{2cm}|p{2cm}|p{2cm}|p{2cm}|p{2cm}|p{2cm}|}
\caption[Baseline characteristics of for the included participants, according to their previous smoker status versus never smokers]{\textbf{Baseline characteristics of for the included participants, according to their previous smoker status versus never smokers}. Adequate statistical tests depending on the variable type assess whether each variable has a distinct distributions between the "exposed" and the "unexposed" group (with exposure being a previous smoker, and no treatment being a never smoker). We also report the Standardized mean difference (SMD) for each variable.}
\label{table1_previous_never_smoker}\\ \hline 
\hline

variable role & variable name   &      variable modalities    & Overall        & Never smoker             & Previous smoker             & P-Value   & Test                                      & SMD (0,1)   \\ \hline
\hline
\endfirsthead
\multicolumn{8}{@{}l}{\ldots continued}\\\hline
variable role & variable name   &      variable modalities     & Overall        & Never smoker             & Previous smoker             & P-Value   & Test                                      & SMD (0,1)   \\ \hline
\endhead

Sample size & n                                                                            &                                                                & 31893          & 20598          & 11295          &           &                                           &                 \\
Outcome & G factor, mean (SD)                                                          &                                                                & -0.0 (1.1)     & -0.1 (1.1)     & -0.0 (1.1)     & \ensuremath{<}0.001    & Two Sample T-test                         & 0.051           \\
\multirow[t]{16}{*}{Mediators}  & Dmri pc 0, mean (SD)                                                         &                                                                & -0.3 (10.2)    & -0.8 (10.0)    & 0.7 (10.6)     & \ensuremath{<}0.001    & Two Sample T-test                         & 0.148           \\
 & Dmri pc 1, mean (SD)                                                         &                                                                & 0.1 (5.2)      & 0.4 (5.2)      & -0.5 (5.2)     & \ensuremath{<}0.001    & Two Sample T-test                         & -0.167          \\
 & Dmri pc 2, mean (SD)                                                         &                                                                & -0.1 (5.0)     & -0.3 (5.0)     & 0.2 (5.0)      & \ensuremath{<}0.001    & Two Sample T-test                         & 0.100           \\
 & Dmri pc 3, mean (SD)                                                         &                                                                & -0.0 (4.4)     & -0.1 (4.4)     & 0.2 (4.4)      & \ensuremath{<}0.001    & Two Sample T-test                         & 0.087           \\
\bottomrule
\end{longtable}

\begin{longtable}{|p{2.5cm}|p{4cm}|p{3cm}|p{2cm}|p{2cm}|p{2cm}|p{2cm}|p{2cm}|p{2cm}|}
\caption[Baseline characteristics for the included participants, according to their weekly alcohol consumption, higher or lower than 14 units, the NHS recommendation]{\textbf{Baseline characteristics for the included participants, according to their weekly alcohol consumption, higher or lower than 14 units, the NHS recommendation}.  Adequate statistical tests depending on the variable type assess whether each variable has a distinct distributions between the "exposed" and the "unexposed" group (with exposure being consuming more than 14 weekly units of alcohol, and no treatment being consuming less than 14 weekly units of alcohol). We also report the Standardized mean difference (SMD) for each variable.}
\label{table1_alcohol_above_nhs_reco}\\ \hline 
\hline

variable role & variable name   &      variable modalities    & Overall        & Less than 14 weekly alcohol units              & More than 14 weekly alcohol units              & P-Value   & Test                                      & SMD (0,1)   \\ \hline
\hline
\endfirsthead
\multicolumn{8}{@{}l}{\ldots continued}\\\hline
variable role & variable name   &      variable modalities     & Overall        & Less than 14 weekly alcohol units              & More than 14 weekly alcohol units              & P-Value   & Test                                      & SMD (0,1)   \\ \hline
\endhead

Sample size & n                                                                            &                                                                 & 32498          & 21256          & 11242          &           &                                           &             \\
Outcome & G factor, mean (SD)                                                          &                                                                 & -0.0 (1.1)     & 0.0 (1.2)      & -0.1 (1.1)     & \ensuremath{<}0.001    & Two Sample T-test                         & -0.140      \\
\multirow[t]{16}{*}{Mediators}  & Dmri pc 0, mean (SD)                                                         &                                                                 & -0.3 (10.2)    & -0.5 (10.2)    & 0.0 (10.2)     & \ensuremath{<}0.001    & Two Sample T-test                         & 0.050       \\
 & Dmri pc 1, mean (SD)                                                         &                                                                 & 0.1 (5.2)      & 0.5 (5.2)      & -0.7 (5.1)     & \ensuremath{<}0.001    & Two Sample T-test                         & -0.229      \\
 & Dmri pc 2, mean (SD)                                                         &                                                                 & -0.1 (5.0)     & -0.2 (5.1)     & 0.0 (4.9)      & \ensuremath{<}0.001    & Two Sample T-test                         & 0.043       \\
 & Dmri pc 3, mean (SD)                                                         &                                                                 & 0.0 (4.4)      & -0.1 (4.4)     & 0.3 (4.3)      & \ensuremath{<}0.001    & Two Sample T-test                         & 0.097       \\
\bottomrule
\end{longtable}

\begin{longtable}{|p{2.5cm}|p{4cm}|p{3cm}|p{2cm}|p{2cm}|p{2cm}|p{2cm}|p{2cm}|p{2cm}|}
\caption[Baseline characteristics of for the included participants, according to their obesity status]{\textbf{Baseline characteristics of for the included participants, according to their obesity status}.  Adequate statistical tests depending on the variable type assess whether each variable has a distinct distributions between the "exposed" and the "unexposed" group (with exposure being obesity, and no treatment being no obesity). We also report the Standardized mean difference (SMD) for each variable.}
\label{table1_obesity}\\ \hline 
\hline

variable role & variable name   &      variable modalities    & Overall        & No obesity              & Obesity              & P-Value   & Test                                      & SMD (0,1)   \\ \hline
\hline
\endfirsthead
\multicolumn{8}{@{}l}{\ldots continued}\\\hline
variable role & variable name   &      variable modalities     & Overall        & No obesity              & Obesity              & P-Value   & Test                                      & SMD (0,1)   \\ \hline
\endhead

Sample size & n                                                                            &                                                                 & 32498          & 26781          & 5717           &           &                                           &             \\
Outcome & G factor, mean (SD)                                                          &                                                                 & -0.0 (1.1)     & -0.0 (1.1)     & 0.0 (1.2)      & \ensuremath{<}0.001    & Two Sample T-test                         & 0.054       \\
\multirow[t]{16}{*}{Mediators}  & Dmri pc 0, mean (SD)                                                         &                                                                 & -0.3 (10.2)    & -0.3 (10.2)    & -0.4 (10.3)    & 0.251     & Two Sample T-test                         & -0.017      \\
 & Dmri pc 1, mean (SD)                                                         &                                                                 & 0.1 (5.2)      & 0.1 (5.1)      & -0.2 (5.6)     & \ensuremath{<}0.001    & Two Sample T-test                         & -0.059      \\
 & Dmri pc 2, mean (SD)                                                         &                                                                 & -0.1 (5.0)     & -0.4 (4.7)     & 1.4 (6.2)      & \ensuremath{<}0.001    & Two Sample T-test                         & 0.338       \\
 & Dmri pc 3, mean (SD)                                                         &                                                                 & 0.0 (4.4)      & -0.3 (4.3)     & 1.5 (4.3)      & \ensuremath{<}0.001    & Two Sample T-test                         & 0.412       \\
 & Dmri pc 4, mean (SD)                                                         &                                                                 & 0.0 (4.2)      & -0.0 (4.1)     & 0.2 (4.8)      & 0.002     & Two Sample T-test                         & 0.048       \\
\bottomrule
\end{longtable}

\end{landscape}
\end{scriptsize} 
\restoregeometry

\begin{figure}
\centering
\includegraphics[width=\textwidth]{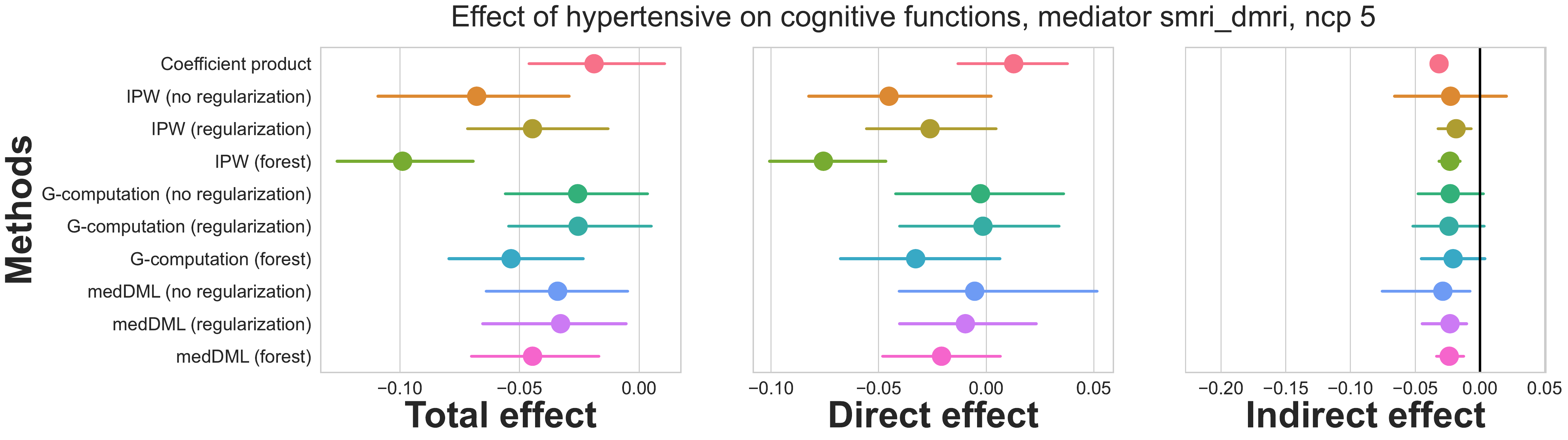}
\includegraphics[width=\textwidth]{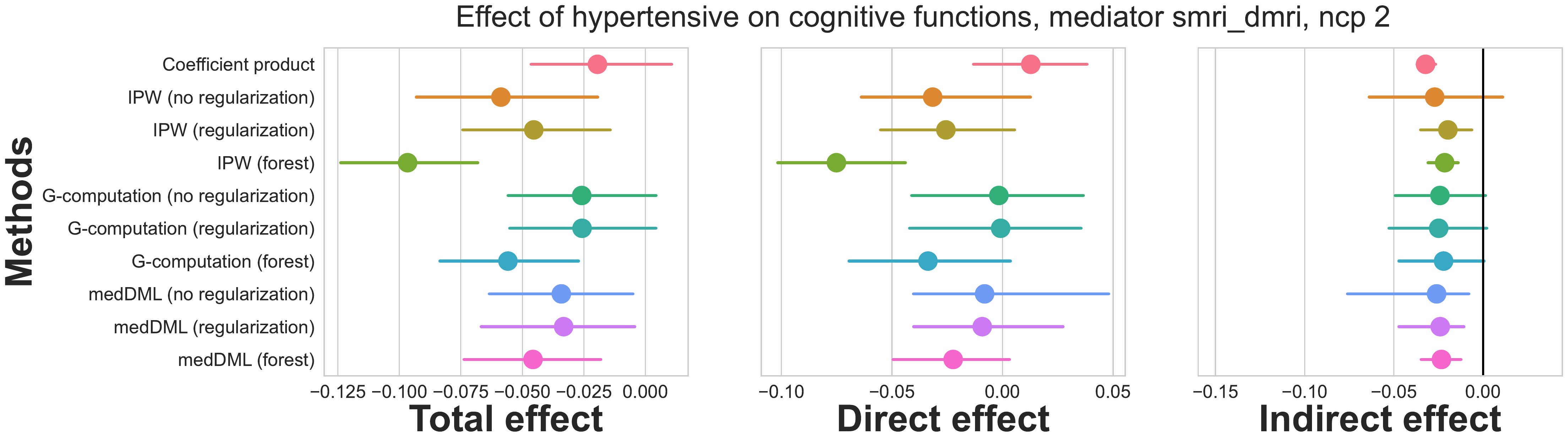}
\caption{\textbf{Causal total, direct and indirect effect of hypertension on cognitive functions, mediated through the brain structure.} The top panel represents results presented in the main text, with 5 components for the imputation of missing values by PCA for confounding variables, and the bottom panel with 2 components. We observe very similar results. We also show the estimates obtained with different estimators, and again, results are quite similar among the different approaches. Each estimate is represented as a mean over 100 bootstrapped samples. The error bar is the 95\% percentile-interval, estimated from the empirical bootstrap distribution.}
\label{diff_ncp_hyper}
\end{figure}

\begin{figure}
\centering
\includegraphics[width=\textwidth]{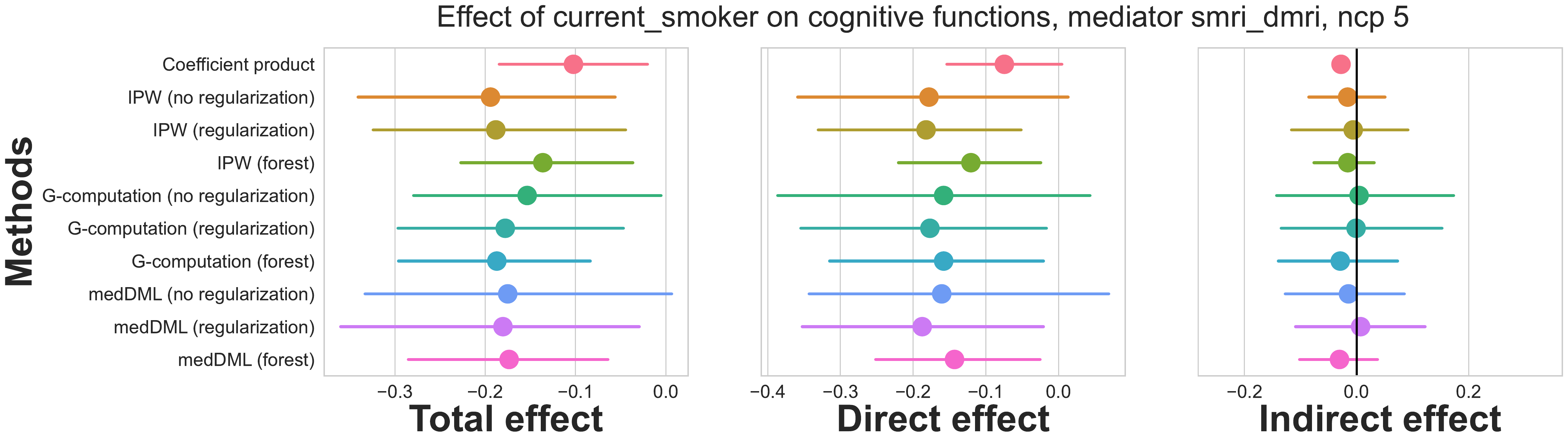}
\includegraphics[width=\textwidth]{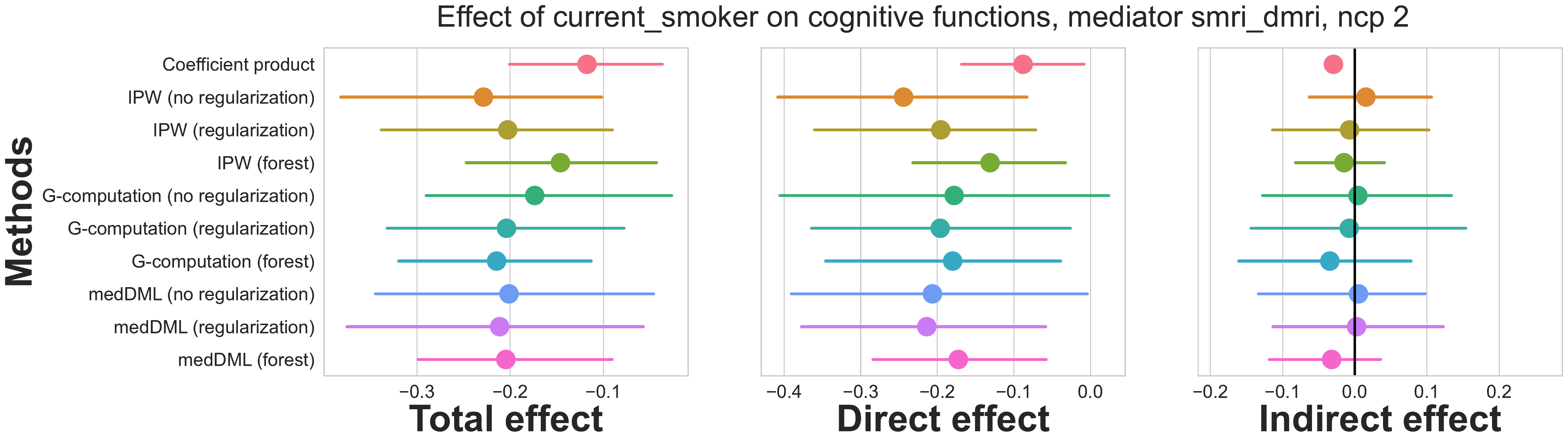}
\caption{\textbf{Causal total, direct and indirect effect of current smoking on cognitive functions, mediated through the brain structure.} The top panel represents results presented in the main text, with 5 components for the imputation of missing values by PCA for confounding variables, and the bottom panel with 2 components. We observe very similar results. We also show the estimates obtained with different estimators, and again, results are quite similar among the different approaches. Each estimate is represented as a mean over 100 bootstrapped samples. The error bar is the 95\% percentile-interval, estimated from the empirical bootstrap distribution.}
\label{diff_ncp_smoke}
\end{figure}

\begin{lstlisting}[caption={Dagitty code to reproduce the causal graph used for the mediation analysis.}, label={dagitty}]
dag {
"brain MRI" [pos="-0.218,1.511"]
"brain structure" [pos="-0.632,1.291"]
"cognitive functions" [outcome,pos="1.301,1.740"]
"college degree" [pos="-1.479,0.467"]
"current depression" [pos="0.466,0.306"]
"current smoking" [exposure,pos="-2.097,1.745"]
"drug use" [pos="-1.333,-0.761"]
"early life factors" [latent,pos="-0.856,-1.068"]
"life long depression" [pos="0.337,-0.074"]
"lifestyle factors" [latent,pos="-1.608,-0.454"]
"maternal smoking" [pos="-1.003,-1.667"]
"number of vehicles" [pos="0.631,-1.689"]
"physical activity" [pos="-2.063,-0.610"]
"screen use" [pos="-1.617,-0.903"]
"social support" [pos="0.967,0.141"]
"socio economic factors" [latent,pos="-0.027,-1.257"]
"townsend index" [pos="0.246,-1.750"]
"weight and height age 10" [pos="-1.598,-1.789"]
U [latent,pos="-2.200,-1.520"]
age [pos="0.954,-1.351"]
alcohol [pos="-1.846,0.848"]
diabetes [pos="-1.041,0.019"]
hypercholesterol [pos="-0.623,0.082"]
hypertension [pos="-0.152,0.146"]
inflammation [pos="0.863,-0.259"]
obesity [pos="1.414,0.370"]
revenue [pos="-0.203,-1.711"]
sex [pos="1.410,-1.112"]
siblings [pos="-0.592,-1.633"]
"brain structure" -> "brain MRI"
"brain structure" -> "cognitive functions"
"college degree" -> "brain structure"
"college degree" -> "cognitive functions"
"current depression" -> "brain structure"
"current depression" -> "cognitive functions"
"current smoking" -> "brain structure"
"current smoking" -> "cognitive functions"
"early life factors" -> "brain structure"
"early life factors" -> "cognitive functions"
"early life factors" -> "college degree"
"early life factors" -> "current depression"
"early life factors" -> "current smoking"
"early life factors" -> "life long depression"
"early life factors" -> "maternal smoking"
"early life factors" -> "social support"
"early life factors" -> "weight and height age 10"
"early life factors" -> alcohol
"early life factors" -> diabetes
"early life factors" -> hypercholesterol
"early life factors" -> hypertension
"early life factors" -> inflammation
"early life factors" -> obesity
"early life factors" -> siblings
"life long depression" -> "brain structure"
"life long depression" -> "cognitive functions"
"lifestyle factors" -> "brain structure"
"lifestyle factors" -> "cognitive functions"
"lifestyle factors" -> "college degree"
"lifestyle factors" -> "current depression"
"lifestyle factors" -> "current smoking"
"lifestyle factors" -> "drug use"
"lifestyle factors" -> "life long depression"
"lifestyle factors" -> "physical activity"
"lifestyle factors" -> "screen use"
"lifestyle factors" -> "social support"
"lifestyle factors" -> alcohol
"lifestyle factors" -> diabetes
"lifestyle factors" -> hypercholesterol
"lifestyle factors" -> hypertension
"lifestyle factors" -> inflammation
"lifestyle factors" -> obesity
"social support" -> "brain structure"
"social support" -> "cognitive functions"
"socio economic factors" -> "brain structure"
"socio economic factors" -> "cognitive functions"
"socio economic factors" -> "college degree"
"socio economic factors" -> "current depression"
"socio economic factors" -> "current smoking"
"socio economic factors" -> "life long depression"
"socio economic factors" -> "number of vehicles"
"socio economic factors" -> "social support"
"socio economic factors" -> "townsend index"
"socio economic factors" -> alcohol
"socio economic factors" -> diabetes
"socio economic factors" -> hypercholesterol
"socio economic factors" -> hypertension
"socio economic factors" -> inflammation
"socio economic factors" -> obesity
"socio economic factors" -> revenue
U -> "current smoking"
U -> "early life factors"
U -> "lifestyle factors"
U -> "socio economic factors"
age -> "brain structure"
age -> "cognitive functions"
age -> "college degree"
age -> "current depression"
age -> "current smoking"
age -> "life long depression"
age -> "social support"
age -> alcohol
age -> diabetes
age -> hypercholesterol
age -> hypertension
age -> inflammation
age -> obesity
alcohol -> "brain structure"
alcohol -> "cognitive functions"
diabetes -> "brain structure"
diabetes -> "cognitive functions"
hypercholesterol -> "brain structure"
hypercholesterol -> "cognitive functions"
hypertension -> "brain structure"
hypertension -> "cognitive functions"
inflammation -> "brain structure"
inflammation -> "cognitive functions"
obesity -> "brain structure"
obesity -> "cognitive functions"
sex -> "brain structure"
sex -> "cognitive functions"
sex -> "college degree"
sex -> "current depression"
sex -> "current smoking"
sex -> "life long depression"
sex -> "social support"
sex -> alcohol
sex -> diabetes
sex -> hypercholesterol
sex -> hypertension
sex -> inflammation
sex -> obesity
}
\end{lstlisting}

\end{document}